\documentclass{aa}  

\usepackage{txfonts}
\usepackage{amsmath}
\usepackage{amssymb}
\usepackage{mathrsfs}
\usepackage{courier}
\usepackage{titlesec}
\usepackage{graphicx}
\usepackage{float}
\usepackage{fancyvrb}
\usepackage{caption}
\usepackage{xcolor}
\usepackage{caption,subcaption}
\usepackage{geometry}
\usepackage{siunitx}
\usepackage{pdflscape}
\usepackage{upgreek}
\usepackage{hyperref}
\hypersetup{colorlinks=true, citecolor=blue}
\DeclareFontEncoding{LS1}{}{}
\DeclareFontSubstitution{LS1}{stix}{m}{n}
\DeclareSymbolFont{symbols4}{LS1}{stixbb}{m}{it}
\DeclareMathSymbol{\hexagonblack}   {\mathord}{symbols4}{"DE}
\DeclareMathSymbol{\pentagonblack}   {\mathord}{symbols4}{"DA}

\begin{document} 

   \title{Physical properties of strong 1 < z < 3 Balmer and Paschen line emitters observed with JWST}
   
   \titlerunning{1<z<3 galaxies with JWST}
   
   \author{L.-M. Seillé
          \inst{1}
          \and
          V. Buat\inst{1}
          \and
          V. Fern{\'a}ndez\inst{2}
          \and
          M. Boquien\inst{3}
          \and
          Y. Roehlly\inst{1}
          \and
          A. Boselli\inst{1}
          \and
          A. Calabr{\`o}\inst{4}
          \and
          R. O. Amor{\'i}n\inst{5, 6}
           \and 
           P. Arrabal Haro\inst{7}
           \and
          B. E. Backhaus\inst{8}
          \and
          M. B. Bagley\inst{9}
          \and
          D. Burgarella\inst{1}
          \and
          N. J. Cleri\inst{10, 11} 
          \and
          M. Dickinson\inst{7}
          \and
          S. L. Finkelstein\inst{9}
          \and
          N. P. Hathi\inst{12}
          \and
          B. W. Holwerda\inst{13}
          \and
          J. S. Kartaltepe\inst{14}
          \and
          A. M. Koekemoer\inst{12}
          \and
          L. Napolitano\inst{4, 15}
          \and
          F. Pacucci\inst{16, 17}
          \and
          C. Papovich\inst{10}
          \and
          N. Pirzkal\inst{18}
          \and
          C. Robertson\inst{13}
          \and
          L. Y. A. Yung\inst{12}
          }

   \institute{Aix Marseille Univ, CNRS, CNES, LAM, Marseille, France\\
              \email{lise-marie.seille@lam.fr}
              \and Michigan Institute for Data Science, University of Michigan, 500 Church Street, Ann Arbor, MI 48109, US
              \and Université Côte d'Azur, Observatoire de la Côte d'Azur, CNRS, Laboratoire Lagrange, 06000, Nice, France
              \and INAF - Osservatorio Astronomico di Roma, via Frascati 33, 00078, Monte Porzio Catone, Italy
              \and ARAID Foundation. Centro de Estudios de Física del Cosmos de Aragón (CEFCA), Unidad Asociada al CSIC, Plaza San Juan 1, E–44001 Teruel, Spain
              \and Departamento de Astronomía, Universidad de La Serena, Avda. Juan Cisternas 1200, La Serena, Chile 
              \and NSF's National Optical-Infrared Astronomy Research Laboratory, 950 N. Cherry Ave., Tucson, AZ 85719, USA
              \and Department of Physics, 196 Auditorium Road, Unit 3046, University of Connecticut, Storrs, CT 06269
              \and Department of Astronomy, The University of Texas at Austin, Austin, TX, USA
              \and Department of Physics and Astronomy, Texas A\&M University, College Station, TX, 77843-4242 USA
              \and George P. and Cynthia Woods Mitchell Institute for Fundamental Physics and Astronomy, Texas A\&M University, College Station, TX, 77843-4242 USA 
              \and Space Telescope Science Institute 3700 San Martin Drive, Baltimore, MD 21218, USA 
              \and Department of Physics, University of Louisville, Natural Science Building 102, Louisville KY 40292, USA
              \and Laboratory for Multiwavelength Astrophysics, School of Physics and Astronomy, Rochester Institute of Technology, 84 Lomb Memorial Drive, Rochester, NY 14623, USA
              \and Dipartimento di Fisica, Università di Roma Sapienza, Città Universitaria di Roma - Sapienza, Piazzale Aldo Moro, 2, 00185, Roma, Italy 
              \and Center for Astrophysics $\vert$ Harvard \& Smithsonian, Cambridge, MA 02138, USA
              \and Black Hole Initiative, Harvard University, Cambridge, MA 02138, USA
              \and ESA/AURA Space Telescope Science Institute
        }

   \date{}

 \abstract
   {The ultraviolet continuum traces young stars while the near-infrared unveils older stellar populations and dust-obscured regions. Balmer emission lines provide insights into gas properties and young stellar objects but are highly affected by dust attenuation. The near-infrared Paschen lines suffer less dust attenuation and can be used to measure star formation rates (SFRs) in star-forming regions obscured by dust clouds.} 
   {We present a new way of combining spectro-photometric data in order to test the robustness of the SFRs and stellar mass estimates of star-forming sources observed with JWST. We also aim to quantify the amount of differential attenuation between the interstellar medium and the birth clouds with the use of Paschen emission lines.}
   {We select 13 sources between redshifts 1 and 3 observed with HST, JWST/NIRCam and NIRSpec based on the availability of at least one Balmer and one Paschen line with S/N $\geq$ 5. With a newly developed version of CIGALE, we fit their hydrogen line equivalent widths (EWs) and photometric data. We assess the impacts of the removal of spectroscopic data by comparing the quality of the fits of the spectro-photometric data to those with photometric data only. We compare the single (BC03) vs binary (BPASS) stellar population models in the fitting process of spectro-photometric data. We derive the differential attenuation and explore different attenuation recipes by fitting spectro-photometric data with BC03. For each stellar model and for each input dataset (with and without EWs), we quantify the deviation on the SFRs and stellar masses from the "standard" choice. }
   {The combination of spectro-photometric data provides robust constraints on the physical properties of galaxies, with a significant reduction in the uncertainties compared to using only photometric data. On average, the SFRs are overestimated and the stellar masses are underestimated when EWs are not included as input data. We find a major contribution of the H$\alpha$ emission line to the broadband photometric measurements of our sources, and a trend of increasing contribution with specific SFR. Using the BPASS models has a significant impact on the derived SFRs and stellar masses, with SFRs being higher by an average of 0.13 dex and stellar masses being lower by an average of 0.18 dex compared to BC03. We show that a flexible attenuation recipe provides more accurate estimates of the dust attenuation parameters, especially the differential attenuation. 
   Finally, we reconstruct the total effective attenuation curves of the most dust-obscured galaxies in our sample. }
   {}


\maketitle
%
\section{Introduction}

The ultraviolet (UV) to near-infrared (NIR)  spectral energy distribution enables the characterisation of stellar age distributions and the assessment of star formation histories (SFHs) of galaxies \citep[e.g.][]{salim05, salim07}. The UV continuum is a tracer of young stars, while the infrared (IR) unveils the older stellar populations and dust-obscured regions within galaxies. Dust grains preferentially absorb shorter-wavelength light, leading to dimming effects that are prominent in the UV and optical regions. In contrast, the NIR is much less affected by dust attenuation making it essential to derive unbiased star formation rates (SFRs) and stellar masses \citep[e.g.][]{calzetti, leja}. Emission lines, such as those arising from ionised hydrogen, provide crucial insights into gas properties and very young ($\leq$ 10 Myrs) stellar objects \citep{kennicutt98, boselli09}. A wide spectral coverage of both continuum and emission line features thus enables robust constraints on SFRs, gas metallicity, and ionisation conditions \citep{shivaei, reddy}. 

Optical and NIR emission lines, driven primarily by stars with lifetimes of 3 to 10 Myrs serve as robust tracers of SFRs \citep{kennicutt09}. Notably, recombination lines of hydrogen, such as the Balmer series, are very valuable tools for tracing star formation activities due to their independence on the exact physical conditions of the gas \citep{osterbrock}. However, these diagnostic lines are affected by dust attenuation. A correction can be made by using the Balmer decrement (H$\alpha$/H$\beta$) but this method is hindered by the spatial resolution of the emission and attenuation \citep{kennicutt, keel, robertson} as the latter may not be uniform over the emission region and high optical depths to Balmer emission. 


Paschen lines, which are found in the NIR and are thus suffering less dust attenuation than the optical Balmer lines, can also be used to measure SFRs in star-forming regions obscured by dust clouds \citep{Alonso,calzetti07,kessler, cleri}. When combined with Balmer lines, they can also be used to constrain nebular reddening and also provide insights into the complex interplay between stars, gas, and dust within galaxies \citep{calzetti,prescott,reddy}.

For the last two decades, astronomers gathered data from ground-based multi-object NIR spectrographs, such as Keck/MOSFIRE and VLT/KMOS, and space-based facilities like the HST/WFC3 grism, which provided extensive measurements of hydrogen recombination lines and nebular reddening for thousands of galaxies up to a redshift of z $\sim$ 2.6 \citep[e.g.][]{kashino, reddy15, shivaei, rezaee21, battisti22}. While using these instruments, only a handful of Paschen lines have been robustly detected in z > 2 lensed galaxies \citep{papovich, finkelstein11}. With the advent of the \emph{James Webb} Space Telescope \citep[JWST, ][]{gardner, gardner_23} equipped with its NIRSpec instrument, Paschen lines are now becoming accessible for intermediate redshift sources ($Pa\alpha$ is detectable until z $\sim$ 1.7 and $Pa\beta$ until z $\sim$ 3).

The comparison of the combination of stellar, nebular and dust models predicting the full spectral energy distribution (SED), with observed broadband fluxes from both continuum and line emissions  has proven to be highly effective in deducing physical parameters within populations of star-forming galaxies \citep{boselli16, fossati, buat18, villa, tacchella, larson, Arrabal_b}. Even so, the addition of spectroscopic data implies a careful consideration in the choice of models and their impact on the output parameters. 

Combining nebular and stellar emissions also requires to account for a different attenuation recipe for both the continuum and the line emission. This differential attenuation can be effectively modelled with the theoretical framework of \cite{charlot} as it allows for two independent attenuation recipes, one for the interstellar medium and one for the current birth clouds. 


Emission line fluxes often suffer from slit losses which needs to be addressed by either using the photometric data on the same spatial aperture or estimating the emission line fluxes beyond the slit through other methods. In this work, we propose to fit simultaneously photometric broadband fluxes and emission line equivalent widths (EWs) to get around the issue posed by slit losses. \citet{boselli16} has shown the feasibility and robustness of this spectro-photometric combination. We develop a new method and apply it on a sample of 13 galaxies observed by JWST in the Cosmic Evolution Early Release Science (CEERS) field. These 1 < z < 3 sources are selected to have at least one Balmer line and one Paschen line detected as well as ancillary data from the Hubble Space Telescope (HST) to cover the rest-frame UV range. We assess the feasibility and reliability of combining spectro-photometric data to derive parameters such as the SFR, stellar mass and dust attenuation parameters. We quantify the impact of including the binary stellar populations from the BPASS models \citep{eldridge, stanway} into our models on the aforementioned output parameters. We compare the quality of the same fit done on spectro-photometric and photometric data only to quantify the uncertainty on the SFR and stellar mass in the absence of spectroscopic input data. Finally, we measure the impact of using a flexible attenuation recipe inspired by the original  \cite{charlot} (CF00) model on the SED fitting output parameters.

Our paper is organised as follows. The selection of the sample used for this work is presented in section \ref{sec:data}. The theoretical framework linking line ratios and nebular dust reddening is outlined in Section \ref{sec:line-ratios}. Section \ref{sec:SED-fitting} is dedicated to our SED fitting process. We present our results on the impact of the choice of data and models in Section \ref{sec:impact-data}. The results pertaining to attenuation are shown in Section \ref{sec:attenuation}. Finally, we discuss the impact of the choice of models of stellar populations, input data and attenuation recipes in Section \ref{sec:impact-models}. We adopt a cosmology with H$_0$=70~km/s/Mpc, $\Omega_\Lambda$=0.7, and $\Omega_m$=0.3.

\section{Data and sample\label{sec:data}}

The data for this work are taken from the Cosmic Evolution Early Release Science Survey (CEERS; ERS 1345, PI: S. Finkelstein) in the CANDELS survey \citep{grogin, koekemoer} of the Extended Groth (EGS, \cite{davis_07, noeske}) field. The main features of the CEERS program are presented in \cite{Arrabal} and \cite{finkelstein}, and will be described in more detail in Finkelstein et al. (in prep.). For each object in this study, we need spectroscopic data to trace the hydrogen line emission of the gas ionised by the very young stars and as much photometric data as we can gather to trace the stellar continuum flux.

\subsection{Spectroscopy\label{ssec:spectroscopy}}

\begin{figure}
\centering
    \includegraphics[height=7cm, width=9 cm]{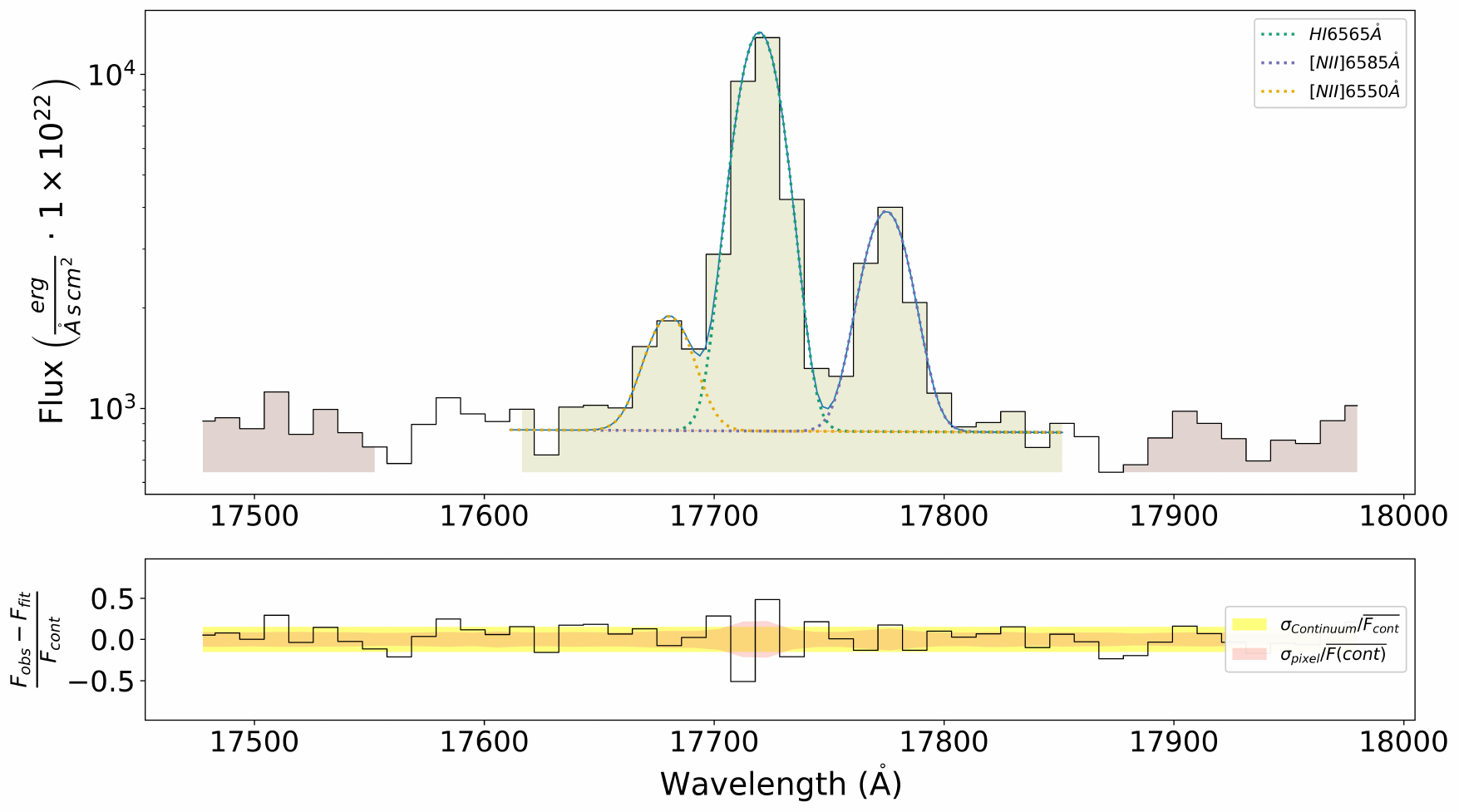}
      \caption{Example of the \textsc{LiMe} fitting process on the H$\alpha$ + [NII] doublet of ID 6563 (z $\sim$ 1.7). The top figure shows the Gaussian profile fittings while the lower figure shows the residual between the observation and the theoretical fits. The red shade in the bottom plot shows the calibration pixel error while the yellow shade shows the standard deviation from the continua bands. The gap between the red and yellow regions is intended to avoid potential contamination of uncertain line wings and artefacts.}
      \label{lime}
\end{figure}

The galaxies studied in this work have been observed with JWST/NIRSpec \citep{jakobsen} multi-object shutter configurations. These were taken with the Micro Shutter Array (MSA; \cite{ferruit}) of size 0.2" $\times$ 0.46" during the CEERS epoch 2 observations in December 2022.

These NIRSpec observations are split into six different MSA pointings, each of them observed with the G140M/F100LP, G235M/F170LP and G395M/F290LP medium resolution (R = 1000) gratings. The total wavelength coverage of the NIRSpec instrument is from $\sim$ 1 to $\sim$ 5.2 $\mu$m. The MSA is configured to use three shutter slitlets thus enabling a three point nodding pattern, shifting the pointing by a shutter length plus the size of the bar between the shutters in each direction for background subtraction. The total exposure time per grating is 3107s for each MSA pointing.

The spectral reduction of NIRSpec observations is described in \cite{Arrabal_b} and will be further explained in Arrabal Haro et al. (in prep.). Here, we choose to focus on the main processing steps based on the STScI Calibration Pipeline \footnote{\url{https://jwst-pipeline.readthedocs.io}} and on the Calibration Reference Data System (CRDS) mapping 1061.

The first step is to correct the raw images for the detector 1/f noise, dark current, bias and "snowball" contaminations (trails produced by cosmic rays). Count-rate maps are then used to create two-dimensional (2D) spectra for each source. 

The second step is to subtract the background by using the three point nodding pattern, apply the flat-field correction, and calibrate the photometry and wavelength of these spectra. They are then resampled in case of spectral trace distortions. The multiple background subtracted 2D spectra (one for each nod position) are then combined to create the final 2D spectrum of the source. 

A post-combination inspection check is done on the reduced 2D spectra in which hot pixels, detector gaps and reduction artefacts are masked. The noise spectrum is automatically calculated by the JWST pipeline and then corrected as described in \cite{Arrabal_b} to take into account the spectral resampling. In the pipeline a slit loss correction is performed by default in the pathloss step.

The line fluxes and EWs are measured using the beta release of \textsc{LiMe} \citep[][accepted for publication in A\&A]{fernandez, fernandez_24}. The fluxes are measured assuming a Gaussian distribution and taking into consideration the pixel uncertainty produced by the NIRSpec reduction described in \cite{Arrabal_b} for the CEERS 0.7 data release. The reported flux uncertainty comes from the standard error definition of the least-squares trust-region method. This uncertainty is propagated to the equivalent width calculation. In this computation the linear continuum flux and its standard deviation are derived from two adjacent continua bands (see Fig \ref{lime} for an example of line fitting). 

These continua regions are manually adjusted for every line to guarantee a representative linear continuum measurement and to avoid artefacts. We take special care to confirm that the pixel uncertainty was of the same order than the flux standard deviation from the adjacent continua. The \textsc{LiMe} fittings for the 0.7 release spectra used in this work can be found at the interactive repository \footnote{\url{https://ceers-data.streamlit.app/}} using the corresponding NIRSpec MSA IDs. This site will be updated as the JWST pipeline and flux calibration keep improving. Additionally, this manuscript line measurements can be found on the supplementary online material as a \emph{FITS} file, where each page contains the measurements from one galaxy. The complete measurements description can be found in the online documentation. \footnote{\url{https://lime-stable.readthedocs.io/en/latest/index.html}} 

\subsection{Photometry}

\begin{table}
\caption{Available photometric bands}
\centering
\begin{tabular}{|c|c|c|c|c|c|}
\hline
  \multicolumn{1}{|c|}{ID} &
   \multicolumn{1}{c|}{IRAC 3}&
   \multicolumn{1}{c|}{IRAC 4}&
   \multicolumn{1}{c|}{MIPS 1}&
   \multicolumn{1}{c|}{SCUBA 450}&
   \multicolumn{1}{c|}{SCUBA 850}\\
\hline
23542 & \checkmark & \checkmark & \checkmark & $\times$ & $\times$ \\
8736 & \checkmark & \checkmark & \checkmark & $\times$ & $\times$ \\
8515 & $\times$ & \checkmark & $\times$ & $\times$ & $\times$ \\
5430 & \checkmark & \checkmark & \checkmark & \checkmark & $\times$ \\
10293 & \checkmark & $\times$ & $\times$ & $\times$ & $\times$ \\
6563 & \checkmark & \checkmark & \checkmark & \checkmark & $\times$ \\
5409 & $\times$ & \checkmark & \checkmark & $\times$ & $\times$ \\
5300 & $\times$ & $\times$ & $\times$ & $\times$ & $\times$\\
3788 & $\times$ & $\times$ & \checkmark & \checkmark & $\times$ \\
8588 & \checkmark & $\times$ & \checkmark & $\times$ & \checkmark \\
8710 & $\times$ & $\times$ & $\times$ & $\times$ & $\times$ \\
16991 & $\times$ & $\times$ & $\times$ & $\times$ & $\times$\\
18294 & $\times$ & $\times$ & $\times$ & $\times$ & $\times$\\
\hline
\end{tabular}
\tablefoot{All galaxies are detected in CFHT u, g, HST F606W, F814W and all NIRCam filters.}
\label{sample_phot}
\end{table}

Our sample galaxies were also observed with JWST/Near Infrared Camera \citep[NIRCam,][]{rieke_m} in December 2022. The ten pointings and the reduction steps are described in \cite{bagley}. We adopt the multi-band SExtractor catalog described in \cite{finkelstein}. This catalog includes all available HST/ACS and WFC3 data (v1.9) and all seven JWST/NIRCam bands (v0.51).

Data from the Canada–France–Hawaii Telescope (CFHT)/Megacam and from the \emph{Spitzer}/InfraRed Array Camera (IRAC) channels are taken from the multi-wavelength photometric catalog of the EGS field of \cite{stefanon}.

Fluxes from the Multiband Imaging Photometer for \emph{Spitzer}/MIPS, 24$\mu$m band, the 450 and 850 $\mu$m \emph{James Clerk Maxwell} Telescope (JCMT)/SCUBA-2 bands
are retrieved from the Super-deblended Catalog of the EGS field of \cite[][accepted for publication in A\&A ]{lebail}.

In cases where two data points are overlapping due to their wavelength proximity (CFHT/Megacam r with HST F606W, CFHT/Megacam i and z with HST F814W, NIRCam F115W with HST F125W, NIRCam F150W with HST F140W and F160W, NIRCam F356W with \emph{Spitzer}/IRAC 1 and NIRCam F444W with \emph{Spitzer}/IRAC 2), only the flux with the best signal to noise ratio (S/N), (i.e. HST over CFHT, NIRCam over HST and IRAC) is kept in the SED fitting process. 

The final photometry catalog includes measurements over the full CEERS NIRCam wavelength range in the F115W, F150W, F200W, F277W, F356W, F410M, and F444W filters, HST/CANDELS ACS\_WFC3 F606W and F814W bands, CFHT/Megacam u* and $g^{\prime}$, \emph{Spitzer}/IRAC channels 3 and 4, \emph{Spitzer}/MIPS 24$\mu$m band, JCMT/SCUBA-2 450 and 850 $\mu$m bands.
In the catalog, we only keep values with a S/N $\geq 2$ for CFHT, HST, IRAC and MIPS and with a S/N $\geq 1$ for PACS and SCUBA-2.

\subsection{Sample selection}

Galaxies included in our study are selected based on several criteria. The first criterion is to have data from NIRCam and NIRSpec gratings since the spatial coverage of both instruments does not completely overlap (this gives us 126 sources). As we also combine NIRCam data with photometry obtained by other facilities, we need all of them to be in agreement. 
The overlaps between HST/NIRCam are checked to only keep galaxies for which fluxes measured in the same aperture of nearby bands (HST/F160W and NIRCam/F150W) were within $1\sigma$ (122 objects remain at this stage). We present the photometric data used in the study in Table \ref{sample_phot}.

Balmer and Paschen lines are necessary as they probe stars located in vastly different environments so we require our objects to have at least one Balmer and one Paschen line with an equivalent width S/N $\geq 5$ (other Balmer and Paschen lines with a S/N $\geq 3$ are also added, see Table \ref{sample_lines}). Getting the lines to compute the ratio of a Paschen line to a Balmer line (Pa/H hereafter) described in Section \ref{ssec:line-ratios} restricts our sample to the redshift range 1 < z < 3. All CEERS spectra are inspected to identify and secure a spectroscopic redshift between 1 and 3 which amounts to 40 unique sources. The requirement of a S/N $\geq 5$ for at least one Balmer line and one Paschen line further restricts our sample down to 17 objects. 

Finally, we want to avoid galaxies with an active galactic nucleus (AGN) as their presence would require much more complex models beyond the scope of this paper. To check if our sources could host AGNs, we compute the ratio of the [NII]-6585 over H$\alpha$ if the [NII]-6585 line is detected (a non-detection of the [NII]-6585 line is indicated by a cross in Table \ref{sample_lines}). [NII]-6585/H$\alpha$ is a linear function of the nebular metallicity and it presents a saturation point so a further increase in the [NII]-6585/H$\alpha$ value is only due to AGN contribution \citep{stasinska}. Following the authors' prescription, we only keep galaxies with log([NII]-6585/H$\alpha$) < -0.2 (four objects were dropped at this stage). Some sources (IDs 8515, 8588 and 8710) are considered "hidden AGNs" by \cite{calabro23} following their diagnostics based on IR emission lines. These AGNs diagnostics are BPT-like diagrams \citep{baldwin} of ratios of [CI]-9850, [PII] 1.188$\mu$m or [FeII] 1.257$\mu$m to Paschen lines (either Pa$\beta$ or Pa$\gamma$) versus a ratio of [SIII]-9530 to Pa$\beta$ or Pa$\gamma$. As we do not detect [PII] or [CI] in these three galaxies and two of them have a [FeII] detection with a S/N < 3 $\sigma$ ([FeII] is not detected for 8588), we decide to still include them as their exact nature as AGNs still needs to be confirmed. Using all of these criteria, we end up with 13 galaxies in our sample as requiring an observation with NIRCam and the NIRSpec gratings drastically reduces the number of available sources.

\begin{table*}
\caption{Characteristics of the objects}
\centering
\begin{tabular}{|c|c|c|c|c|}
\hline
  \multicolumn{1}{|c|}{ID} &
  \multicolumn{1}{c|}{Redshift} &
  \multicolumn{1}{c|}{log NII(6585)/H$\alpha$} &
  \multicolumn{1}{c|}{Fitted emission line EWs} & 
  \multicolumn{1}{c|}{Symbol}\\
\hline
23542 & 1.277 & -0.78 & 
H$\alpha$, H$\beta$, Pa$\alpha$, Pa$\beta$, Pa$\gamma$ & \Large $\star$\\
8736 & 1.553 & -0.66 & 
H$\alpha$, H$\beta$, \textcolor{red}{H$\gamma$}, Pa$\alpha$, \textcolor{red}{Pa$\beta$} & \Large $\blacktriangle$\\
8515 & 1.567 & -0.77 & 
H$\alpha$, H$\beta$, H$\gamma$, H$\delta$, Pa$\alpha$, Pa$\beta$, \textcolor{red}{Pa$\gamma$} & \Large $\blacktriangleleft$\\
5430 & 1.676 & -0.31 & 
H$\alpha$, H$\beta$, Pa$\beta$ , \textcolor{red}{Pa$\gamma$} & \\
10293 & 1.676 & -0.53 & 
H$\alpha$, H$\beta$, Pa$\alpha$ & \Large $\blacktriangleright$\\
6563 & 1.699 & -0.63 & 
H$\alpha$, H$\beta$,  H$\gamma$, Pa$\alpha$, Pa$\beta$, Pa$\gamma$ & \Large $\blacktriangledown$\\
5409 & 1.699 & $\times$ & 
H$\alpha$, H$\beta$, Pa$\alpha$, Pa$\beta$ & \Large $\medbullet$\\
5300 & 2.136 & $\times$ & 
H$\alpha$, H$\beta$, H$\gamma$, H$\delta$, Pa$\beta$, Pa$\gamma$ & \Large $\blacksquare$\\ 
3788 & 2.295 & -1.23 & 
H$\alpha$, H$\beta$, H$\gamma$, H$\delta$, Pa$\beta$, Pa$\gamma$ & $\hexagonblack$\\
8588 & 2.336 & -0.35 & 
H$\alpha$, H$\beta$, \textcolor{red}{H$\gamma$}, Pa$\beta$, Pa$\gamma$ & \Large $\times$\\
8710 & 2.337 & -1.05 & 
H$\alpha$, H$\beta$, H$\gamma$, Pa$\gamma$ & \\
16991 & 2.540 & -1.28 & 
H$\alpha$, \textcolor{red}{H$\beta$}, H$\delta$, \textcolor{red}{Pa$\beta$}, Pa$\gamma$ & \\
18294 & 2.635 & \textcolor{red}{-1.05} & 
H$\alpha$, H$\beta$, \textcolor{red}{H$\gamma$}, Pa$\beta$ & \Large $\blacklozenge$\\
\hline\end{tabular}
\tablefoot{The spectroscopic redshifts are the mean of the estimated spectroscopic redshift of each line. Crosses indicate a non detection of the [NII]-6585 line. Emission line EWs used in the fits for each source are indicated with those having a S/N $\leq 5$ in red. A symbol is assigned to each object appearing in Fig \ref{flux_ratios}.}
\label{sample_lines}
\end{table*}

\section{Line ratios as an indicator of differential attenuation\label{sec:line-ratios}}


\begin{figure}
\centering
    \includegraphics[width=\columnwidth]{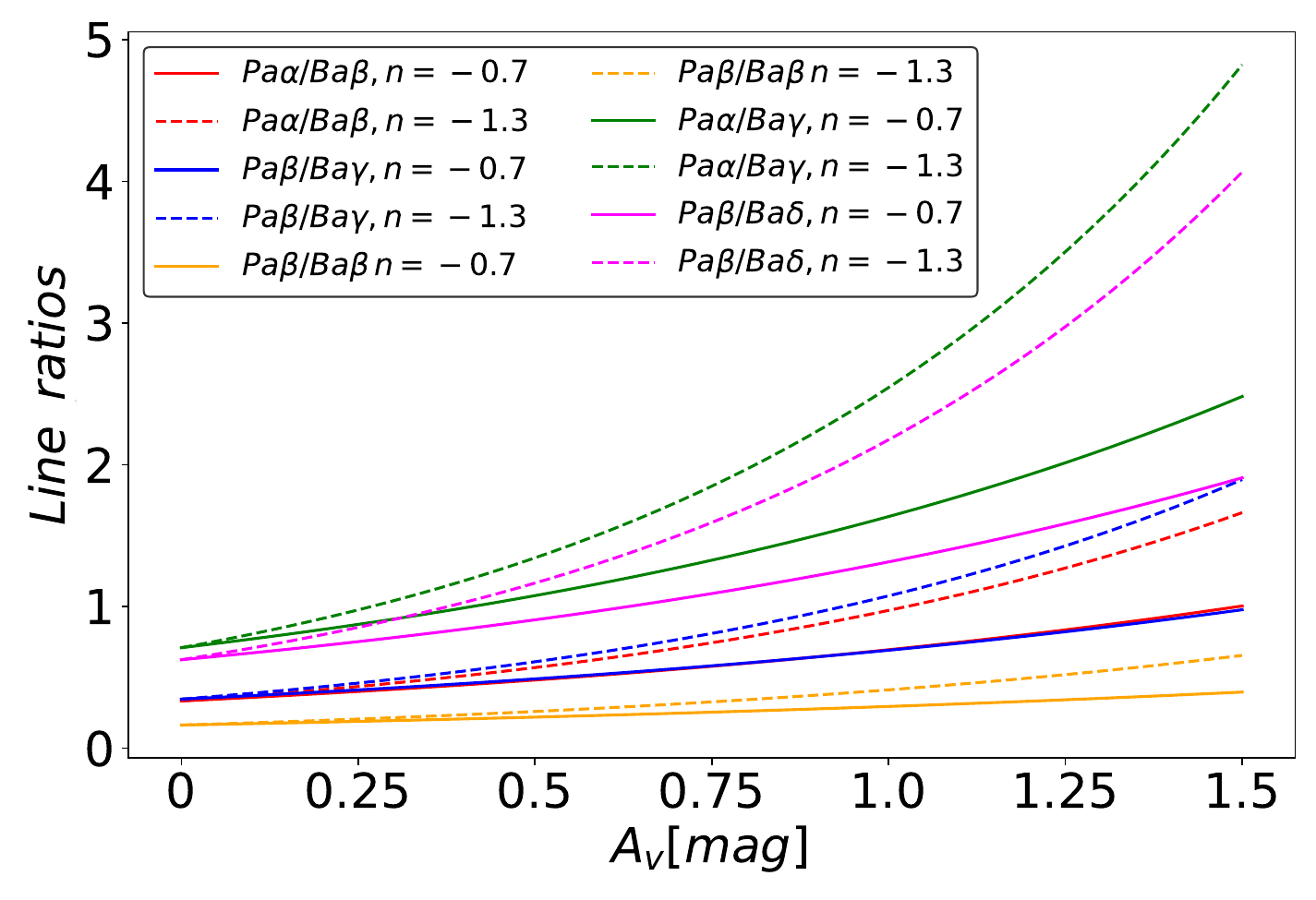}
      \caption{Theoretical line ratios for a case B recombination computed following the BCs component of the \cite{charlot} attenuation recipe and varying the exponents presented in Eq \ref{eq:AvBC}. $n^{\mathrm{BC}}$ = -0.7 corresponds to a shallower curve (solid lines) while $n^{\mathrm{BC}}$ = -1.3 corresponds to a steeper curve (dashed lines).}
      \label{theo_ratios}
\end{figure}

\subsection{Theoretical line ratios\label{ssec:line-ratios}}

Within our sample, we are able to get multiple Balmer and Paschen lines. In this section, we compute theoretical ratios  of lines, i.e. ratios expected for a case B recombination and a temperature of 10 000 $K$ \citep{osterbrock}, as well as model the reddening of the stellar emission following a CF00 attenuation recipe. The key feature of this model is the computation of two complementary recipes, one for the birth clouds (BCs) and one for the interstellar medium (ISM) In this way, we are able to properly account for an age-dependent attenuation where not only the total amount of dust attenuation changes as a function of the stellar age but also the way the stellar light is obscured, as originally introduced by CF00.

In this model, young stars up to an age of 10 Myrs are surrounded by BCs which progressively dissipate over time. Dust attenuation works differentially, with an efficient attenuation in the BCs combined with a lower attenuation in the surrounding ISM. The double component attenuation CF00 recipe is critical as it means that the emission lines and UV light emitted by young stars are attenuated by both the BCs and the ISM. However, once stars have migrated outside the BCs, the emitting radiation is subject only to the diffuse ISM attenuation. Throughout this study, we refer to the results or models of dust attenuation as "recipes" and not "curves". Indeed, the latter would suggest that a unique or universal result exists while the real effective result depends on the SFH and attenuation parameters of each individual galaxy. 

We utilise the CF00 model in which two different power-law attenuation recipes are used to compute the total attenuation, $A_\uplambda$:  

\begin{eqnarray}
  A_\uplambda^{\mathrm{BC}} & = &A_\mathrm{V}^{\mathrm{BC}} (\uplambda/0.55)^{n^{\mathrm{BC}}}, \label{eq:AvBC}\\
  A_\uplambda^{\mathrm{ISM}} & = &A_\mathrm{V}^{\mathrm{ISM}} (\uplambda/0.55)^{n^{\mathrm{ISM}}},\label{eq:AvISM}\\
  A_\uplambda & = & A_\uplambda^{\mathrm{BC}} + A_\uplambda^{\mathrm{ISM}}\label{eq:Av}
\end{eqnarray}
where $n^{\mathrm{ISM}}$ is the slope of the attenuation curve for the ISM and $n^{\mathrm{BC}}$ the slope of the attenuation curve for the birth clouds. Both power-law exponents are fixed to -0.7 in the original recipe of CF00. However, these parameters have been shown to vary among galaxies \citep[e.g.][]{buat12, chevallard, kriek, battisti, lofaro, salim18, trayford, pantoni, boquien22} so we decide to keep these exponents as free parameters. The attenuation for the birth clouds, $A_\mathrm{V}^{\mathrm{BC}}$, is computed using the $\mu$ parameter:

\begin{equation}
    \mu = \frac{A_\mathrm{V}^{\mathrm{ISM}}}{A_\mathrm{V}^{\mathrm{ISM}}+A_\mathrm{V}^{\mathrm{BC}}} \label{eq:mu}
\end{equation}

In the absence of dust, the flux ratios of the different recombination lines are fixed, for a given temperature and electron density. In this case, comparing the observed line ratios with the intrinsic ones allows us to infer the reddening coming from the nebular regions.

We use the Balmer decrement (H$\alpha$/H$\beta$) as it is available for all our sources with fluxes measured with a high S/N. For the Pa/H lines ratio, we have many different combinations available thanks to the large wavelength coverage of the three gratings. We check which combination gives us the best chance at discriminating between two different exponents for a power law which follows Eq \ref{eq:AvBC} of the CF00 recipe by plotting the Pa/H ratios at our disposal in Fig \ref{theo_ratios}. The two ratios with the widest gap are Pa$\alpha$/H$\gamma$ and Pa$\beta$/H$\delta$, they are however, only available for a few sources in our sample (see Table \ref{sample_lines}). Because of this issue, we settle for the two next best ratios: Pa$\alpha$/H$\beta$ and Pa$\beta$/H$\gamma$ (in cases where both ratios are available, both will be included in the analysis). Finally we note that the Pa$\beta$/H$\beta$ ratio needs very high levels of attenuation to start being discriminant between both exponents, so we are not using this ratio in this section.

\subsection{Analysis of observed line ratios\label{ssec:analysis-line-ratios}}

We want to ascertain if some objects in our sample have hidden attenuation before going further into our analysis. Indeed, a galaxy with a high Pa/H ratio compared to a Balmer decrement close to 2.86 would indicate a hidden and highly attenuated component. To do this, we compare the observed line fluxes ratios to the theoretical ones. However, this comparison comes with a few caveats, the first one being slit losses. Following \cite[][submitted to A\&A]{napolitano} we assume that the slit losses affect all lines equally. The second issue comes from the underlying absorption lines of the stellar continuum, especially for H$\beta$, H$\gamma$ and H$\delta$ which have an absorption line EW $\sim$ 5\AA (see Table \ref{absorption}). We correct the fluxes for these absorptions by computing the absorbed flux from the best model of CIGALE (Sect.\ref{ssec:absorption lines}).   

\begin{table*}
\caption{Absorption line corrections in Angstroms}
\centering
\begin{tabular}{|c|c|c|c|c|c|c|c|}
\hline
  \multicolumn{1}{|c|}{Object ID} &
  \multicolumn{1}{c|}{Pa$\alpha$}&
  \multicolumn{1}{c|}{Pa$\beta$}&
   \multicolumn{1}{c|}{Pa$\gamma$}&
  \multicolumn{1}{c|}{H$\alpha$} &
  \multicolumn{1}{c|}{H$\beta$} &
  \multicolumn{1}{c|}{H$\gamma$} &
  \multicolumn{1}{c|}{H$\delta$} \\
\hline
23542 & 0 & 0 & 0 & 2.5 & 5.0 & 6.5 & 7.0 \\
8736 & 0 & 2.0 & 2.0 & 2.0 & 4.5 & 5.0 & 7.0 \\
8515 & 0 & 2.5 & 2.0 & 2.0 & 6.0 & 6.5 & 7.0 \\
5430 &  0 & 3.0 & 2.0 & 2.5 & 5.0 & 6.0 & 7.0 \\
10293 & 0 & 0 & 0 & 3.0 & 5.0 & 5.5 & 6.5 \\
6563 & 2.0 & 3.0 & 3.0 & 3.0 & 6.0 & 5.5 & 6.0 \\
5409 & 0 & 0 & 0 & 4.0 & 6.5 & 5.5 & 7.0 \\
5300 & 0 & 0 & 0 & 1.5 & 4.5 & 4.0 & 4.5 \\
3788 & 0 & 1.0 & 2.0 & 1.5 & 4.0 & 4.5 & 5.0 \\
8588 & 1.0 & 3.0 & 2.5 & 2.5 & 6.0 & 6.0 & 7.0\\
8710 & 0 & 0 & 0 & 1.5 & 4.5 & 4.0 & 7.0\\
16991 & 0 & 0 & 0 & 2.0 & 5.0 & 5.0 & 5.5 \\
18294 & 0 & 2.0 & 1.0 & 2.0 & 6.5 & 6.5 & 7.0 \\
Average & 0.20 (0.08) & 1.30 (0.23) & 1.10 (0.19) &  2.30 (0.27)&  5.30 (0.27) &  5.40 (0.28) & 6.40 (0.28) \\
\hline\end{tabular}
\tablefoot{The absorption line EWs measurements are rounded to the upper half Angstrom and have an uncertainty of 0.5 $\AA$.}
\label{absorption}
\end{table*}

\begin{figure*}[!htbp]
\includegraphics[width=0.5\textwidth]{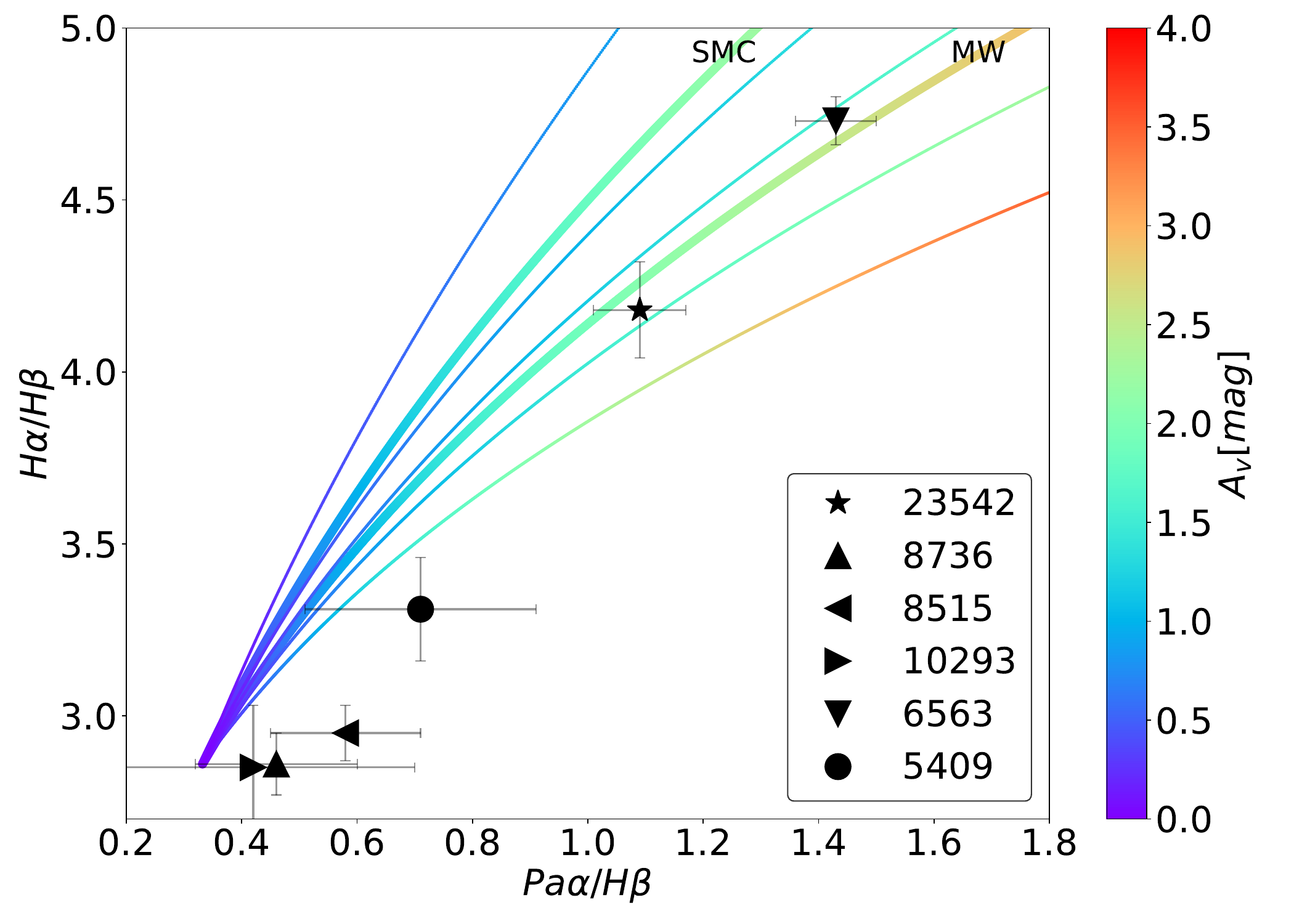}
\includegraphics[width=0.5\textwidth]{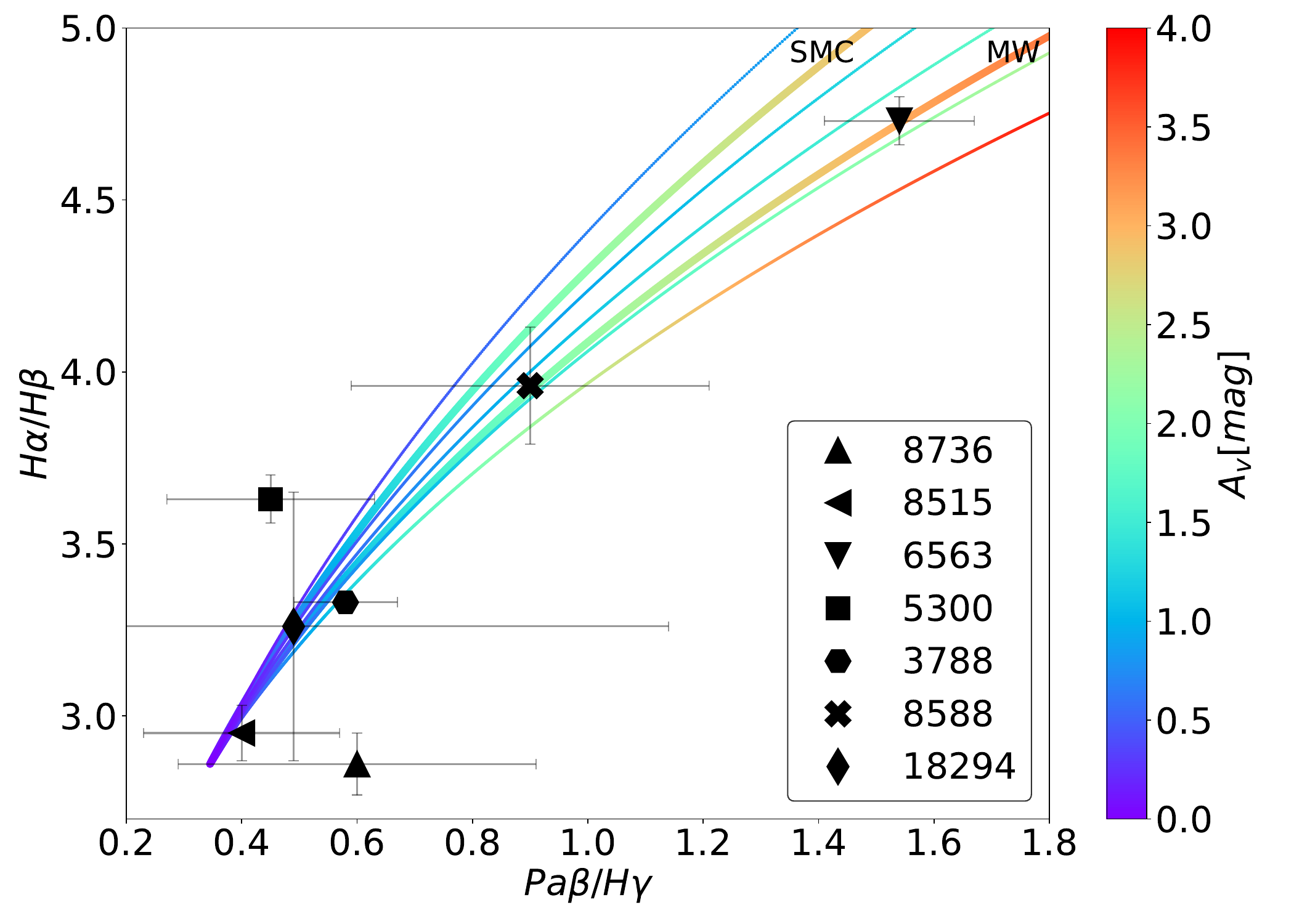}
\caption{Observed fluxes ratios corrected for underlying stellar absorption. The thin coloured curves are power laws following the CF00 framework with varying exponents, from top to bottom: -2.0, -1.3, -1.0, -0.7, -0.4.}
\label{flux_ratios}
\end{figure*}

We plot H$\alpha$/H$\beta$ against Pa$\alpha$/H$\beta$ or Pa$\beta$/H$\gamma$, depending on the redshift and availability of the lines along with the theoretical predictions for each model along with the evolution of the attenuation in Fig \ref{flux_ratios}. For reference, we also add a Milky Way \citep{cardelli} and a Small Magellanic Cloud \citep{gordon} attenuation recipe on top of several power laws with varying exponents. 

We find that our objects encounter at most a very moderate hidden attenuation, a result comparable to what \cite{reddy} found for a sample of similar redshift CEERS sources. Taking into account the uncertainties on the flux measurements, our galaxies have either a very low attenuation or are compatible with the attenuation model presented in Sect \ref{ssec:line-ratios}. In the left panel of Fig \ref{flux_ratios}, four objects are located to the right of the model lines (IDs 8736, 8515, 10293 and 5409). The H$\alpha$/H$\beta$ of IDs 8736 and 10293 shows that both objects are consistent with the models and their Pa/H ratio is compatible with this result within less than 1$\sigma$ (0.8 and 0.5 respectively). ID 5409 has a large uncertainty on its Pa/H ratio which also makes this galaxy compatible with the models within 1 $\sigma$. Finally, the Pa$\alpha$/H$\beta$ ratio of ID 8515 is compatible with the models within 2 $\sigma$ while its other Pa/H ratio, Pa$\beta$/H$\gamma$, is compatible with the models and a very low amount of attenuation within 0.1 $\sigma$ (see the right panel of Fig \ref{flux_ratios}). From Fig \ref{flux_ratios}, we conclude that the modified CF00 recipe is able to reproduce the attenuation processes at work in obscured galaxies. 


\begin{figure*}
\centering
\begin{subfigure}{.24\linewidth}
  \centering
  \includegraphics[width=\textwidth]{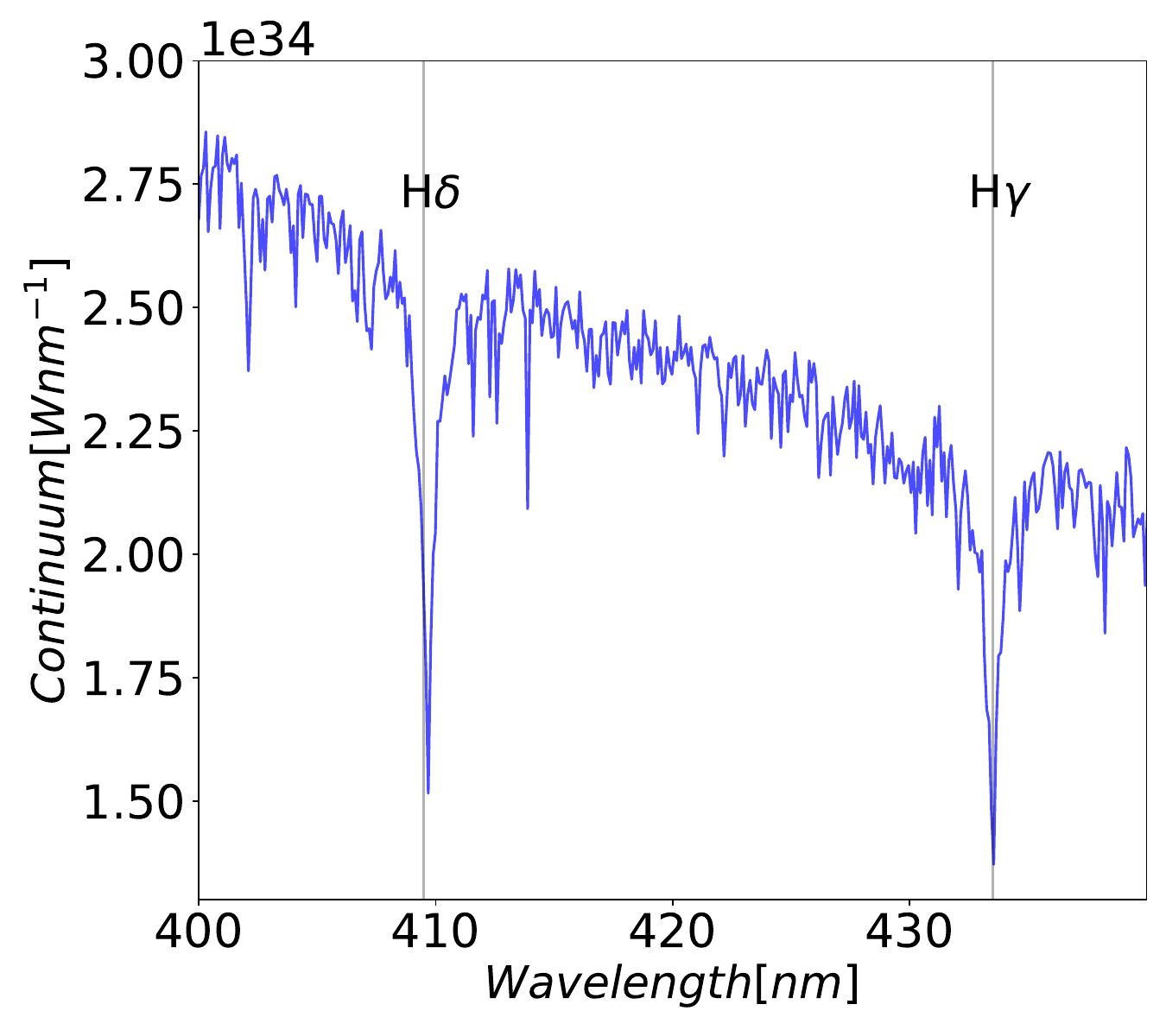}
\end{subfigure}
\begin{subfigure}{.24\linewidth}
  \centering
  \includegraphics[width=\textwidth]{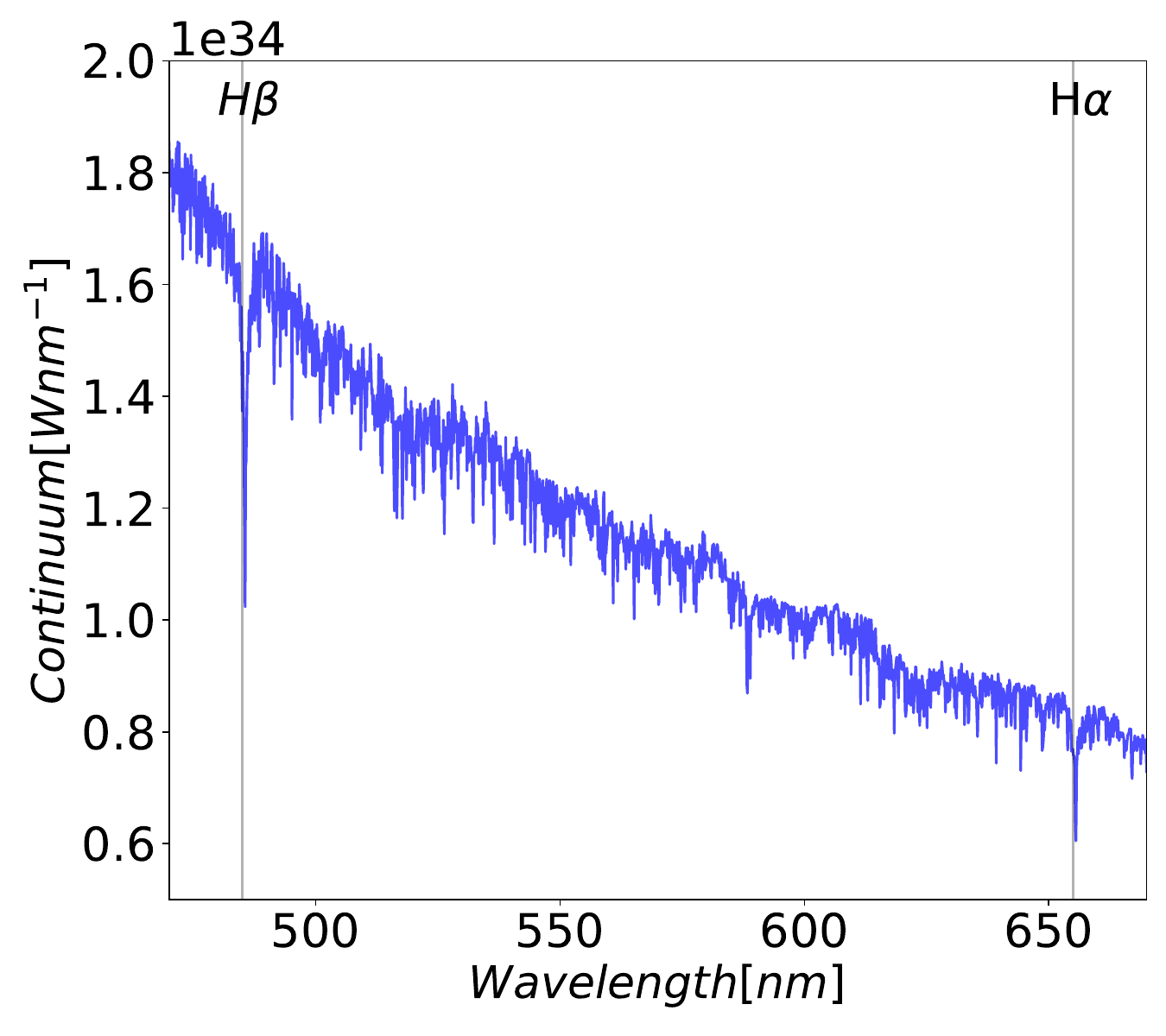}
\end{subfigure}
\begin{subfigure}{.24\linewidth}
  \centering
  \includegraphics[width=\textwidth]{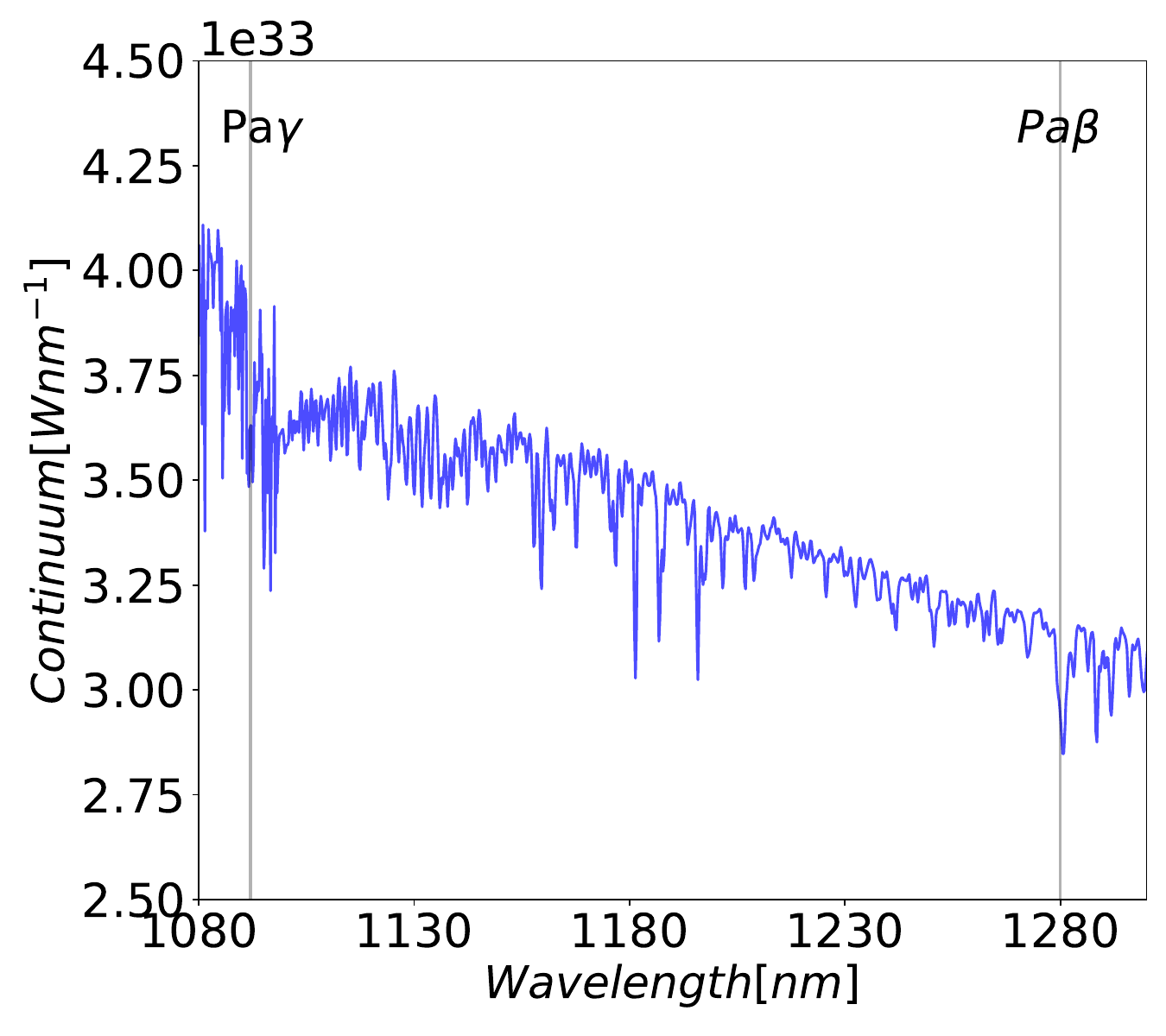}
\end{subfigure}
\begin{subfigure}{.24\linewidth}
  \centering
  \includegraphics[width=\textwidth]{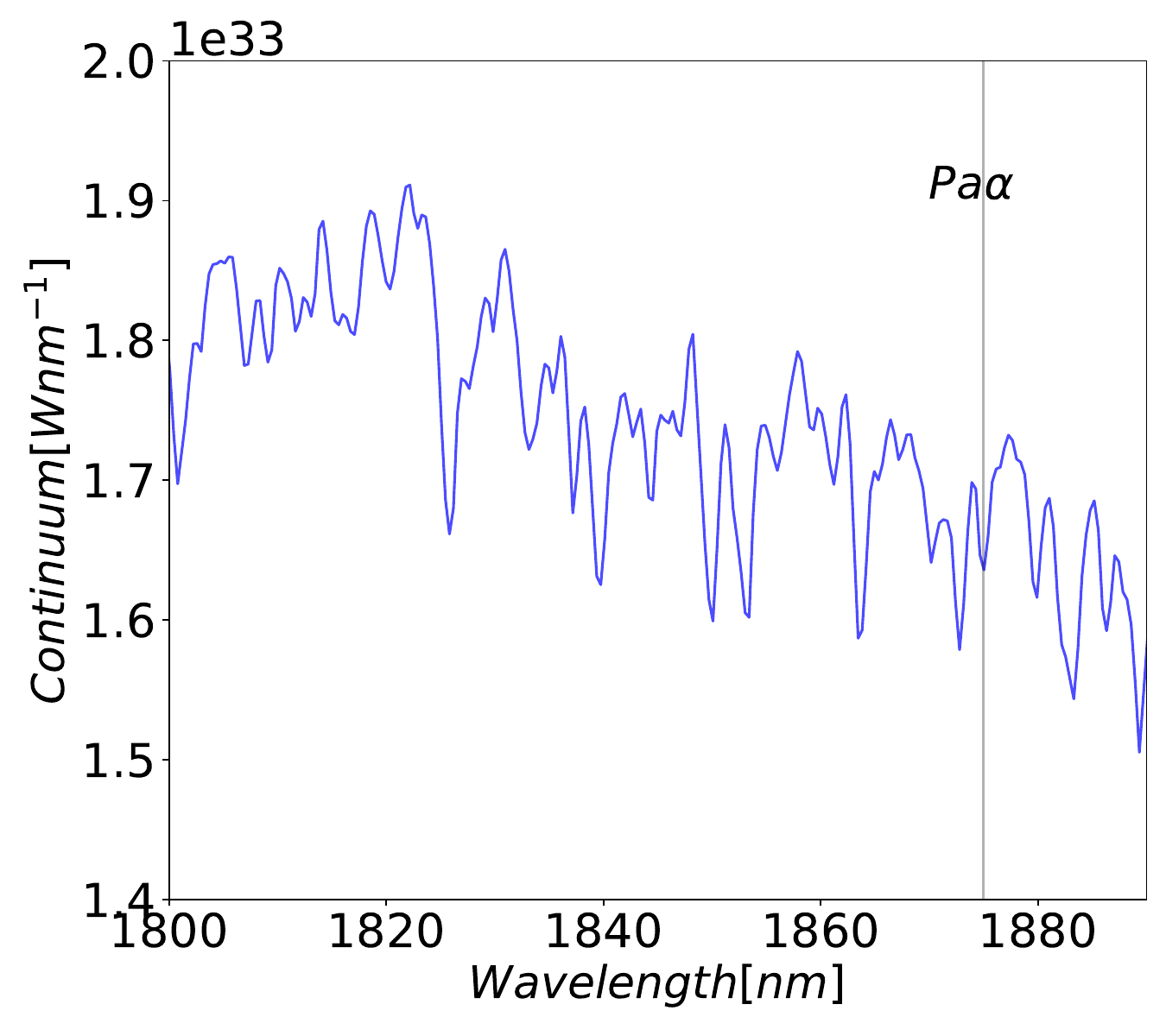}
\end{subfigure}%
\caption{Snapshots of the stellar continuum of the best model from the fit of spectro-photometric data made with BPASS for ID 5430. All absorption line EWs for this object can be found in Table \ref{absorption}. }
\label{absorption_plot}
\end{figure*}

\subsection{Absorption line EWs \label{ssec:absorption lines}}

One common difficulty of measuring hydrogen emission line EWs is the presence of an underlying absorption line. In our fitting method, presented in Sect.\ref{ssec:EW}, we reproduce the EW measurements carried out on observed spectra (Sect.\ref{ssec:spectroscopy}), avoiding any correction for an underlying absorption either on observed or model spectra.

Absorption lines are part of the stellar emission templates used to fit the galaxies and in this section we  make use of the best fit model  to compute seven hydrogen absorption line EWs for each galaxy of our sample. We use the binary models of BPASS as even the resolution of the high resolution models from \cite{bruzual} (BC03) is not enough to accurately model these absorption lines. We show an example of the best model of ID 5430 in Fig \ref{absorption_plot} and the values of the EWs are reported in Table \ref{absorption}.

In this Table, we notice a trend of decreasing EWs with the increase of the wavelength of the line. It is in agreement with previous studies \citep[e.g.][]{groves, dominguez, valentino, calabro}. We find that the EWs are highly dependant on the galaxy and its specific SFH and care should be taken when correcting with averaged EWs. 

\cite{dominguez} and \cite{valentino} did not find any absorption EWs for Paschen lines while \cite{calabro} finds slightly higher values (no uncertainties are given) than our average measurements for Pa$\gamma$ (2.50 $\AA$ compared to 1.10 $\pm$ 0.19 $\AA$) and Pa$\beta$ (2.00 $\AA$ compared to 1.30 $\pm$ 0.23 $\AA$) and no absorption for Pa$\alpha$ (against 0.20 $\pm$ 0.08 $\AA$). Part of these differences may be explained by the sample selection  as their study is focusing solely on  massive dusty starbursts at z $<1$.

Our average value for H$\alpha$  (2.30 $\pm$ 0.27 $\AA$) agrees with the value reported by \cite{dominguez} and  \cite{calabro}(2.70 $\pm$ 0.52 $\AA$, and 2.50 $\AA$ respectively). \cite{valentino} finds a value of 3.50 $\AA$ through SED fitting but do not provide an associated uncertainty.
EWs of H$\gamma$ and H$\delta$ have not been computed in these other studies so we cannot compare them. 


The main driver  of the strength of absorption line EWs is  likely to be the SFH of the galaxy. We find a continuous decrease of the strength of the absorption lines with  increasing wavelengths  from Pa$\alpha$ to H$\delta$. We explain this result by a larger contribution of old stars to the stellar  emission  at longer wavelength, reducing the impact of absorption lines from late B and A stars on the stellar continuum. 

\section{Spectral energy distribution fitting\label{sec:SED-fitting}}

To fit the SEDs from the combination of broadband fluxes and EWs, we use the modelling software CIGALE. For a complete description of CIGALE and its functionalities see \cite{boquien19}. The models are built by successively calling modules, each corresponding to a single physical component or process. CIGALE combines a stellar SED built with an SFH, stellar population models and nebular emission with dust absorption and emission components.  The energy balance between stellar and nebular dust absorption and dust re-emission is conserved. Energetic photons produced by massive stars ionise the surrounding gas which re–emits the energy in the form of a series of emission lines and a nebular continuum. 

The quality of the fit is assessed by the value of the $\chi^2$ and the  values of the physical parameters and their corresponding uncertainties are estimated as the likelihood-weighted means and standard deviations. Below we briefly present the modules we use and our input parameter values which are summarised in Table \ref{SED_param}.

Our SED fitting combines photometric fluxes and EWs as spectroscopic data. When combining integrated photometric data and EWs measured on the NIRSpec spectra we assume that these EWs  are representative of the whole galaxies. 

Along with the SED fitting process  with CIGALE it is possible to build and analyse a mock catalogue related to the observed dataset\citep{buat14,ciesla}.  CIGALE uses the best-fit model of each of the objects  obtained with the SED-fitting procedure. The flux densities and  EWs  of the mock SEDs are then computed by randomly picking a flux or EW  value from the normal distribution generated using  the best model value and the error of the input data as standard deviation. CIGALE is then run on this artificial catalogue with the same configuration as for the first run  in order to compare the exact values of the physical parameters corresponding to the artificial SEDs to the parameters estimated by the code with the probability density function of each parameter. Below we use this comparison to check the robustness of the output SED fitting parameters of interest for our study and our ability to constrain them. 


\begin{table*}
\caption{CIGALE modules and input parameters}
\centering
\begin{tabular}{l r r}
\hline
\hline
  \multicolumn{1}{c}{Parameter} &
  \multicolumn{1}{c}{ Symbol} &
  \multicolumn{1}{c}{Range}
  \\
\hline
\multicolumn{1}{c}{Delayed SFH}\\
\hline
  Age of the main population & $age_{\mathrm{main}}$ & 2000, 3000, 4000, 5000\\
  e-folding timescale of the delayed SFH & $\tau_{\mathrm{main}}$ & 500, 2000, 5000, 8000\\
  Age of the burst & $age_{\mathrm{burst}}$ & 10, 20, 50, 100\\
  Burst stellar mass fraction & $f_{\mathrm{burst}}$ & 0.0, 0.1, 0.15, 0.2\\
\hline
\multicolumn{1}{c}{Dust attenuation}\\
\hline
  V-band attenuation in the ISM & $A_\mathrm{V}^{\mathrm{ISM}}$ & 0 to 1.6 with 0.1 bins\\
  $A_\mathrm{V}^{\mathrm{ISM}}$ / ($A_\mathrm{V}^{\mathrm{ISM}}$ + $A_\mathrm{V}^{\mathrm{BC}}$) & $\mu$ & 0 to 1.0 with 0.1 bins\\
  Power-law slope of dust attenuation in the BCs & $n^{\mathrm{BC}}$ & 0.7, 1.0, 1.3\\
  Power-law slope of dust attenuation in the ISM & $n^{\mathrm{ISM}}$ & 0.4 to 1.5 with 0.1 bins\\
\hline
\multicolumn{1}{c}{Dust emission}\\
\hline
  Mass fraction of PAHs &  $q_\mathrm{PAH}$ & 0.47, 1.12, 3.90\\
  Minimum radiation field & $U_\mathrm{min}$ &  1.0, 5.0, 10.0\\
  Fraction illuminated & $\gamma$ & 0.02, 0.05, 0.1\\
\hline
\hline
\end{tabular}
\label{SED_param}
\end{table*}

\subsection{Stellar population models and star formation history\label{ssec:stellar-pops}}


We first  define an SFH  to compute the stellar spectrum and we choose a delayed exponential SFH in the form of $t \times \exp(-t/\tau)$ where $\tau$ is the e-folding time of the main stellar population.
We add an on-going burst of constant star formation,  defined with the mass fraction created during the burst  ($f_{\rm burst}$ from 0 to 20$\%$) and with its age ($age_{\rm burst}$ from 10 to 100 Myr). We refer to the age of the onset of  the delayed star formation as $age_{\rm main}$, this parameter ranges from 2000 Myrs to 5000 Myrs and is always lower than the age of the universe at the redshift of the galaxy.

We consider two  models of stellar populations: the widely used BC03 models and the BPASS models (version 2.2)  which account for binary stars \citep{eldridge,stanway}. The latter has been shown to be very efficient when fitting UV and optical photometry with spectroscopic data \citep[e.g.][]{larson, bolamperti, rezaee, tacchella, lecroq} as they were built with a focus on best reproducing the emission from young and massive stars \citep[e.g.][submitted to ARA\&A]{eldridge, stanway, reddy_22, marchant}.  The BPASS binary models have known issues when fitting stellar populations older than a few gigayears \citep{eldridge, stanway, han}, the first one being that models remain bluer compared to single star models. Two reasons are given by \cite{eldridge} to explain this effect: the fixed binary fraction regardless of the mass of the star and the width of the time bins at late ages combined with a relatively sparse sampling of stellar masses. Binary models also take a longer time to reach the quiescence \citep{eldridge}. This leads to an underestimation of the number of old and small binary stars. 

In both cases, we choose the Initial Mass Function of \cite{chabrier} and we assume that the metallicity is fixed to the solar value: $Z = 0.02$.

\subsection{Dust attenuation and emission\label{ssec:dust-attenuation-emission}}

To account for the dust attenuation, we apply the model presented in Sect. \ref{ssec:line-ratios} which corresponds to the \emph{modified\_CF00} module in CIGALE. In this module, $A_\mathrm{V}^{\mathrm{ISM}}$, $\mu$ , $n^{\mathrm{BC}}$ and $n^{\mathrm{ISM}}$ (Eq.~\ref{eq:AvBC}, \ref{eq:AvISM}, and \ref{eq:mu}) are free parameters. We sample the parameter space of $A_\mathrm{V}^{\mathrm{ISM}}$, $\mu$ and $n^{\mathrm{ISM}}$ very thoroughly (with 0.1 step). 
$n^{\mathrm{ISM}}$ ranges from -0.4 to -1.5  to represent the broad distributions of values found from both observations \citep[e.g.][]{lofaro, buat19, pantoni, seille} and simulations  \citep[e.g.][]{chevallard, roebuck, trayford}. The tests we perform on mock catalogues show that $n^{\mathrm{BC}}$ is not constrained. We thus introduce only three values for $n^{\mathrm{BC}}$: the widely used value of -1.3 \citep{charlot, dacunha, chevallard}, -0.7 as proposed in the original CF00 recipe \citep{charlot} and an in-between value of -1.0.

To model the dust re-emission of the five sources that have far-IR data, we use the models from \cite{draine07}, updated in \cite{draine14} which can be found in the \emph{dl2014} module in CIGALE. These models reproduce the diffuse dust emission heated by the general stellar population which means that the dust is illuminated with a single radiation field $U_\mathrm{min}$. The models also include a component to recreate the emission of the dust tightly linked to star-forming regions. In that case, the dust is illuminated with a variable radiation field intensity superior to $U_\mathrm{min}$
The dust mass fraction of dust linked to the star–forming regions (respectively diffuse emission) is referred to as $\gamma$ (respectively 1 - $\gamma$). The polycyclic aromatic hydrocarbon (PAH) abundance, $q_\mathrm{PAH}$ is also a free parameter in the models.

\begin{figure*}
\centering
\begin{subfigure}{.5\textwidth}
  \centering
  \includegraphics[width=10cm]{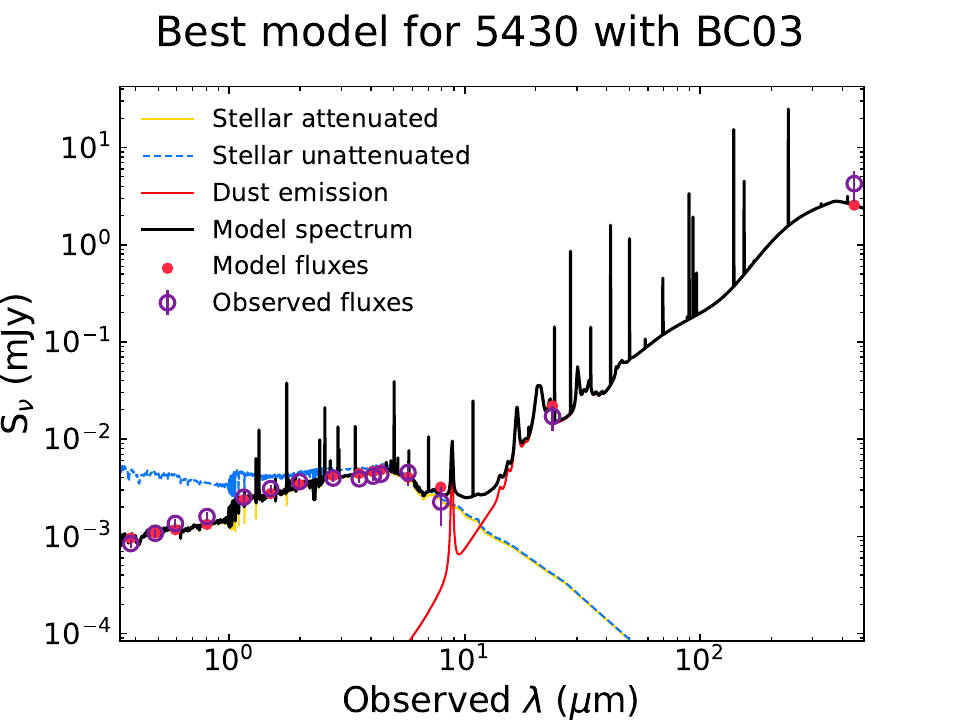}
\end{subfigure}%
\begin{subfigure}{.5\textwidth}
  \centering
  \includegraphics[width=10cm]{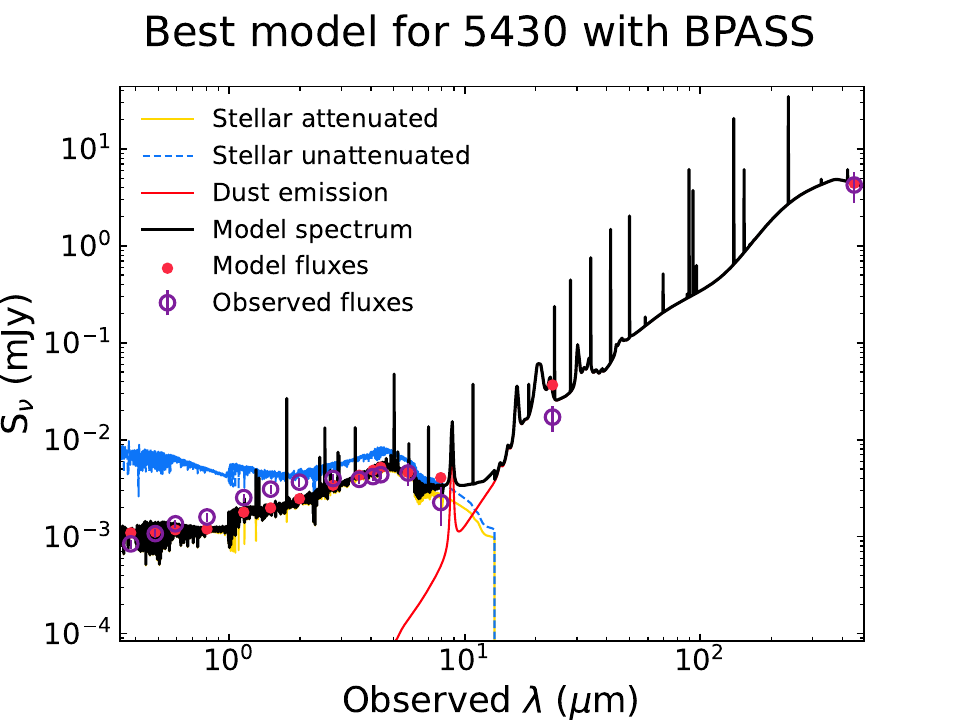}
\end{subfigure}
\caption{Comparison of the SEDs of the galaxy ID 5430 fitted with BC03 (left panel) and with BPASS (right panel).}
\label{SEDs_5430}
\end{figure*}

\subsection{Nebular emission and EWs\label{ssec:EW}}

CIGALE models the emission of the ionised gas with CLOUDY 17.01 \citep{ferland98, ferland17} with the configuration described in \cite{villa} and \cite{theulé}. 129 nebular emission lines intensities are calculated and re-scaled with the number of Lyman continuum photons from the stellar emission of the galaxy. We are only interested in hydrogen lines which makes the impact of the input parameters on the emission line intensities negligible. Therefore we decide to fix all parameters pertaining to the nebular emission to the values mentioned below. The radiation field intensity is given by the ionisation parameter:
\begin{equation}
    logU = log(\frac{n_{\gamma}}{n_H})
\end{equation}
where $n_{\gamma}$ is the number density of photons capable of ionising hydrogen and $n_H$ the number density of hydrogen. The ionisation parameter is fixed to -2.0, a value typical of extra galactic HII regions \citep{osterbrock_06}. These configurations are also parameterised according to the gas-phase metallicity, Z$_{gas}$, which is fixed to the value of the stellar metallicity (Z=0.02). The electron density is set to 100 cm$^{-3}$, a typical density for the ionised parts of a diffuse nebula \citep{osterbrock_06}. The ionising spectrum is modelled  as a constant burst of star formation  of 10 Myrs \citep{boquien19, villa} computed with either BC03 or BPASS models.

We model the Paschen and Balmer line EWs \footnote{The EWs computation used in this work is not included in the current released version of CIGALE, 2022.1. The newest version of the EWs computation is available on request.} so as to reproduce the measurements described in Sect \ref{ssec:spectroscopy}. For each emission line of each galaxy, the input wavelengths  (a six values "array") defined with \textsc{LiMe}  are used to delimit the line region and two nearby featureless continua on each side of the line. The code then performs a linear fit of the overall continuum which will be subtracted from the line. \textsc{LiMe} deblends the H$\alpha$ and [NII] doublet of our galaxy sample (see Fig \ref{lime}) and these lines are also considered separately in CIGALE. Finally as pointed out in Sect \ref{ssec:analysis-line-ratios} these EWs are not corrected for underlying absorption as they are computed and thus affected in the same way as the EWs measured with \textsc{LiMe}.

\section{Impact of the choice of data and models on the fits\label{sec:impact-data}}



\subsection{Fit of the spectro-photometric data of BC03 vs BPASS}

\begin{figure}
\centering
   \includegraphics[width=\columnwidth]{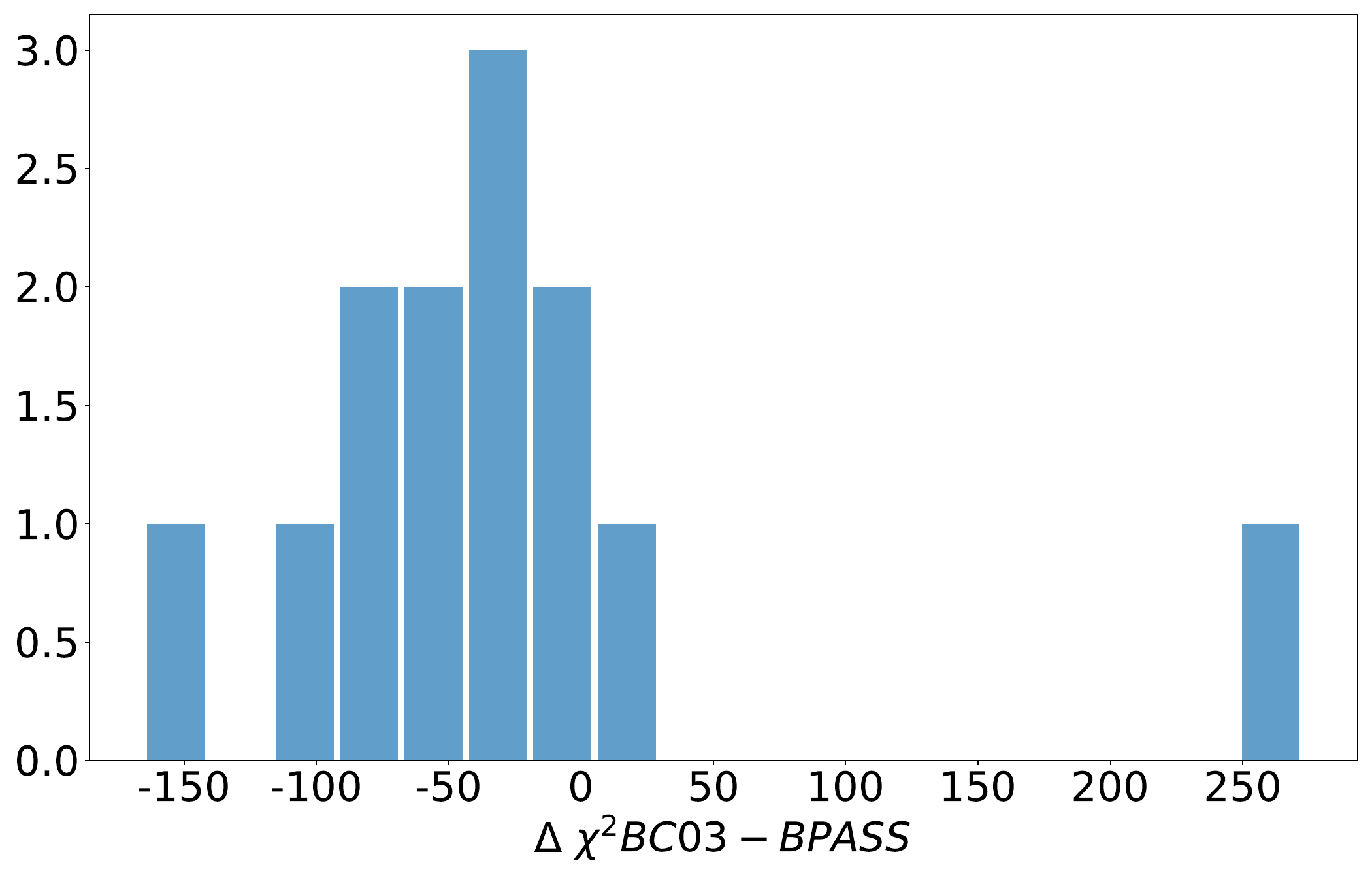}
      \caption{Difference of the $\chi^{2}$ between the BC03 and BPASS stellar population models for all galaxies in our sample. A negative value favors BC03 while a positive one favors BPASS.}
      \label{chi2}
\end{figure}

\begin{figure}
\centering
    \includegraphics[width=\linewidth]{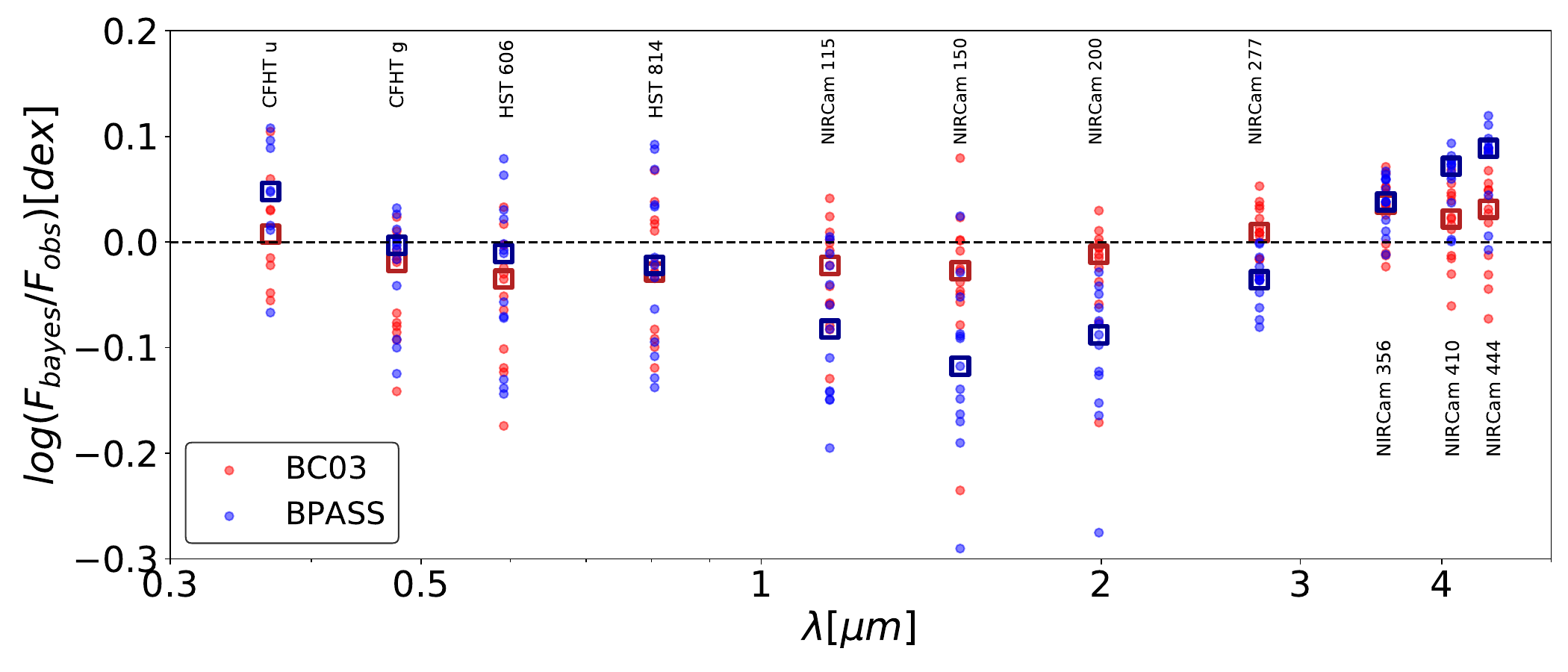}
      \caption{Quality assessment of the photometry fitting for the fits of the spectro-photometric data with BC03 (red) and BPASS (blue). For each model, the median of each filter is represented as a square.}
      \label{quali-phot}
\end{figure}

In Sect.~\ref{ssec:stellar-pops}, we outlined the widely acknowledged performance of the BPASS models in reproducing the emission of young stars but also some caveats which could impact the results of the SED fitting process. In this  subsection, we compare the quality of the fits of the spectro-photometric data made with either BPASS or the more all-purpose BC03 models. 

In Fig \ref{SEDs_5430} we show a representative example of fits performed with BPASS and BC03. The photometric data appear to be better fitted with the BC03 models. We check if we can statistically (from the quality of the fit) select the best model of stellar populations to fit each object from our sample. For this, we use the Bayesian Information Criterium (BIC, adapted from \cite{salmon}), the best model corresponding to the lowest  BIC. The two fits being performed with the same number of free parameters, the difference between the two BIC, $\Delta BIC$,  is equal to the difference between the two best $\chi^{2}$ obtained for the  fits  with BC03 and BPASS. The $\chi^{2}$ is computed as:

\begin{equation}
    \chi^{2} = \sum_{i} \left( \frac{f_i-\alpha \times m_i}{\sigma_i}\right)^{2} + \sum_{j} \left(\frac{f_j-m_j}{\sigma_j}\right)^{2}
\end{equation}

with $\alpha$ a rescaling parameter \citep[see][for a detailed explanation]{boquien19}, \textit{$f_i$} and \textit{$m_i$} being the observed and model fluxes, \textit{$f_j$} and \textit{$m_j$} being the observed and model EWs and $\sigma$ the corresponding observational uncertainties. 

We adopt the limit of $|\Delta BIC|$ > 6 to put a strong preference on a model, a positive preference being defined as  2< $|\Delta BIC|$ < 6 \citep{salmon}. The $\Delta\chi^{2}$ distribution is shown in Fig \ref{chi2}. Eleven objects present a strong preference (and one, ID 10293, a positive preference) for the BC03 models while only  one  galaxy (ID 23542) is  found with  a strong preference for the BPASS models. For this object the photometric data is  better fitted with the BC03 models (see the SEDs in Appendix \ref{sec:appendix}) but the EWs are better fitted with the BPASS models. The uncertainties on the measured EWs are very small for this galaxy ($\sim10$ times lower than for the other sources) which leads to a higher impact of spectroscopic data on the calculation of the $\chi^2$ value.

We  present a more detailed comparison of the quality of the fit of the photometric data with  BPASS and  BC03 models in Fig \ref{quali-phot}. For each filter and both models we compute the ratio between the bayesian estimate of the flux given by CIGALE over the measured flux. While the CFHT and HST fluxes do not show a significant difference of quality between both models, there is a clear trend for the four bluest NIRCam filters to underestimate the model fluxes with respect to the observed values when the binary BPASS models are used. The CFHT and HST filters cover the rest-frame  UV to optical spectrum while the JWST filters cover the  rest-frame optical to NIR range. As pointed out in Sect. \ref{ssec:stellar-pops}, the BPASS models are not optimised to reproduce older stars which we suspect to cause the underestimation of the model fluxes in these specific filters, a result supported by \cite{osborne}. According to this test on our full sample we conclude that the BC03 models better reproduce the photometric measurements.

\begin{figure}
\centering
    \includegraphics[width=\columnwidth]{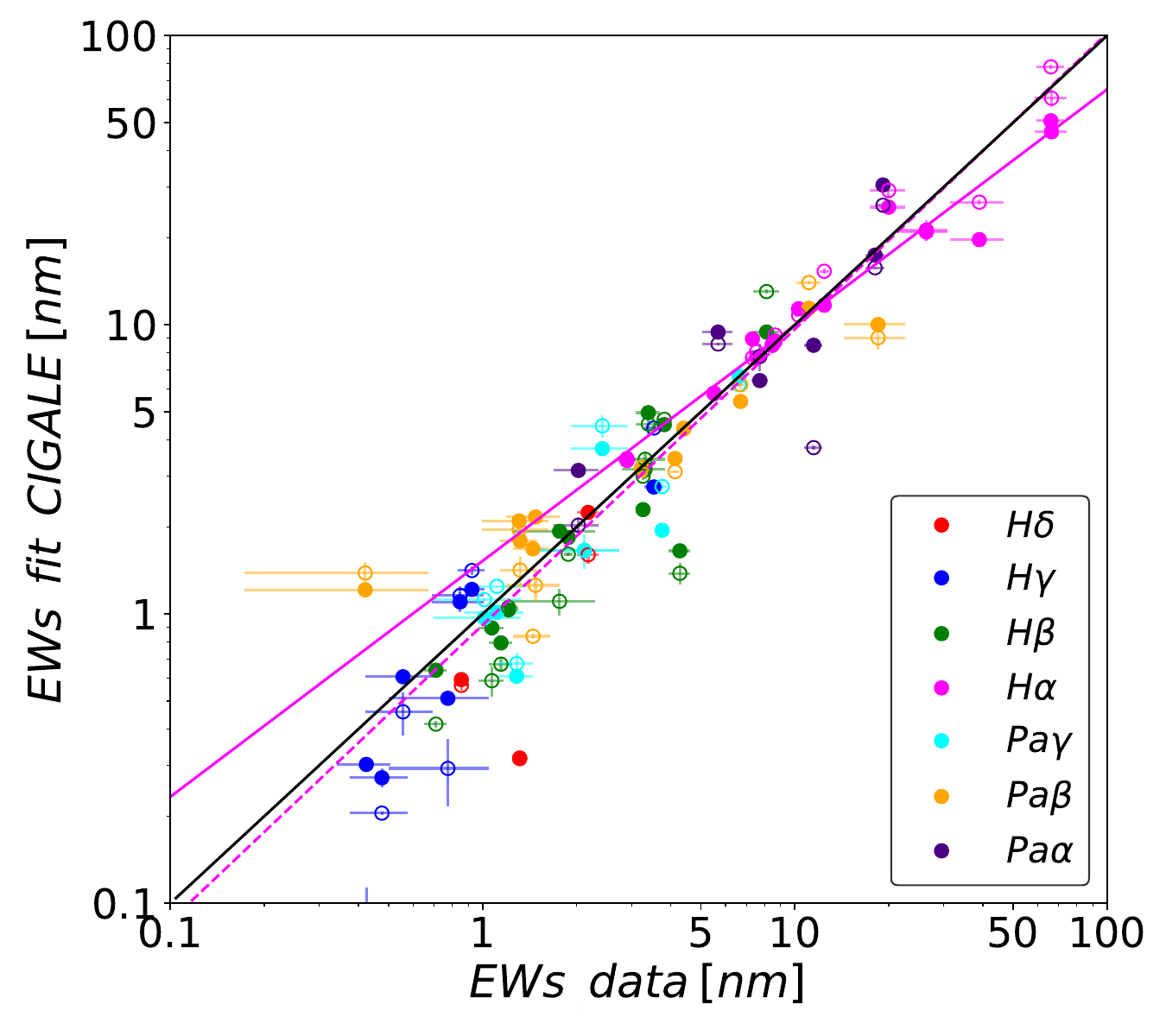}
      \caption{Quality assessment of the EWs fitting for the fits of the spectro-photometric data. Filled dots are used for BC03 fits and empty dots for BPASS fits. The linear fit of the H$\alpha$ points is plotted as a solid magenta line for BC03, dashed magenta for BPASS and the black solid line is the 1:1 line.}
      \label{EWs_fit}
\end{figure}

We also compare the quality of the fit of the spectroscopic data for both BC03 and BPASS models by comparing the EWs measured on the observed spectra to their bayesian estimation with  the CIGALE fit. As can be seen in Fig \ref{EWs_fit} both models are equally good at reproducing the observed EW larger than $\sim 10$ \AA while fainter emission lines (EW < 10 \AA) tend to be slightly underestimated. 

To conclude both stellar models are able to fit well the EWs but photometric data are better fitted with BC03 models.

\subsection{Impact of fitting only photometric data and line emission contributions\label{ssec:impact-fit-data}}

When a substantial amount of nebular emission contaminates the broadband data \citep{anders, zackrisson, pacifici15}, it can lead to large  uncertainties on the results of the SED fitting as the code lacks the information on the relative contribution of the line to the broadband emission. Notorious examples of erroneous interpretations due to intense  nebular emission are broadband measurements made with \emph{Spitzer}/IRAC \citep{fazio} interpreted as a Balmer break \citep[e.g.][]{chary, zackrisson, vanderwel} and more recently measurements  with JWST which mimic a Lyman break leading to incorrect estimates of redshifts \citep[e.g.][submitted to ApJ]{donnan, Arrabal_b, zavala, davis} .
Below we focus on the consequences of fitting only photometric data which accounts for the vast majority of studies making use of SED fitting. We also measure the contribution of line emission in some of the NIRCam broadband filters obtained with the fits of both photometric fluxes and EWs.
For these studies we fit our 13 galaxies with both BC03 and BPASS models despite the results of the BIC test. We choose to use both as more and more studies include the BPASS models. 
\begin{table*}
\caption{H$\alpha$ contribution to the NIRCam filter}
\centering
\begin{tabular}{|c|c|c|c|c|c|}
\hline
  \multicolumn{1}{|c|}{ID} &
   \multicolumn{1}{|c|}{Redshift} &
   \multicolumn{1}{|c|}{sSFR\_BC03 ($10^{-9}$$yr^{-1}$)} &
   \multicolumn{1}{c|}{Contribution\_BC03} &
   \multicolumn{1}{|c|}{sSFR\_BPASS ($10^{-9}$$yr^{-1}$)} &
   \multicolumn{1}{c|}{Contribution\_BPASS}\\
\hline
23542 &  1.277 & 2.65 & 42\% & 6.34 & 52\% \\
5430 & 1.676 & 3.39 & 43\% & 6.28 & 51\% \\
10293 & 1.676 & 1.07 & 10\% & 6.20 & 51\% \\
6563 & 1.699 & 1.31 & 41\% & 6.28 & 51\% \\
5409 & 1.699 & 1.35 & 22\% &  0.60 & 10\%  \\
5300 & 2.136 & 7.04 & 45\% & 5.24 & 38\%  \\
3788 & 2.295 & 22.7 & 68\% & 28.5 & 76\% \\
8588 & 2.336 & 2.83 & 20\% & 2.07 & 13\% \\
8710 & 2.337 & 6.09 & 40\% & 4.70 & 24\%  \\
Average & & 5.40 & 36\% & 7.40 & 42\%  \\
\hline
\end{tabular}
\tablefoot{Contribution of nebular emission of the NIRCam filter containing the H$\alpha$ line computed with the BC03 and BPASS models.}
\label{conta}
\end{table*}

\begin{figure}
\centering
    \includegraphics[width=\columnwidth]{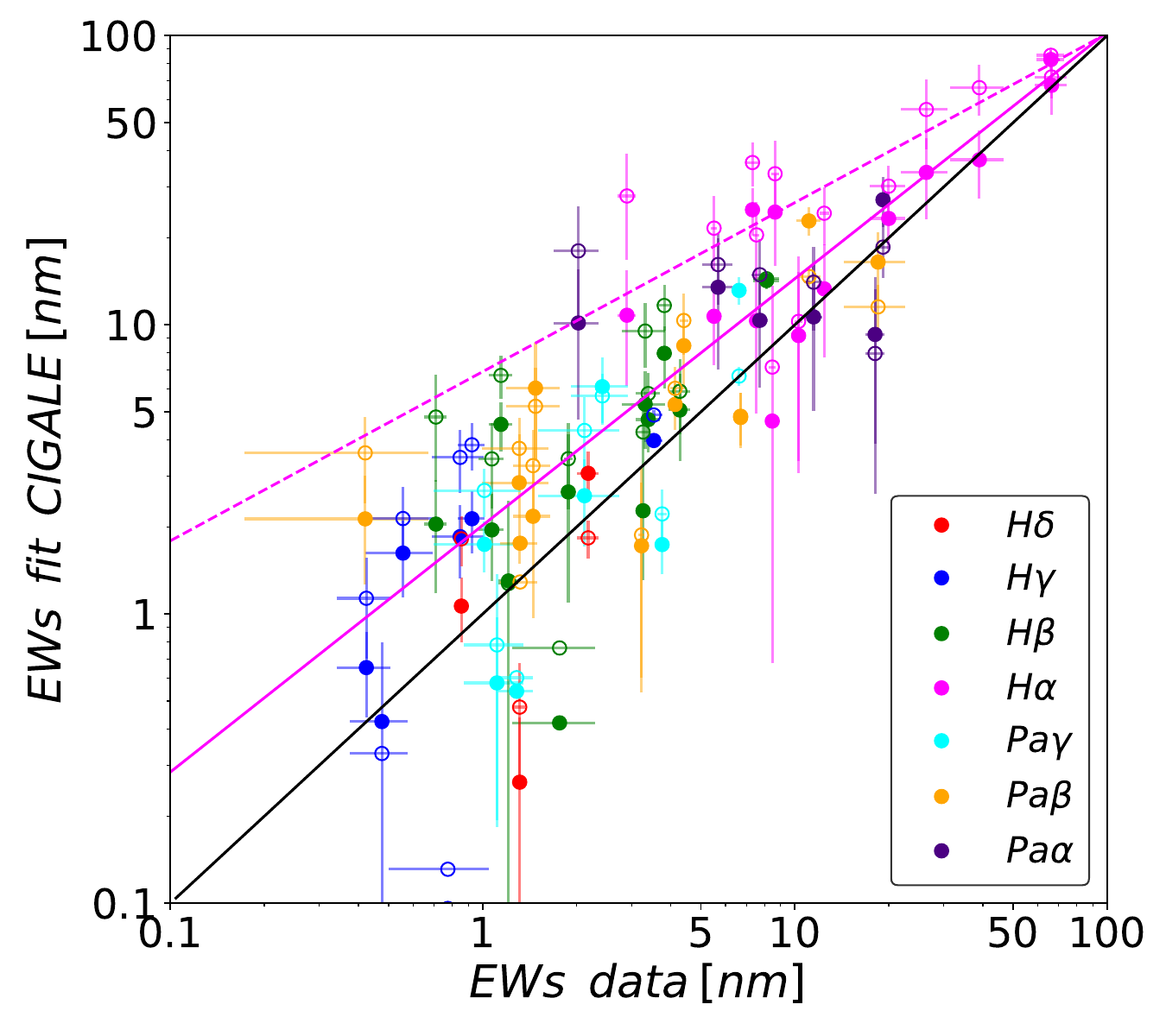}
      \caption{Quality assessment of the EW estimations when fitting photometric data only. Filled dots are used for BC03 fits and empty dots for BPASS fits. The linear fit of the H$\alpha$ points is plotted as a solid magenta line for BC03, dashed magenta for BPASS and the black solid line is the 1:1 line.}
      \label{EWs_fit_cont}
\end{figure}

\subsubsection{Fitting only photometric data}

We start by looking at the quality of the fit of photometric data only. We obtain the same trend as  in Fig \ref{quali-phot} i.e. an overestimation of the fluxes in the four bluest NIRCam filters with the BPASS models. Next we compare  the estimations of the EWs of hydrogen recombination  lines with the measured EWs and we present the results in Fig \ref{EWs_fit_cont}. As expected these estimations are less accurate than the ones obtained when fitting simultaneously photometric fluxes and EWs (Fig \ref{EWs_fit}). Both stellar population models tend to  overestimate  the EWs of the lines as CIGALE is favouring stronger bursts of star formation compared to the ones obtained  when fitting the spectro-photometric data: the stellar mass fraction created during  the burst increases from an average value of 0.05 to 0.13.  This overestimation is much more prevalent with  the BPASS models. This larger contribution of the lines we find for the fits with BPASS (average of $\sim$ 45\%) compared to those made with BC03 (average of $\sim$ 15\%) is explained by the addition of a more intense  and younger burst of star formation with median values of $f_{\rm burst}$ and  $age_{\rm burst}$  equal to 0.2 and 50 Myrs for BPASS against  0.1 and 100 Myrs  for BC03. 

\subsubsection{Contribution of line emission in broad-band filters}

\begin{figure}
\centering
    \includegraphics[width=\columnwidth]{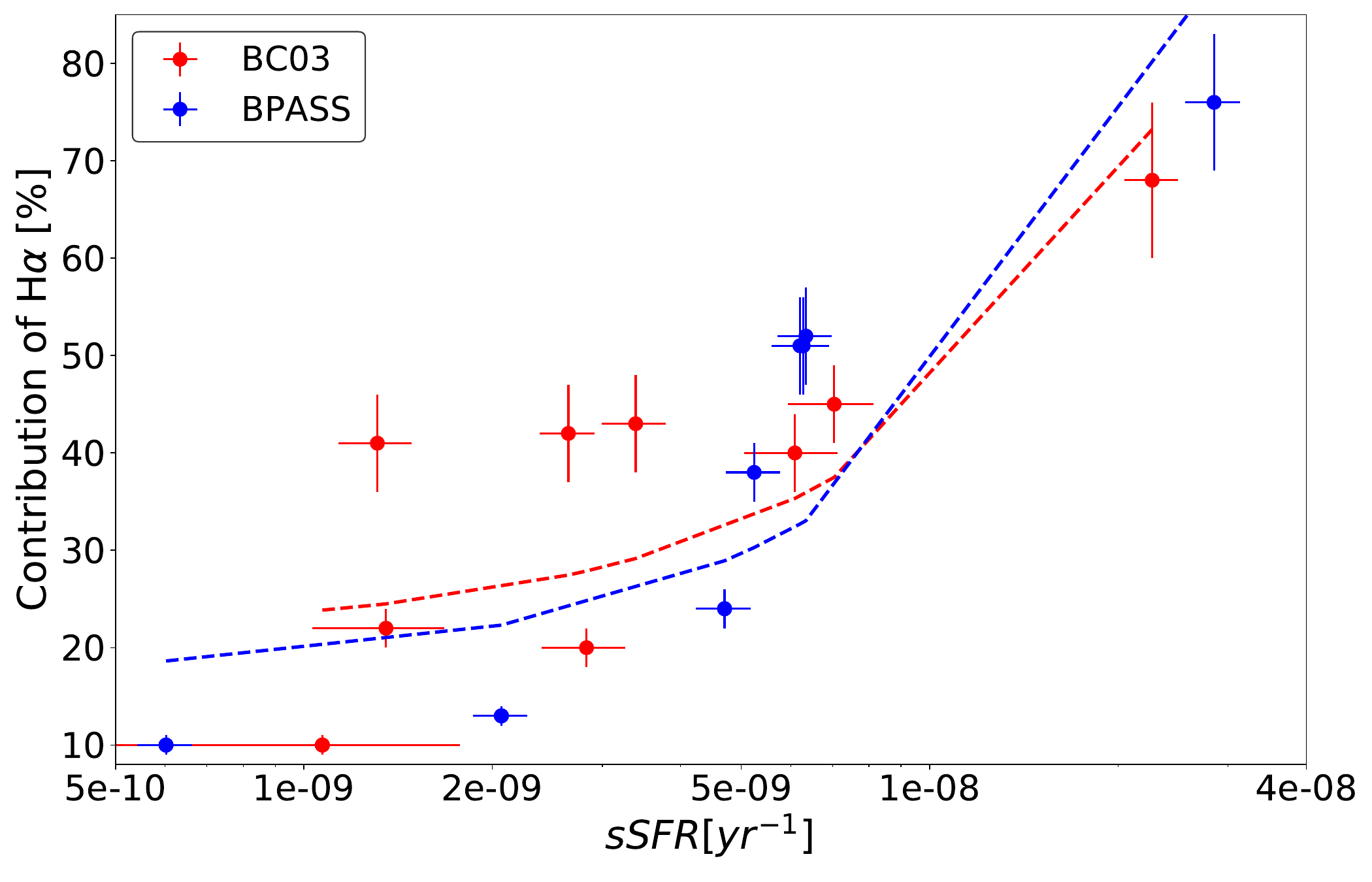}
      \caption{Correlation between the sSFR and the level of contamination in the NIRCam filter containing the H$\alpha$ line. The error on the y-axis comes from the uncertainty on the attenuation. The dashed lines represent the best fit of each sample of data points.}
      \label{conta_trend}
\end{figure}

To estimate the contribution of the emissions lines to the flux measured in each of our broadband  filters, we go back to the fit  of photometric data and EWs and we consider the best model obtained  with the BC03 and BPASS models. We compute the ratio of the flux coming from  the nebular emission and the flux of the continuum   within the broadband filter. At the redshift of our galaxies, strong emission lines (H$\alpha$, H$\beta$, H$\gamma$, [OIII]-4959, [OIII]-5007, [NII]-6548 and [NII]-6583) are lying within the bandpasses of the NIRCam F150W and F200W filters. Out of all these lines, H$\alpha$ is the most intense which makes its contribution easier to track. In Table \ref{conta} we provide the contribution of the nebular emission \footnote{Other lines than these strong emission lines, as well as the nebular continuum, are included by CIGALE in the nebular emission, but their contribution is found negligible.} in the NIRCam broadband filter containing the H$\alpha$ line for the BC03 and BPASS models. The filter to which H$\alpha$ is contributing changes between F150W and F200W depending on the redshift of each source. We plot the percentage of nebular emission contribution in the filter containing H$\alpha$ in Fig \ref{conta_trend} against the specific SFRs (sSFRs) of the galaxies. The sSFRs are obtained by dividing the bayesian estimation of the SFRs averaged over the last 10 Myrs with the bayesian estimation of the stellar mass both obtained from the output parameters of the best model from the SED fitting. Four galaxies (IDs 8515, 8736, 16991 and 18294) are not included since the H$\alpha$ line is redshifted in-between NIRCam filters.

The  contribution  of the  H$\alpha$ line is found to increase with the  sSFR for both models. We find a stronger average contribution from the nebular emission with BPASS (42\%) than with BC03 (36\%) as well as a larger average sSFR of 7.4 $\times$ $10^{-9}$ $yr^{-1}$ for BPASS and 5.4 $\times$ $10^{-9}$ $yr^{-1}$ for BC03. \cite{mckinney} find an average contribution of the H$\alpha$+[NII] doublet of 60\% for their sample composed of hot, dust-obscured galaxies and dusty, star-forming galaxies at 1 < z < 4 and include the [NII] doublet which may explain the higher percentage they found.  We observe a large scatter between the values obtained with BC03 and BPASS. 
This dispersion reflects the inherent  differences between both models already discussed in Sect. \ref{ssec:stellar-pops} such as the more intense emission lines generated by BPASS or the "missing" low mass stars.  We further discuss the impact of the contribution of nebular emission on the stellar mass and SFR with both BPASS and BC03 models in Sect. \ref{sec:impact-models}.

\section{Dust attenuation \label{sec:attenuation}}

In this section, we study the output dust  attenuation parameters and for a few  cases where the amount of attenuation is high enough, we also reconstruct the total effective dust attenuation curve of each object. We  only consider the run performed with the BC03 models since the fit of photometric fluxes  is crucial to measure the dust attenuation  of  the stellar continuum.

\subsection{Dust attenuation parameters\label{ssec:dust_param}}

We start by discussing the output parameters computed by CIGALE: $A_\mathrm{V}$, $n^{\mathrm{ISM}}$ and $\mu$ defined in the CF00 recipe (section \ref{ssec:line-ratios}). The two quantities $n^{\mathrm{ISM}}$ and $\mu$ cannot be accurately estimated if the amount of dust in the ISM is too low. Following \cite{corre} we select a subsample of fitted sources with a required minimum $A_\mathrm{V}>$ 0.3 mag. Six  galaxies fulfill this condition: ID 5409, ID 5430, ID 6563, ID 8588, ID 8710 and ID 23542. Only two objects have $A_\mathrm{V}>$ 1 mag, the four other ones have $A_\mathrm{V}<$ 0.6 mag.

\begin{table}
\caption{Attenuation parameters of our subsample}
\begin{tabular}{|c|c|c|c|c|}
\hline
  \multicolumn{1}{|c|}{ID} &
   \multicolumn{1}{c|}{$A_\mathrm{V}$}&
   \multicolumn{1}{c|}{$n^{\mathrm{ISM}}$}&
    \multicolumn{1}{|c|}{$\mu$} &
    \multicolumn{1}{|c|}{m}\\
\hline
\footnotesize 23542 & \footnotesize 0.46 (0.12) & \footnotesize -0.86 (0.23) &  \footnotesize 0.53 (0.14) & \footnotesize -1.07 (0.09) \\
\footnotesize 5430 & \footnotesize 0.59 (0.01) & \footnotesize -0.77  (0.11) & \footnotesize 0.10 (0.01) & \footnotesize -0.98 (0.13)\\
\footnotesize 6563 & \footnotesize 1.01 (0.26) & \footnotesize -0.86 (0.23) & \footnotesize 0.25 (0.05) & \footnotesize -0.91 (0.11)\\
\footnotesize 5409 & \footnotesize 0.41 (0.03) & \footnotesize -0.66 (0.12) & \footnotesize 0.30  (0.01) & \footnotesize -1.11 (0.11)\\
\footnotesize 8588 &  \footnotesize 1.49 (0.04) & \footnotesize -0.60 (0.03) & \footnotesize 0.70 (0.02) & \footnotesize -0.72 (0.07)\\
\footnotesize 8710 & \footnotesize 0.38 (0.08) & \footnotesize -0.61 (0.19) & \footnotesize 0.35 (0.08) & \footnotesize -0.94 (0.09)\\
\footnotesize Means & \footnotesize 0.72 (0.12) & \footnotesize -0.73 (0.15) & \footnotesize 0.37 (0.08) & \footnotesize -0.96 (0.10)\\ 
\hline
\end{tabular}
\tablefoot{$A_\mathrm{V}$ refers to the total attenuation in the V band, $n^{\mathrm{ISM}}$ is the slope of the attenuation in the ISM (see Eq \ref{eq:AvISM}), $\mu$ is the differential attenuation defined in Eq \ref{eq:mu} and $m$ is the slope of the power-law used to fit our subsample.}
\label{att_param}
\end{table}

In Table \ref{att_param}, we present the results of the SED fitting for the attenuation parameters of our subsample. 
The average values of $\mu$  and $n^{\mathrm{ISM}}$ ($n^{\mathrm{ISM}}$ = -0.73 $\pm$ 0.15 and $\mu$ = 0.37 $\pm$ 0.08) are  in good  agreement with the original values of CF00 ($\mu =0.3$ and $n^{\mathrm{ISM}}=-0.7$).
However, similarly to \cite{battisti, battisti20}, we find that $\mu$ can vary a lot (from 0.1 to 0.7) between objects and fixing it to the standard CF00 value of 0.3 can change the shape of the SED, the SFRs and stellar masses (see Sect \ref{ssec:fits_att}). 


Several studies have shown a link between the redshift and a higher colour excess (as defined with the attenuation recipe of \cite{calzetti}) needed for emission lines than for the stellar continuum \citep[e.g.][]{yoshikawa, whitaker}. For our subsample of galaxies, we conclude that there is no major trend of $\mu$ increasing with redshift between 1 and 3.


\subsection{Derivation of the effective attenuation curves\label{ssec:effective-curves}}

In the CF00 recipe, the interplay between  $A_v^{\mathrm{ISM}}$ and $A_v^{\mathrm{BC}}$ (through $\mu$) depends on the star formation history \citep{charlot}. 
As a consequence, no universal attenuation curve can be defined.

With CIGALE, we can derive  the effective attenuation curve $A_\uplambda/A_\mathrm{V}$ for each galaxy of our subsample, $A_\mathrm{V}$ being the total attenuation in the V band. We select fifteen wavelength bands (from the rest-frame far-UV to the near-IR) as the code is able to compute the attenuation in filters different from the ones used to fit the input data. With this feature in mind we add fifteen custom-made filters to the code. We choose to create very narrow filters (10 nm each) to carefully select which part of the spectrum is going to be sampled. The reason for this choice is two-fold: we can avoid contamination from emission and absorption lines but also pinpoint a specific part of the spectrum. We use the bayesian estimations of CIGALE for fifteen photometric bands to reconstruct the total effective attenuation curve of our subsample of galaxies.

We show the attenuation curves of our subsample. All curves except that of ID 8588 agree well on the UV to optical part of the spectrum but all six of them are much more spread out once we reach the near-IR.


We fit each attenuation curve with a single powerlaw:

\begin{equation}
    A_\uplambda = A_\mathrm{V} (\uplambda/0.55)^{m}
\end{equation}

and we report the resulting exponents in Table \ref{att_param}. We also fit our whole subsample with a single powerlaw and find an exponent of $m$=-0.96 $\pm$ 0.10. By colour-coding each curve with its associated value of $\mu$ we see a trend between the decrease in $\mu$ and the steepening of the curve (i.e. the exponent $m$ is closer to -1.3). A high $\mu$ equals a negligible increase of attenuation for stars in birth clouds compared to the ISM meaning a single powerlaw (for ID 8588). The differential attenuation found for the galaxies  with $\mu$ < 0.35 mag leads to a steeper attenuation law. Using radiative transfer modelling, it has been shown that the effective attenuation curve depends on the amount of dust attenuation \citep[e.g.][]{pierini, saftly, feldmann, roebuck, trayford}. As predicted by the models the IR part of the curves do not follow the model of \cite{calzetti} and ID 8588 is the only galaxy to follow the model in the UV-optical which shows the need for a two-components dust attenuation model. Finally we compared the values of $\mu$ obtained through a fit made with $\mu$, $A_v^{\mathrm{ISM}}$, $n^{\mathrm{ISM}}$ and $n^{\mathrm{BC}}$ being free and a fit made with $\mu$ and $A_v^{\mathrm{ISM}}$ being free and both exponents fixed to -0.96. Both sets of values are consistent within their respective errorbars. 

\begin{figure}
\centering
    \includegraphics[width=\linewidth]{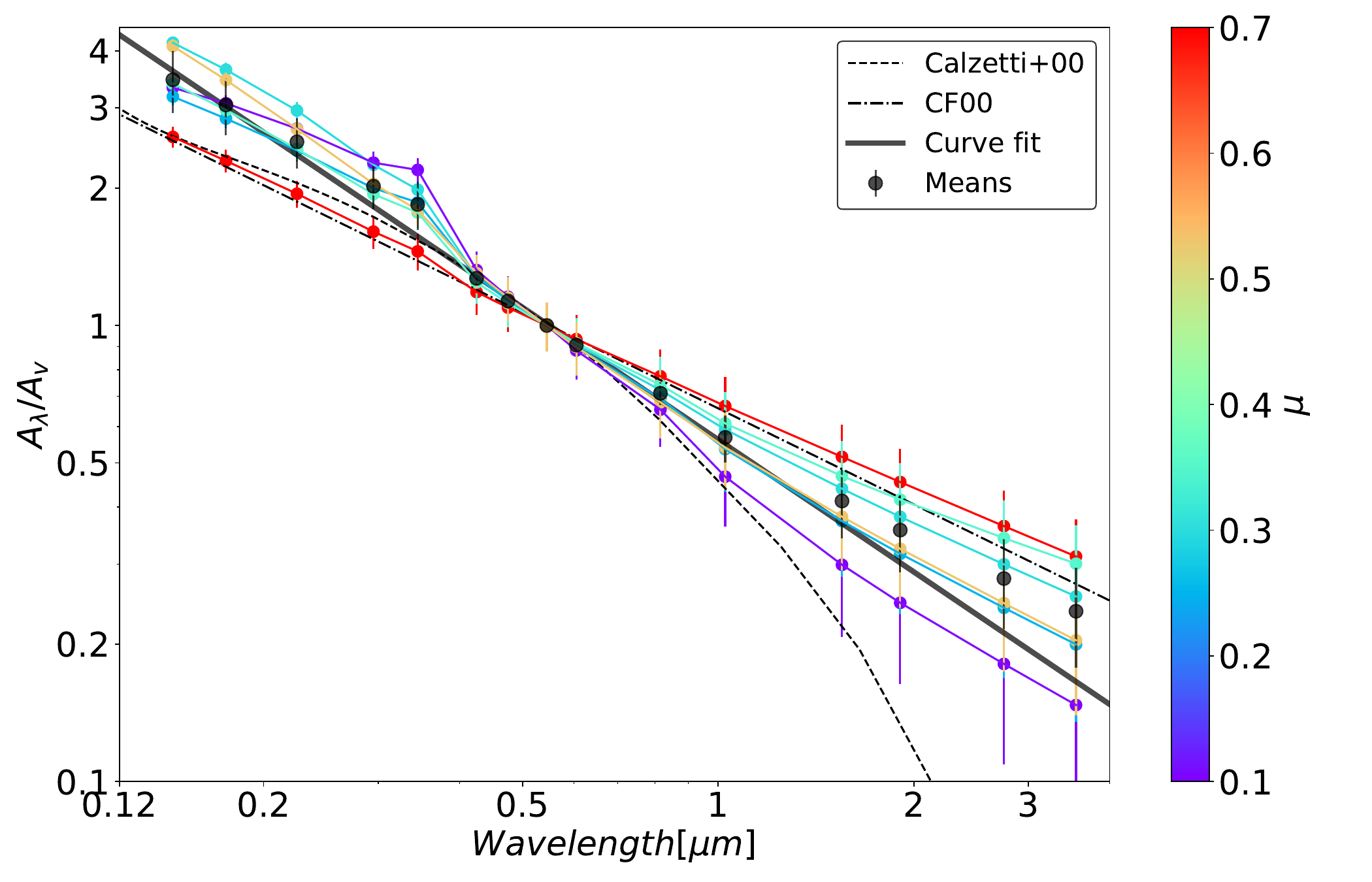}
      \caption{Total attenuation curves of the subsample with a fit of the mean curve in a black solid line. On the rightmost measurement from top to bottom: IDs 8588, 8710, 5409, 23542, 6563 and 5430. The classical CF00 recipe as well as the Calzetti law have been added for comparison. Each total attenuation curve is color-coded with its value of $\mu$ presented in Table \ref{att_param}.}
      \label{att_curves}
\end{figure}

\section{Impact of input data type and models on SFRs and stellar mass determinations\label{sec:impact-models}}

The precision and confidence in constraining the physical parameters for our sample of galaxies through statistical fits of observations rely on two key factors: the nature of the observations (whether photometry alone or in conjunction with EWs) and the spectral libraries used to model them. Changing one or both of these factors can lead to major modifications in the SFH which may result in different output parameters and especially SFRs and stellar masses.

As getting robust SFRs and stellar masses is crucial in numerous contexts \citep[e.g.][submitted]{pacucci, juodzbalis}, we explore the effects of the choice of stellar populations to fit the spectro-photometric data on both aforementioned output parameters. Then we quantify the variation of SFRs and stellar masses between a fit performed with broad-band photometry alone and with spectro-photometric data, the latter being used as our reference fit. Finally we assess the impacts of the change of the attenuation recipe on the SFRs and stellar masses when fitting spectro-photometric data with the BC03 models.

\subsection{Fits with BPASS or BC03 models \label{ssec:fits_stellar_models}}

\begin{figure}
\centering
\begin{subfigure}{0.48\columnwidth}
  \includegraphics[width=\columnwidth]{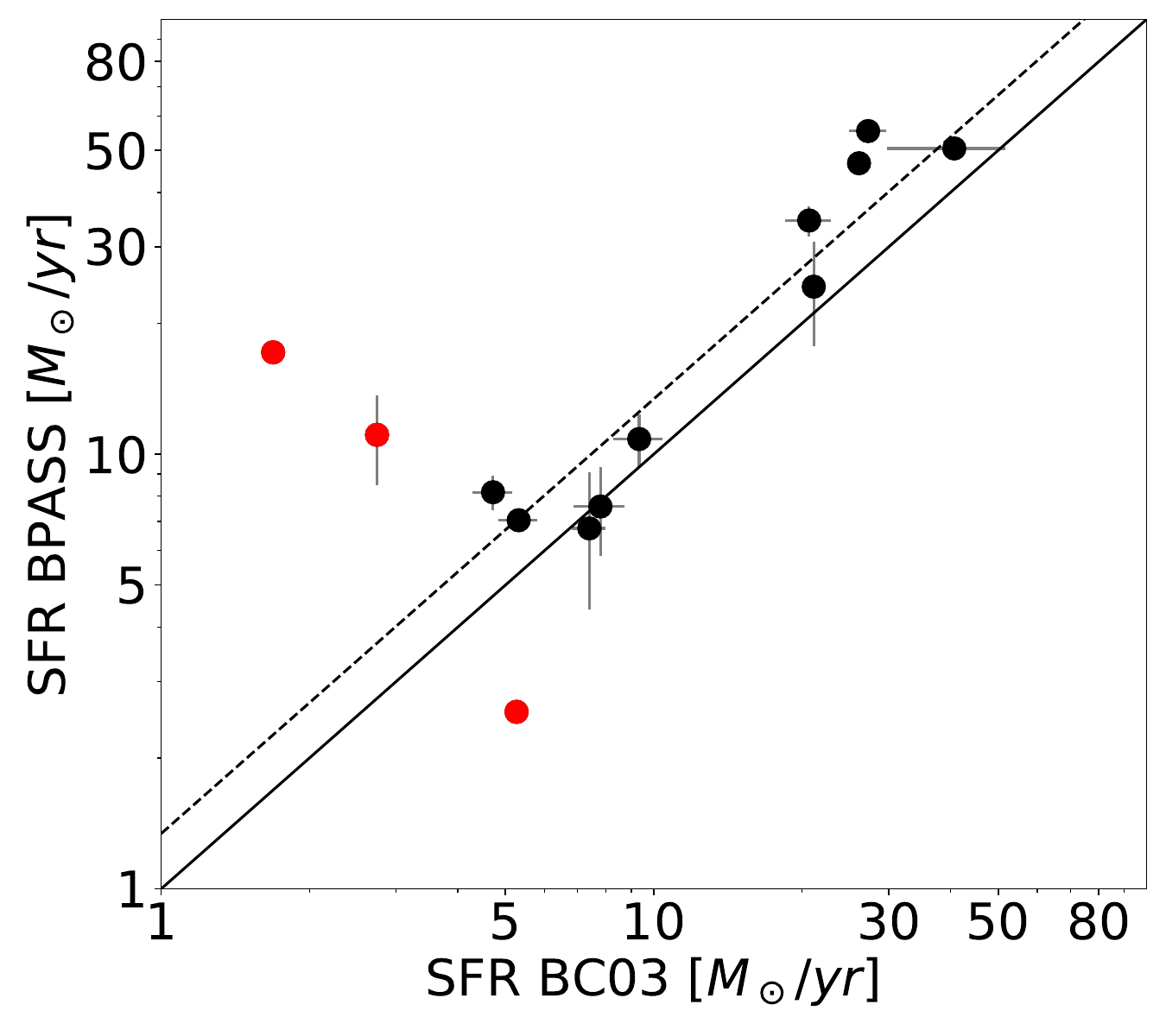}
\end{subfigure}
\begin{subfigure}{0.48\columnwidth}
  \includegraphics[width=\columnwidth]{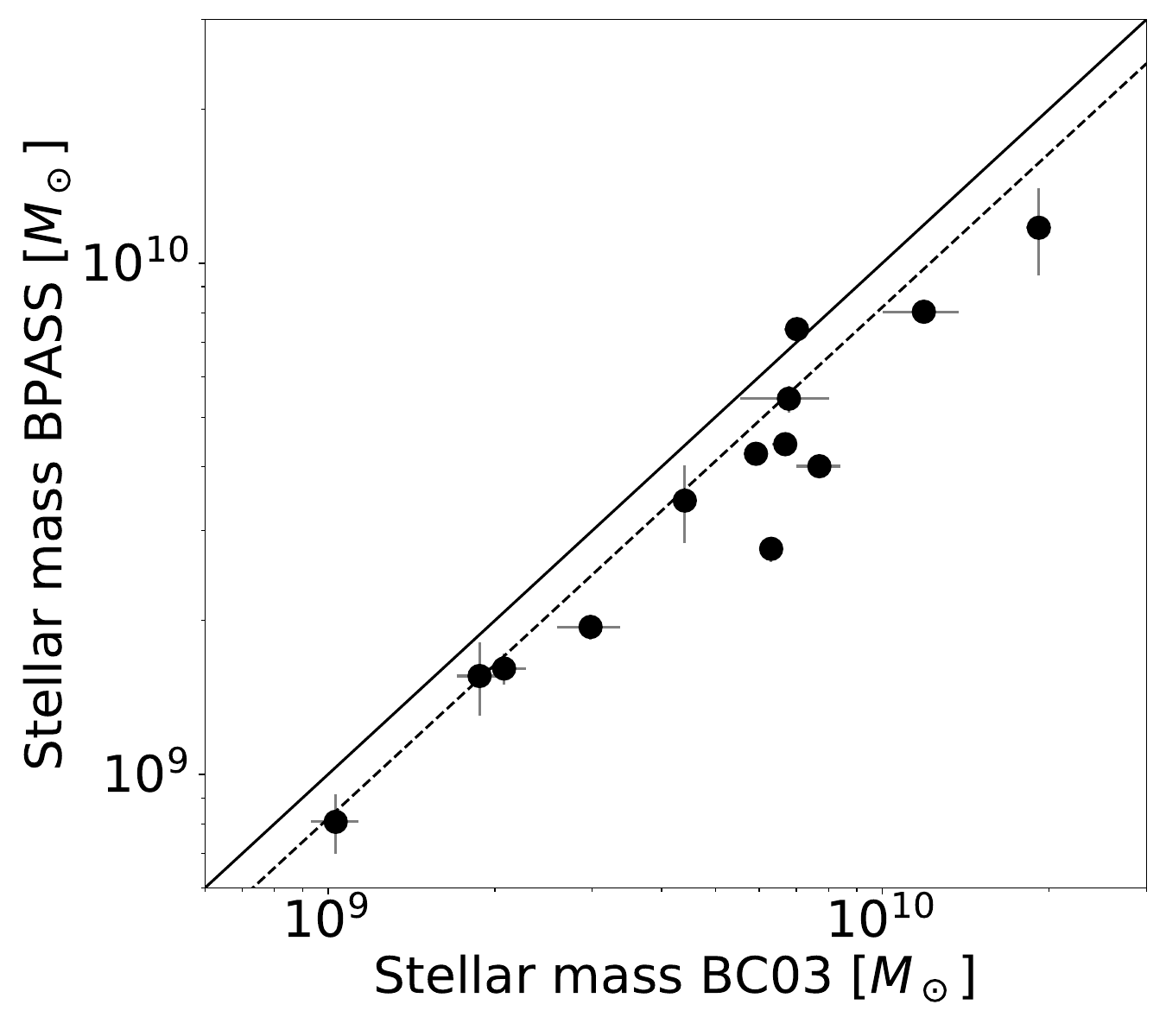}
\end{subfigure}
\caption{Comparison of the SFR (left panel) and stellar mass (right panel) obtained through fitting spectro-photometric data in a run made with BC03 and a run made with BPASS. The solid line represents the 1:1 relation while the dashed line represents the mean deviation of our sample (without the three outliers in the case of the SFR)}.
\label{sfr_mass_bpass}
\end{figure}

Here we compare SFRs and stellar mass estimations obtained with either BPASS or BC03 models. We start with SFRs estimations. Fig \ref{sfr_mass_bpass} shows a clear trend of increased SFRs when using  BPASS  compared to BC03 models.  
Excluding the three outliers (IDs 10293, 8736 and 5409 in red in Fig \ref{sfr_mass_bpass}) the SFRs derived through SED fitting with BPASS are found on average 0.13 dex larger than with BC03. The value of the current SFR is expected to be sensitive to the burst fraction ($f_{burst}$) added to the delayed SFH. The average value of $f_{burst}$ with BPASS is found to be almost double the average $f_{burst}$ value with BC03 (0.15 and 0.08 respectively) which may explain the higher SFRs we get while fitting our sample with the BPASS models. For the three outliers the difference is even higher as one of the two $f_{burst}$ values is found to be null. The two galaxies with a much higher SFR obtained with BPASS are fitted with no burst with BC03 and with a $f_{burst}$ value equal to 0.20 and 0.12  for ID 10293 and ID 8736 respectively with BPASS. Conversely for the galaxy with a  much higher SFR with BC03 (ID 5409): no burst is found with BPASS while a small but positive burst fraction (0.01) is obtained with BC03. All in all  the difference in SFRs obtained with both stellar population models seems to be correlated with the burstiness of the model rather than with an intrinsic difference between single and binary stars. Indeed \cite{eldridge} and \cite{wilkins} found that the SFRs obtained with BPASS are $\sim$ 0.1 dex only lower than those obtained with BC03 when the SFH is fixed.

We find higher stellar masses with  BC03 models as shown in Fig \ref{sfr_mass_bpass}. The mass derived through SED fitting with the BPASS models is on average 0.18 dex lower than with the BC03 models.\cite{osborne} also report lower masses (by 0.17 dex) when using the BPASS models compared to the BC03 models for a sample of CANDELS sources with redshift between 0.7 and 2.3. We attribute these lower masses to the deficit in low mass stars of the BPASS models addressed in Sect \ref{ssec:dust-attenuation-emission}. 


\begin{figure}
\centering
\begin{subfigure}{0.48\columnwidth}
  \includegraphics[width=\columnwidth]{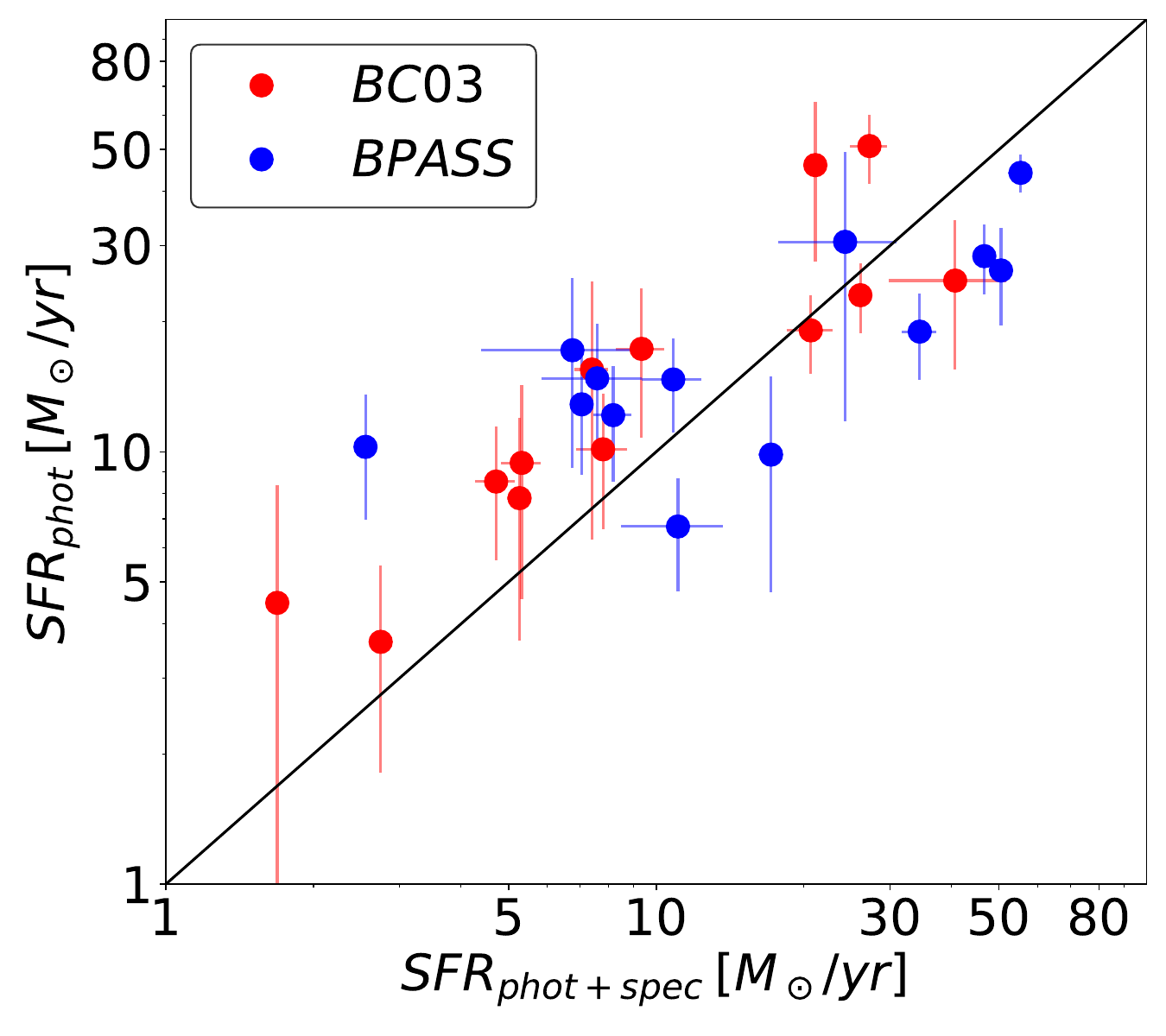}
\end{subfigure}
\begin{subfigure}{0.48\columnwidth}
  \includegraphics[width=\columnwidth]{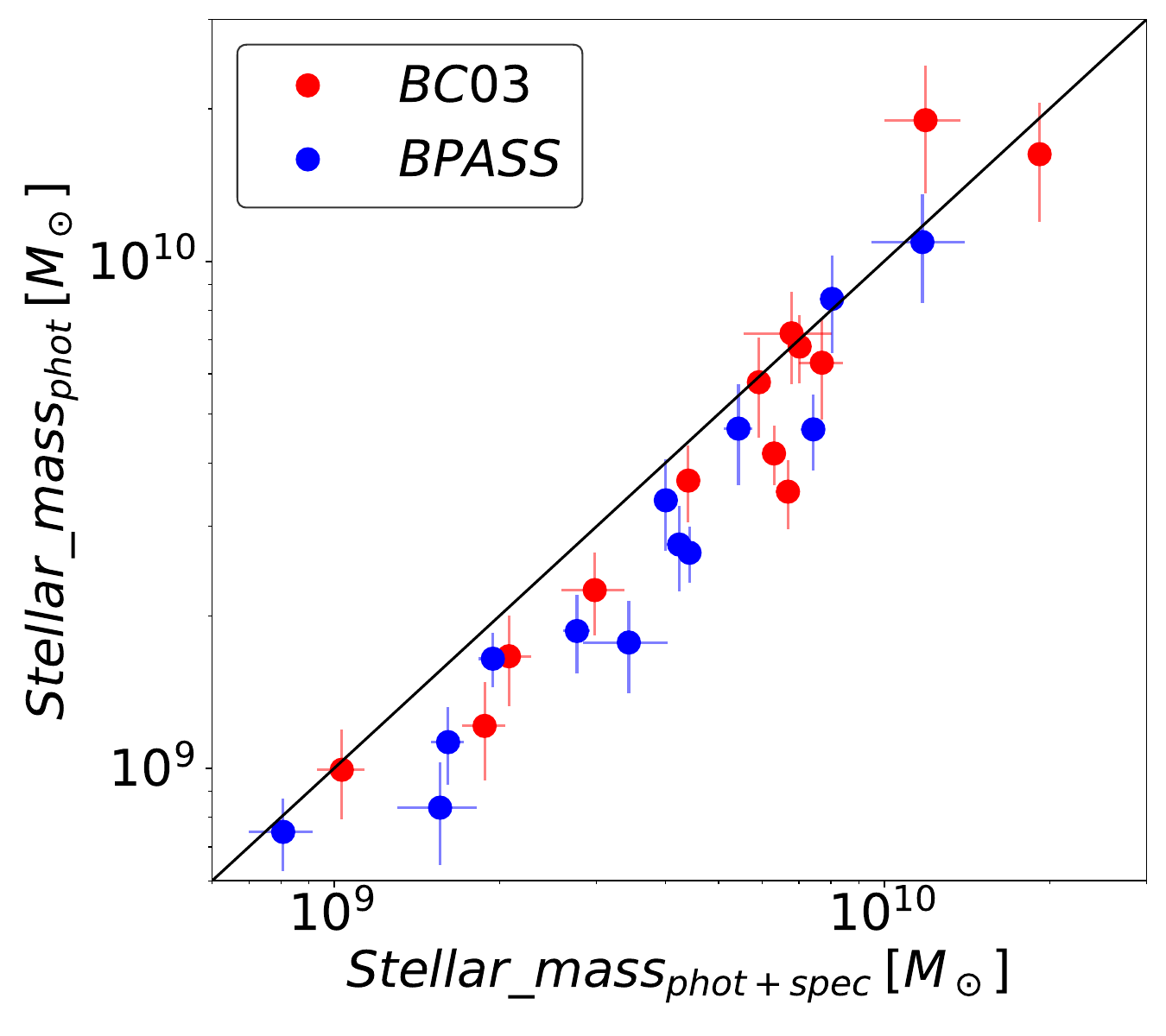}
\end{subfigure}
\caption{Comparison of the SFR (left panel) and stellar mass (right panel) of BC03 and BPASS when fitting photometry only (y-axis) and spectro-photometry (x-axis).}
\label{sfr_mass_cont}
\end{figure}

\subsection{Fits of photometric data or spectro-photometric data \label{ssec:fits_phot}}

Here we assess the impact of using only photometry with both stellar population models on the stellar mass and SFR estimations. In Fig \ref{sfr_mass_cont} we compare the estimates of the SFR obtained with a fit of photometric data to those from a fit of spectro-photometric data with both BC03 and BPASS. The SFR  derived with the fit of photometric data only is on average 0.16 dex higher with BC03 and 0.03 dex higher with BPASS compared to values  obtained with spectro-photometric measurements. Considering the scatter found in this comparison we also provide the maximal and minimal absolute value of the difference in $log(SFR)$. For BC03 (BPASS) the largest discrepancy is of 0.42 (0.60) dex and the lowest is of 0.04 (0.09) dex. Errors on the SFRs also become much lower when fitting the spectro-photometric data. SFRs lower than 30 M$_\odot$/yr tend to be systematically overestimated when only photometric fluxes are fitted by 0.18 dex with BC03 and by 0.09 dex with BPASS. We attribute this increase to the stronger emission lines computed by the code in the absence of spectroscopic data as shown in Fig \ref{EWs_fit_cont}  which are due to a more active star formation process. Our results are in agreement with \cite{pacifici15} who also find a higher SFR (by 0.13 dex) while using BC03 and photometric data only for a sample of 364 3D-HST galaxies at z < 3. 

\begin{figure}
\centering
\begin{subfigure}{0.49\columnwidth}
  \includegraphics[width=\columnwidth]{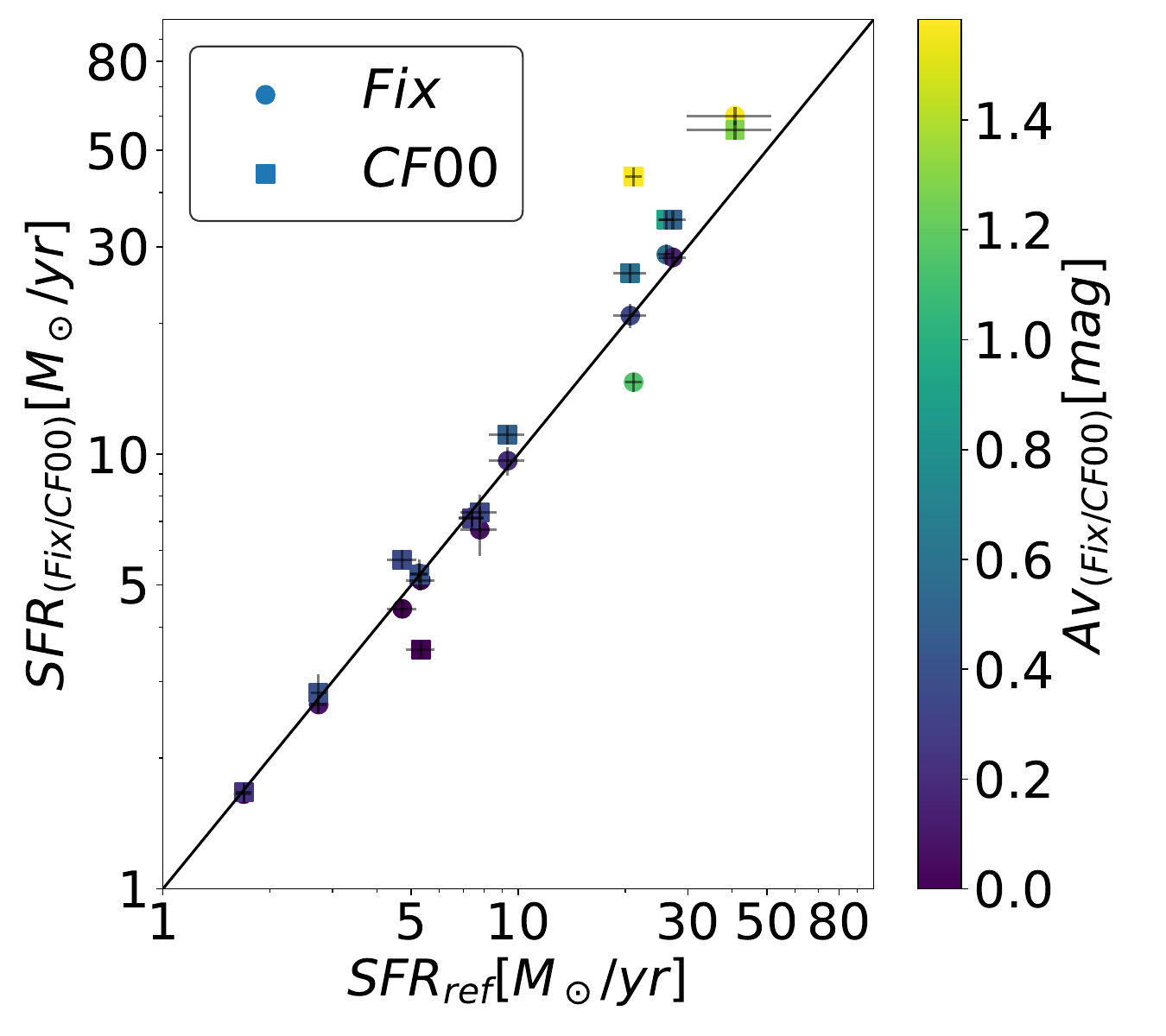}
  \label{sfr_att}
\end{subfigure}
\begin{subfigure}{0.49\columnwidth}
  \includegraphics[width=\columnwidth]{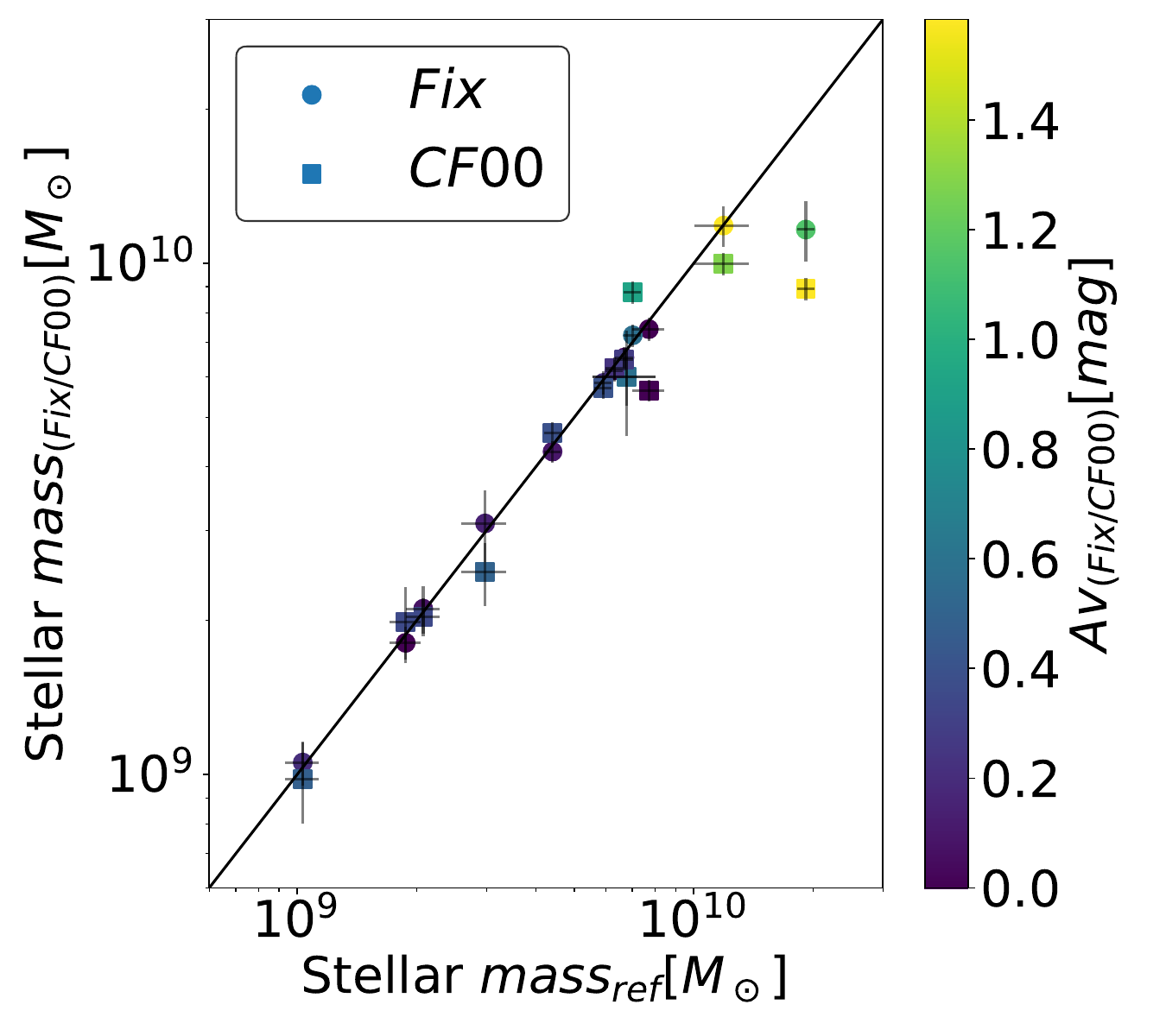}
  \label{mass_att}
\end{subfigure}
\caption{Comparison of the SFR (left panel) and stellar mass (right panel) obtained with a fit in which both exponents are fixed to -0.96 (dots) and a fit following the original CF00 recipe (squares) with respect to the one we get while using the parameters presented in Sect \ref{sec:SED-fitting}.}
\label{param_att}
\end{figure}

The  stellar masses are found larger when EWs are added to the fit by $\sim$ 0.15 dex regardless of the chosen stellar populations as shown in Fig \ref{sfr_mass_cont}. The lower masses obtained with only  photometric data  are due to  an overestimation of the line contribution (see Fig \ref{EWs_fit_cont}): the contribution of young stars to the SED is higher leading to a lower contribution of older stars which make the bulk of the galaxy mass. Our results are in agreement with the findings of \cite{schaerer10} who fits z $\sim$ 7 sources while \cite{pacifici15} finds no noticeable difference in the stellar mass output with the addition of spectroscopic data to the fits for their sample of z < 3 galaxies. 

These findings highlight the importance of considering both spectroscopic and photometric data in SED fitting to obtain accurate estimates of star formation rates and stellar masses, particularly for galaxies with lower star formation rates. They also show the impact of emission line contributions on parameter estimation emphasising that care needs to be taken when considering the results for sources with strong emission lines.


\subsection{Fits with different attenuation recipes\label{ssec:fits_att}}

Dust attenuation has a direct impact on the derived SFRs and to a lesser extent on stellar masses. Any change in the shape of the attenuation curve can affect (albeit in varying degrees) the estimations of these quantities as shown in \cite[e.g.][]{schaerer10, pacifici15, lofaro, buat18, buat19}. In the following analysis we compare the effects of changing the attenuation curve parameters in the framework of the CF00 recipe. 

We test three different attenuation curves with separate runs using the BC03 models only and the full spctro-photometric dataset. We do not include the BPASS models in these tests as the attenuation curves presented in Sect.~\ref{ssec:effective-curves} cannot be reconstructed with the binary stellar population models. The first run called $Run_{ref}$  is performed with the four parameters of the modified CF00 recipe taken free as presented in Table \ref{SED_param}; it will be considered the reference run.
The second run is performed with the original CF00 recipe  ($Run_{CF00}$) in which $\mu$ is fixed to 0.3 and both exponents are equal to -0.7 while the third one makes use of the single average power-law introduced in Sect.~\ref{ssec:effective-curves}; $\mu$ is  taken free and we call this run $Run_{fix}$.

As already noted in Sect.~\ref{ssec:effective-curves} we do not find noticeable changes in $\mu$ between $Run_{ref}$ and $Run_{fix}$. The SFRs are found consistent with a slightly larger scatter with CF00 than with the single power-law as shown in Fig \ref{param_att}.
The agreement between the three stellar mass estimations is very good with a scatter of 0.03 and 0.06 dex for $Run_{fix}$  and $Run_{CF00}$  respectively when they are compared to $Run_{ref}$ (see Fig \ref{param_att}).


From the results shown in this section we conclude that fixing the slope of the attenuation curve (and even $\mu$) is much less impactful on the SFRs and stellar masses than including binary stellar populations or fitting only photometric data.  However this lesser effect could be a result of our sample being mostly composed of objects with low dust content. Indeed the most discrepant objects in Fig \ref{param_att} have an amount of dust $A_v$ > 1 mag.

\section{Conclusions}

In this work we assessed the robustness of the estimates of the SFRs, stellar masses and dust attenuation parameters of 13 sources in the CEERS field observed with HST, JWST/NIRCam and NIRSpec between redshifts 1 and 3. These objects are selected based on the availability of at least one Balmer and one Paschen line with S/N $\geq$ 5 to allow for a solid determination of the amount of differential attenuation at play in our sample. The aforementioned physical parameters are computed using a newly developed version of CIGALE which combines photometric fluxes and emission line EWs as input data. We summarise our main results as follows: 


\begin{itemize}
    \item The absorption line EWs of seven hydrogen lines from Pa$\alpha$ (up to 2 $\AA$) to H$\delta$ (up to 7 $\AA$) are measured. This is the first time an absorption EW for the Pa$\alpha$ line is measured and a steady trend of decreasing EWs with longer wavelengths is found. 
    
    \item The photometric data is better fitted with BC03 compared to BPASS while the quality is similar for EWs. Meanwhile the SED fitting code overestimates the EWs when fitting photometry only (by an average of $\sim$ 15\%) an effect we find to be even more pronounced when using the BPASS models (average of $\sim$ 45\%).

    \item We find higher SFRs (0.13 dex) and lower stellar masses (0.17 dex) with the fit of spectro-photometric data using BPASS compared to BC03. There is an overestimation of the SFRs (0.16 dex with BC03 and 0.03 with BPASS) and an underestimation of stellar masses (0.15 dex) when fitting only photometry compared to fitting spectro-photometric data.
    
    \item From the best model computed by CIGALE for each source we assess the contribution of the H$\alpha$ line emission to the measurements made with the NIRCam filters (F150W or F200W depending on the redshift). We find H$\alpha$ to be a major contributor to the broadband photometric fluxes (average of 36\% and up to 68\%) even more so while using the BPASS models (average of 42\% and up to 76\%).
    
    \item We find that using a double power law as introduced by the CF00 recipe to model the dust attenuation allows for solid constraints on the differential attenuation.  The attenuation parameters derived for a subsample of sources with $A_v$ > 0.3 mag are on average consistent with the CF00 recipe, especially the differential attenuation ($\mu$=0.37 $\pm$ 0.08). The change of flexibility in the dust attenuation recipe has no effect on SFRs and stellar masses but this could be different for more attenuated objects. We reconstruct the effective attenuation curves of our subsample and derive a unique power law having an exponent of -0.96. 
    
\end{itemize}

\begin{acknowledgements}

Lise-Marie Seillé gratefully acknowledges support
from the PNCG; the Programme Investissements d’Avenir; and the Initiative d’excellence A*Midex (doctoral grant AMX-21-HAN-01).

\end{acknowledgements}

%
%
\bibliographystyle{aa}
\bibliography{main.bib}

\onecolumn
\appendix

\section{Showcasing the sample SEDs\label{sec:appendix}}

\begin{figure*}[hbt!]
\begin{subfigure}{.33\linewidth}
  \centering
  \includegraphics[width=1.1\textwidth]{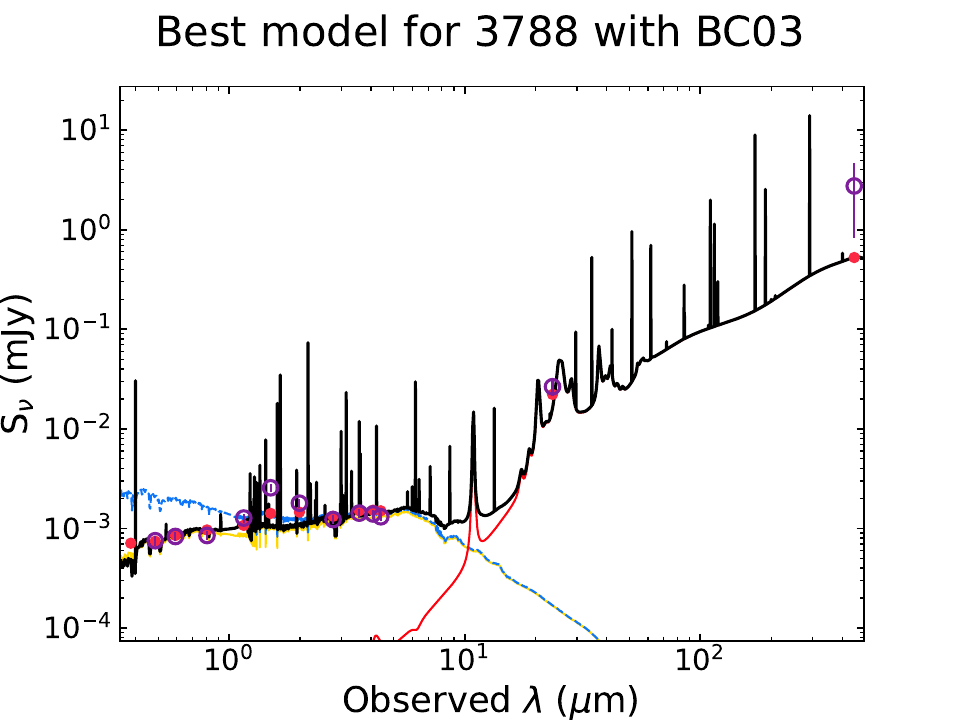}
\end{subfigure}%
\begin{subfigure}{.33\linewidth}
  \centering
  \includegraphics[width=1.1\textwidth]{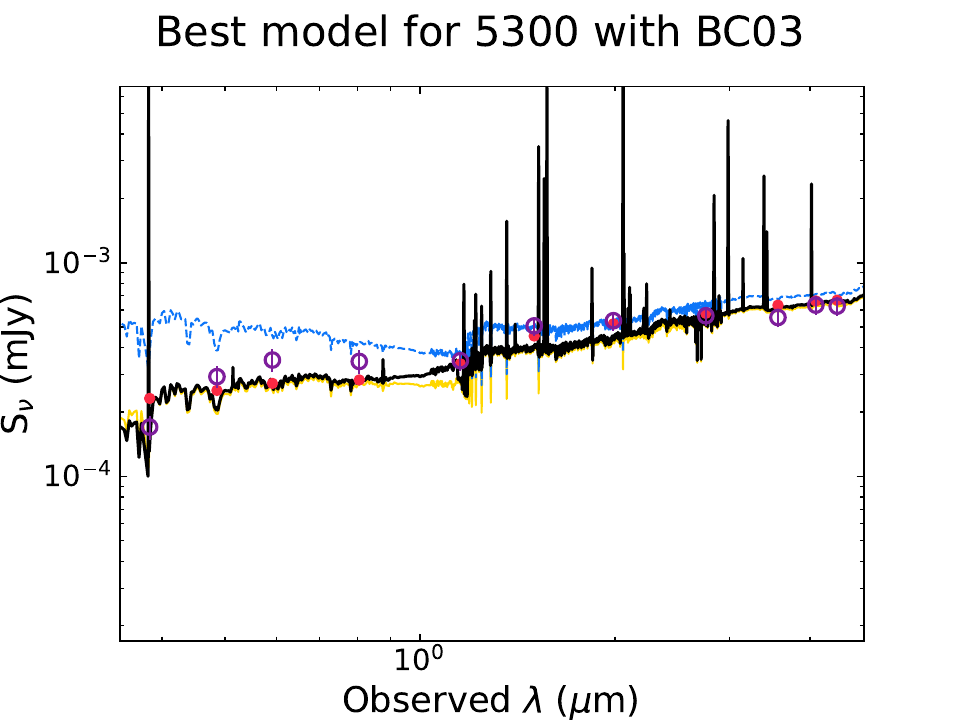}
\end{subfigure}
\begin{subfigure}{.33\linewidth}
  \centering
  \includegraphics[width=1.1\textwidth]{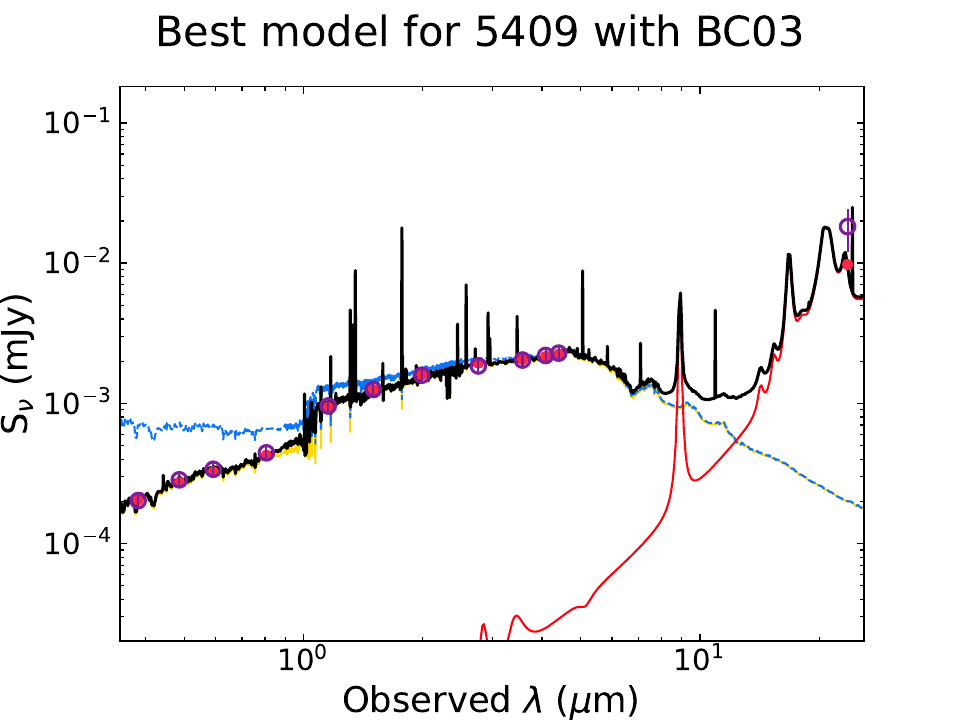}
\end{subfigure}
\begin{subfigure}{.33\linewidth}
  \centering
  \includegraphics[width=1.1\textwidth]{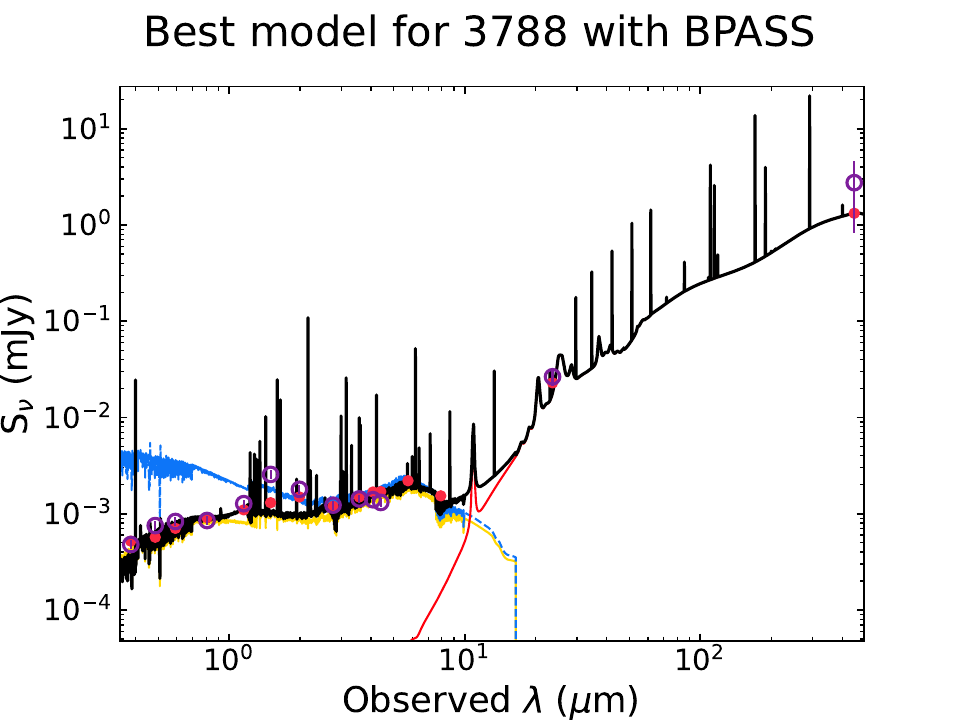}
\end{subfigure}%
\begin{subfigure}{.33\linewidth}
  \centering
  \includegraphics[width=1.1\textwidth]{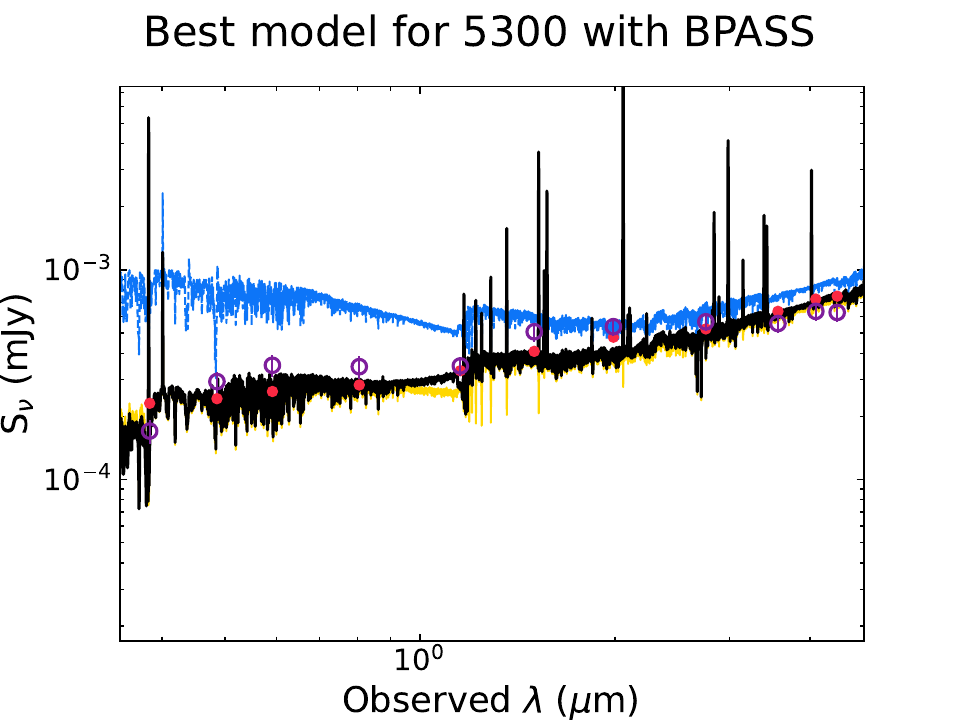}
\end{subfigure}
\begin{subfigure}{.33\linewidth}
  \centering
  \includegraphics[width=1.1\textwidth]{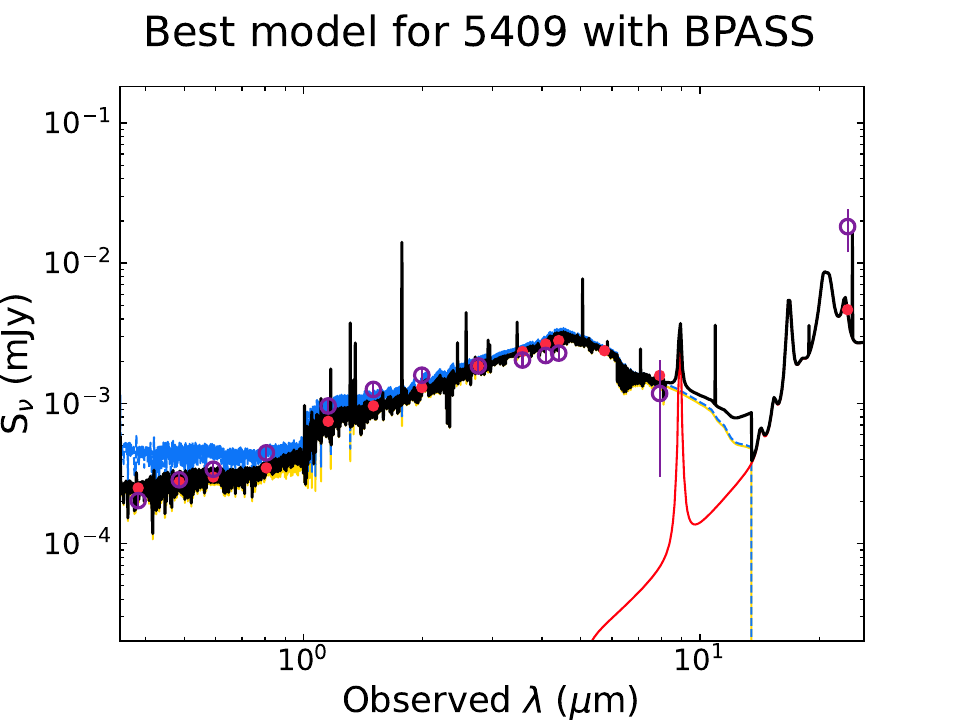}
\end{subfigure}
\begin{subfigure}{.33\linewidth}
  \centering
  \includegraphics[width=1.1\textwidth]{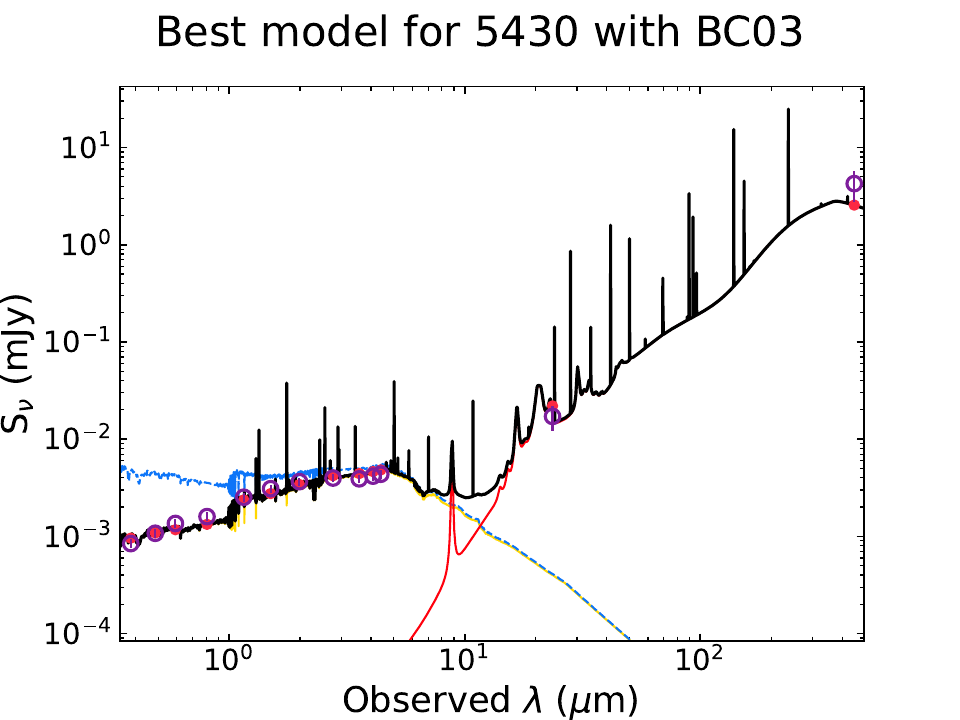}
\end{subfigure}%
\begin{subfigure}{.33\linewidth}
  \centering
  \includegraphics[width=1.1\textwidth]{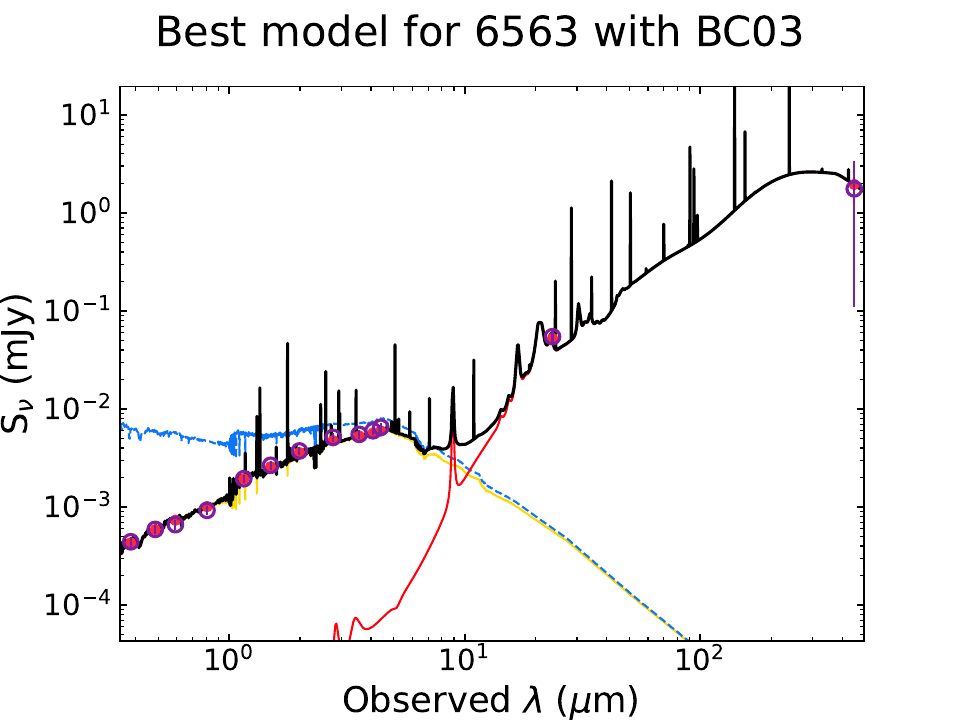}
\end{subfigure}
\begin{subfigure}{.33\linewidth}
  \centering
  \includegraphics[width=1.1\textwidth]{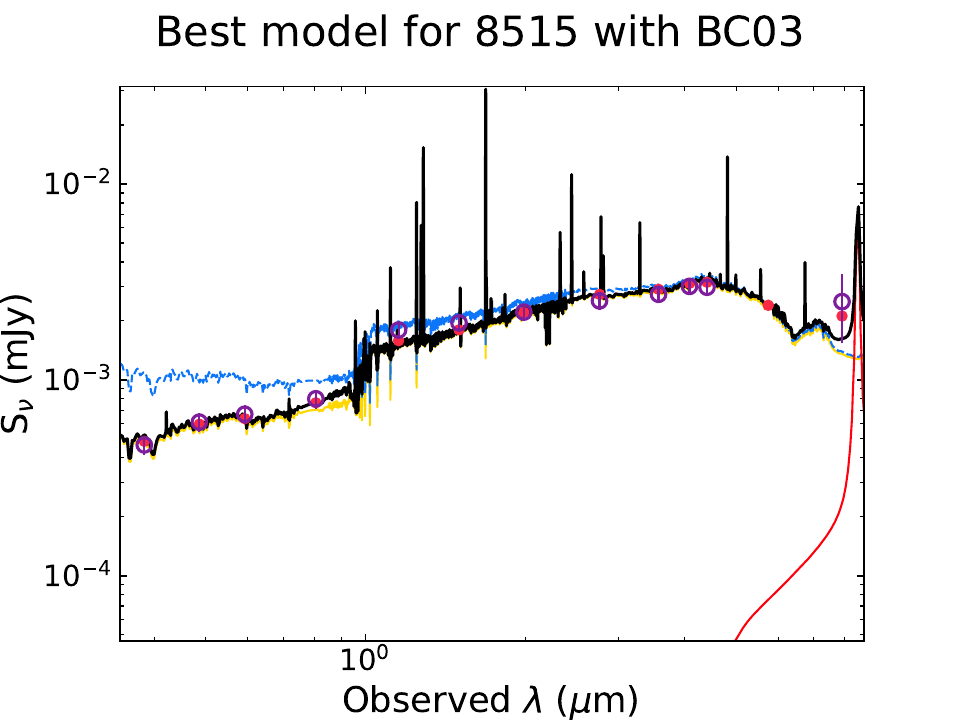}
\end{subfigure}
\begin{subfigure}{.33\linewidth}
  \centering
  \includegraphics[width=1.1\textwidth]{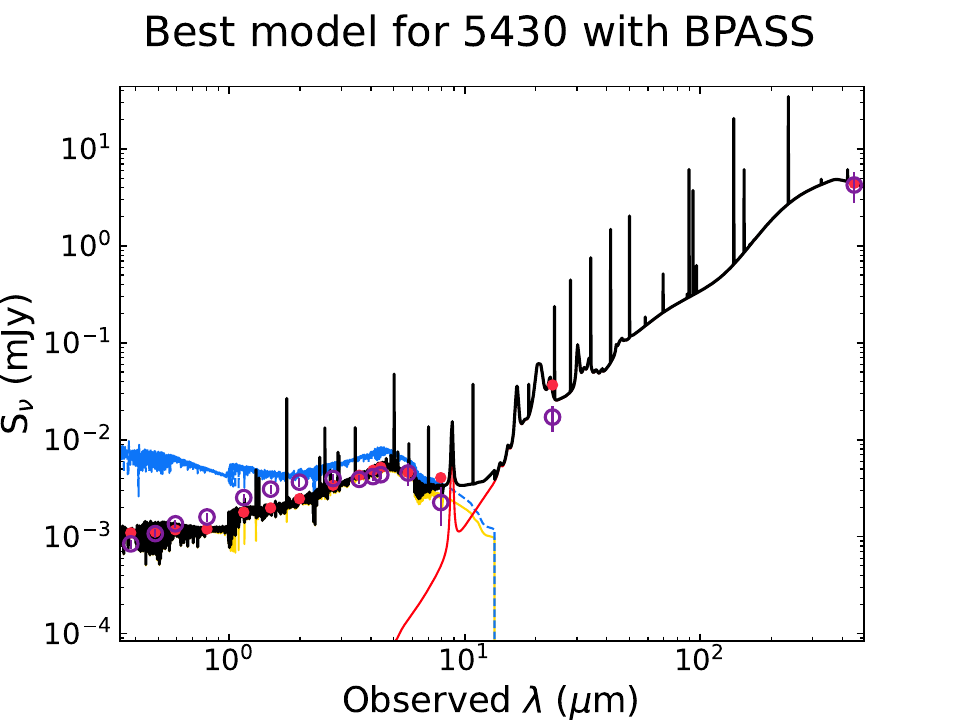}
\end{subfigure}%
\begin{subfigure}{.33\linewidth}
  \centering
  \includegraphics[width=1.1\textwidth]{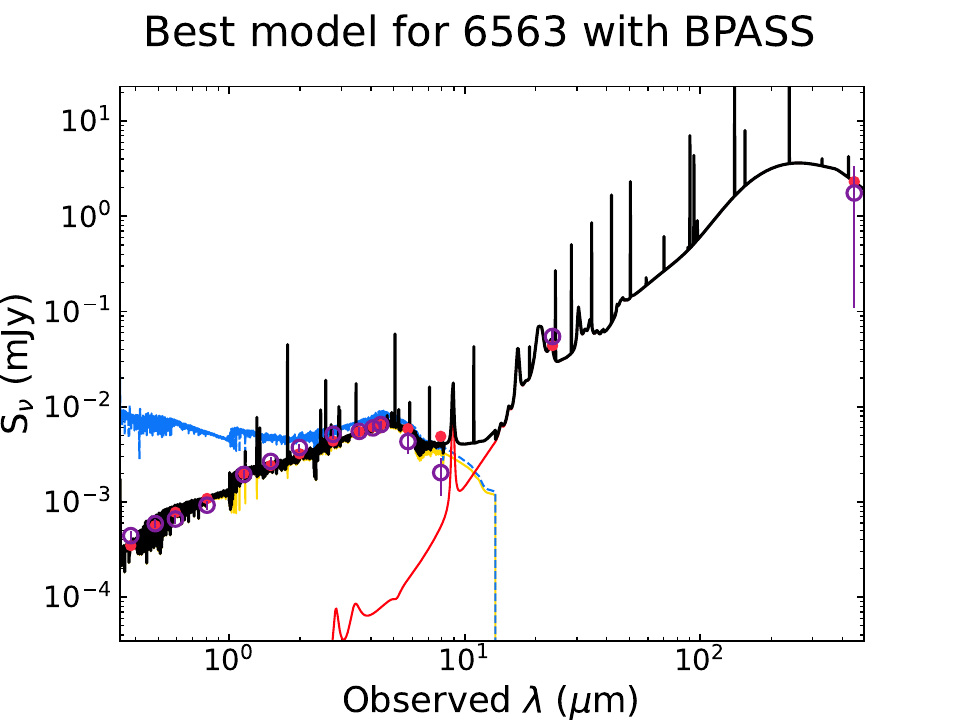}
\end{subfigure}
\begin{subfigure}{.33\linewidth}
  \centering
  \includegraphics[width=1.1\textwidth]{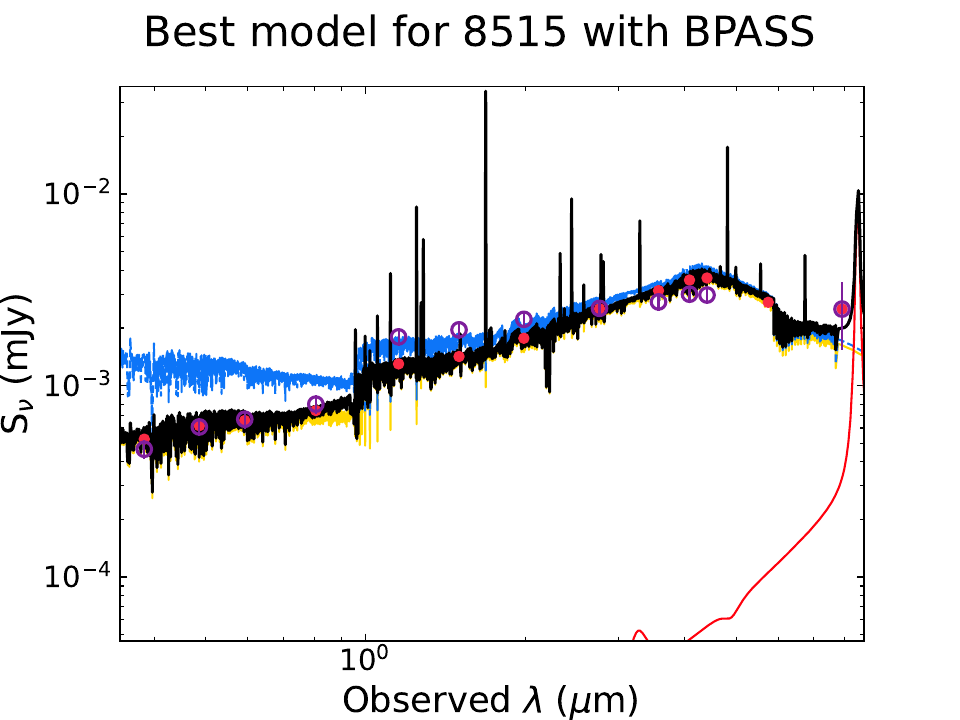}
\end{subfigure}
\caption{SEDs of the spectro-photometric data with the BC03 and BPASS models. The legend is identical to that of Fig \ref{SEDs_5430}.}
\end{figure*}

\begin{figure*}\ContinuedFloat
\centering
\begin{subfigure}{.33\linewidth}
  \centering
  \includegraphics[width=1.1\textwidth]{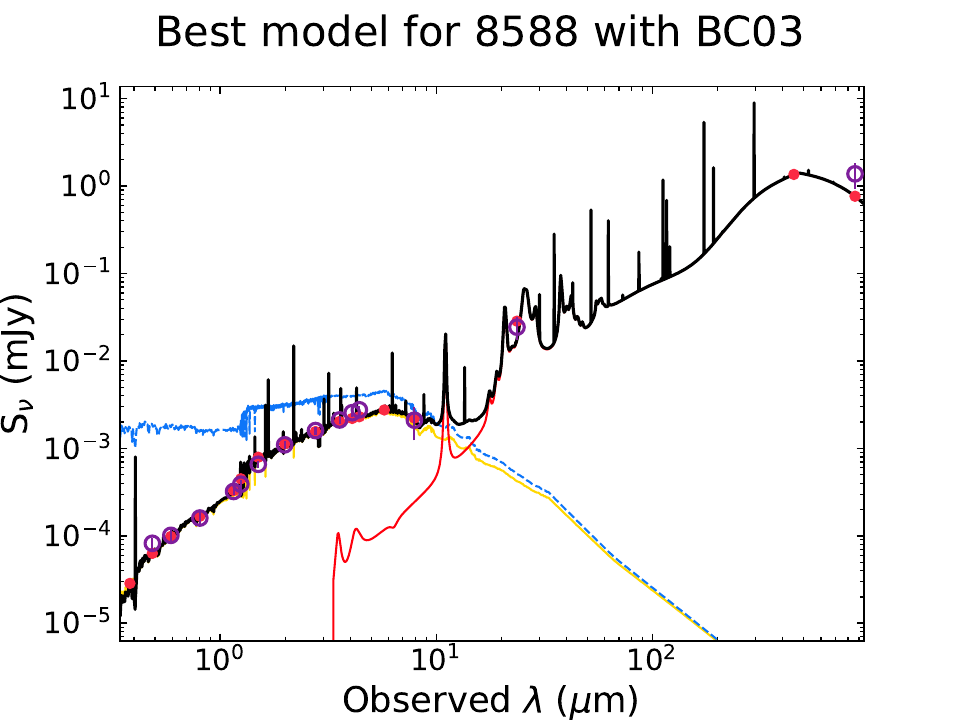}
\end{subfigure}%
\begin{subfigure}{.33\linewidth}
  \centering
  \includegraphics[width=1.1\textwidth]{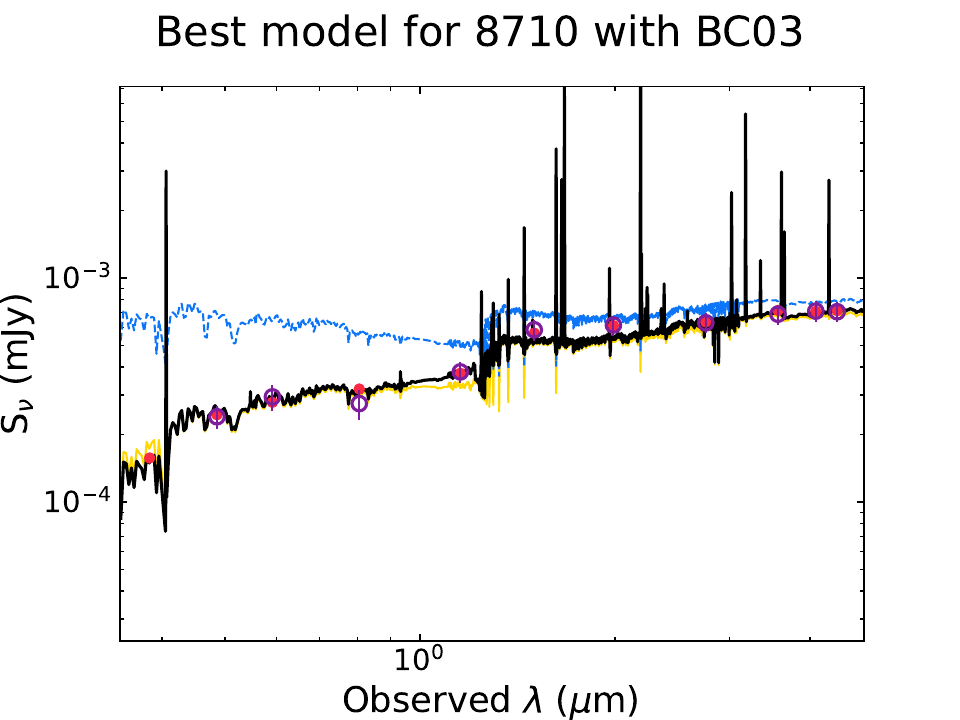}
\end{subfigure}
\begin{subfigure}{.33\linewidth}
  \centering
  \includegraphics[width=1.1\textwidth]{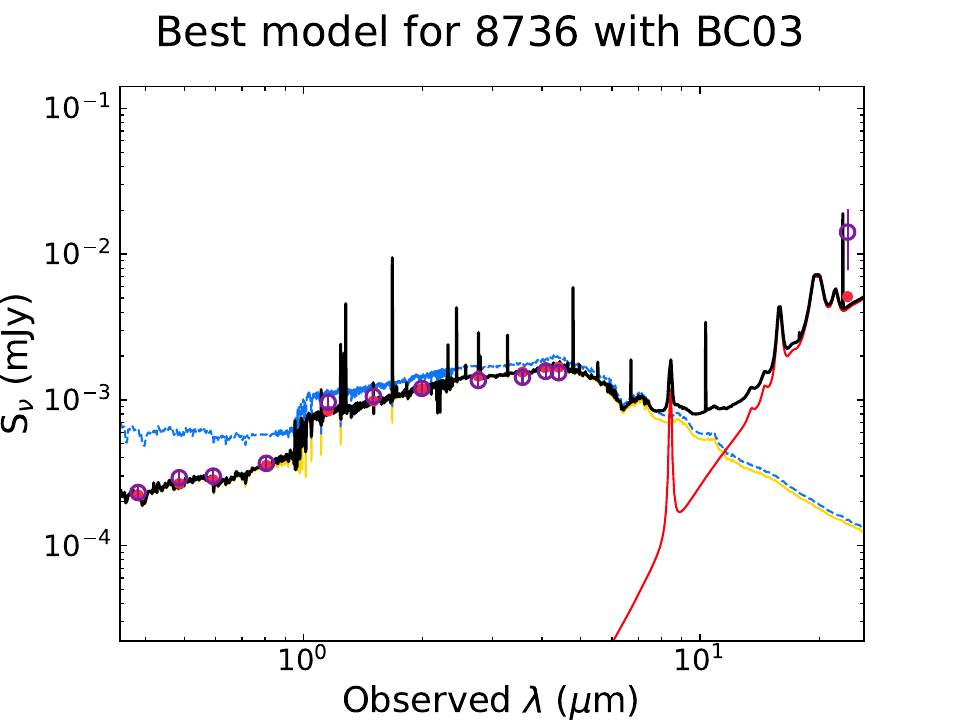}
\end{subfigure}
\begin{subfigure}{.33\linewidth}
  \centering
  \includegraphics[width=1.1\textwidth]{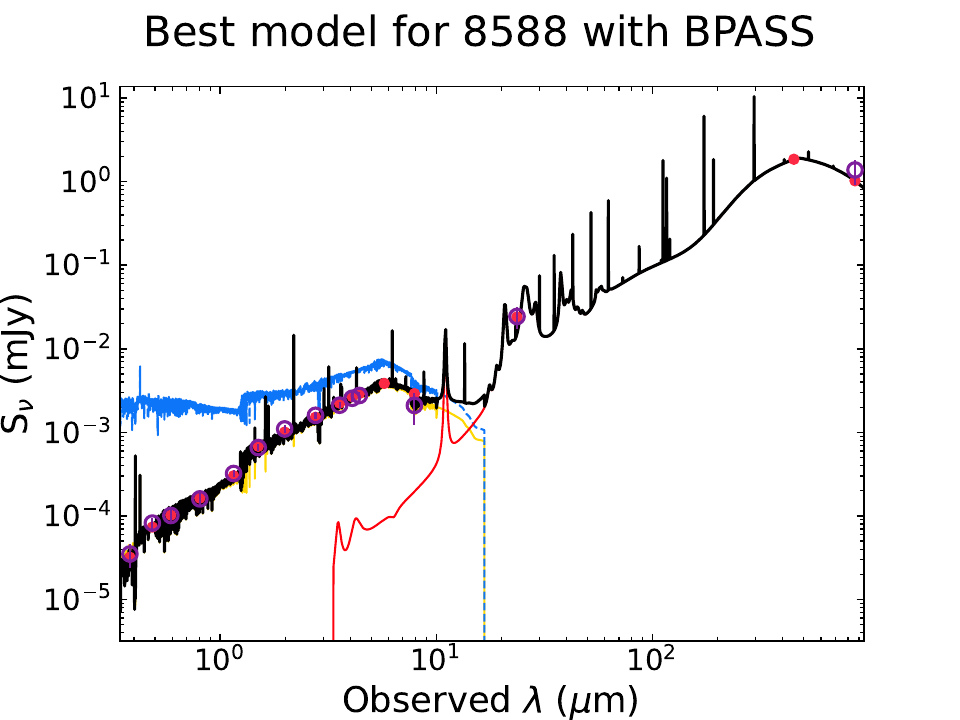}
\end{subfigure}%
\begin{subfigure}{.33\linewidth}
  \centering
  \includegraphics[width=1.1\textwidth]{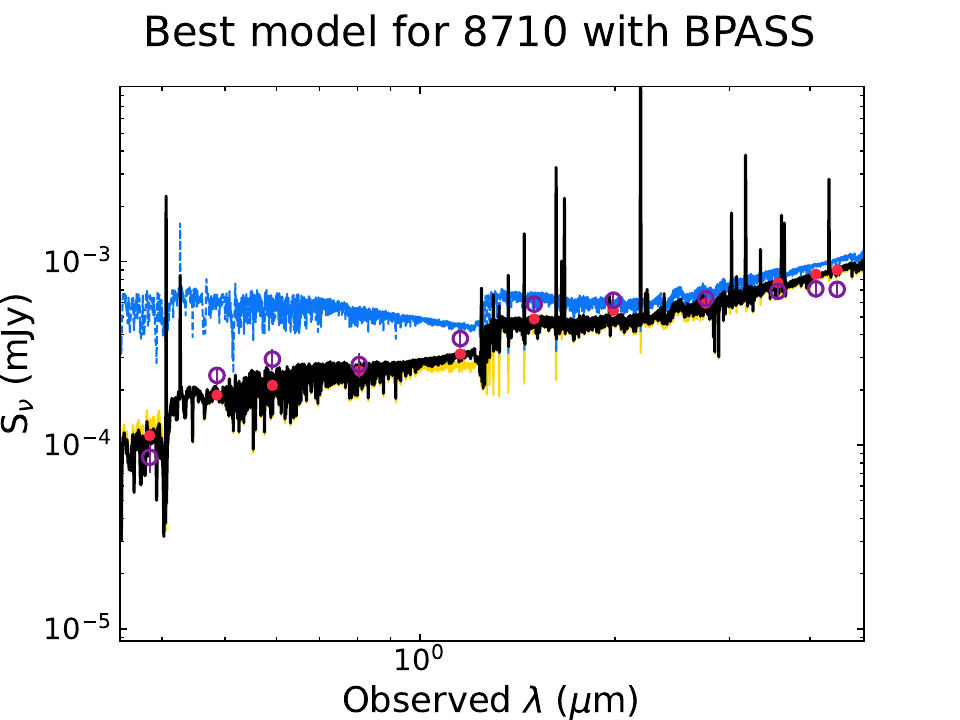}
\end{subfigure}
\begin{subfigure}{.33\linewidth}
  \centering
  \includegraphics[width=1.1\textwidth]{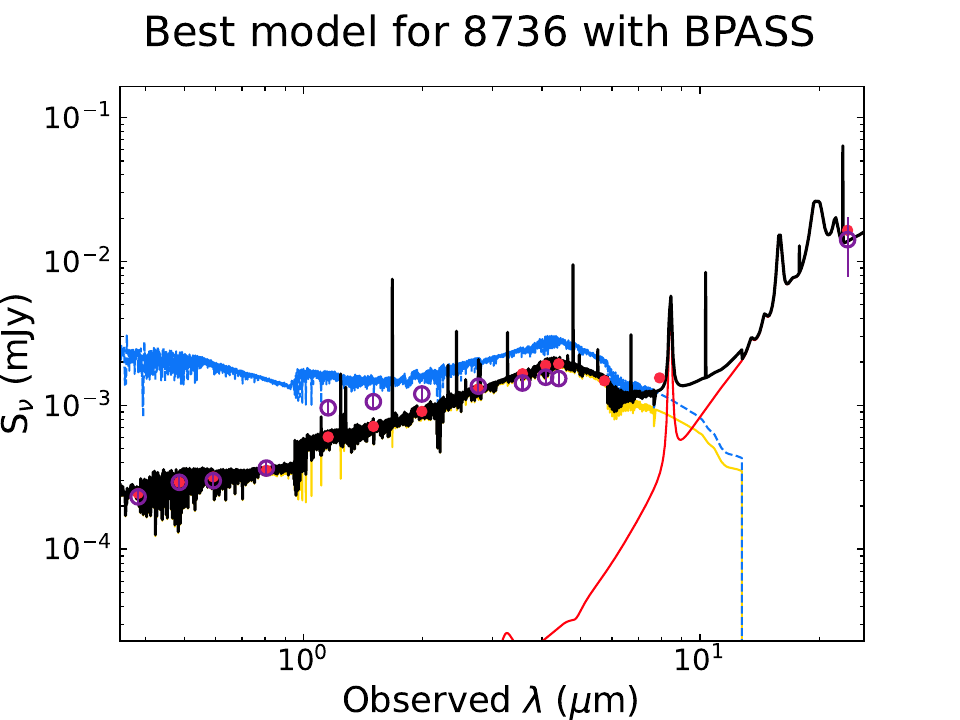}
\end{subfigure}
\begin{subfigure}{.33\linewidth}
  \centering
  \includegraphics[width=1.1\textwidth]{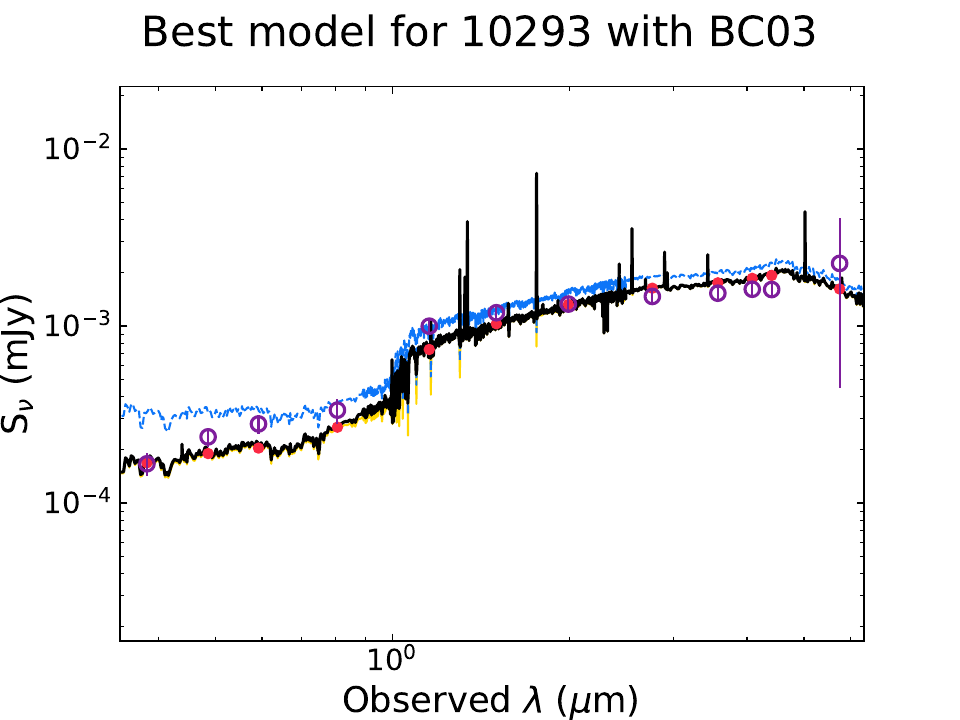}
\end{subfigure}%
\begin{subfigure}{.33\linewidth}
  \centering
  \includegraphics[width=1.1\textwidth]{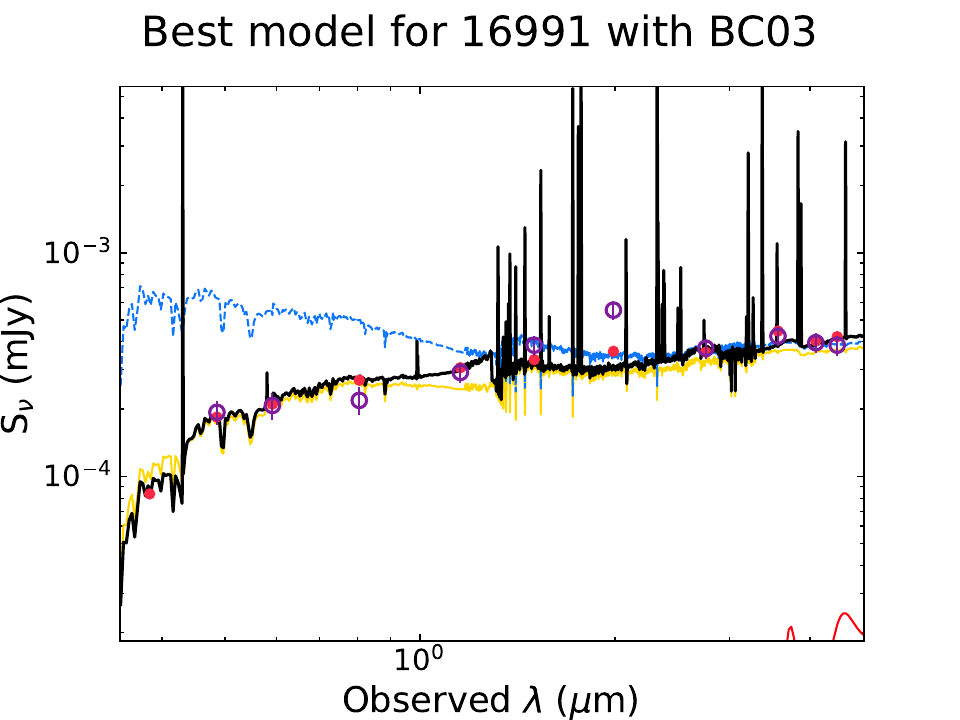}
\end{subfigure}
\begin{subfigure}{.33\linewidth}
  \centering
  \includegraphics[width=1.1\textwidth]{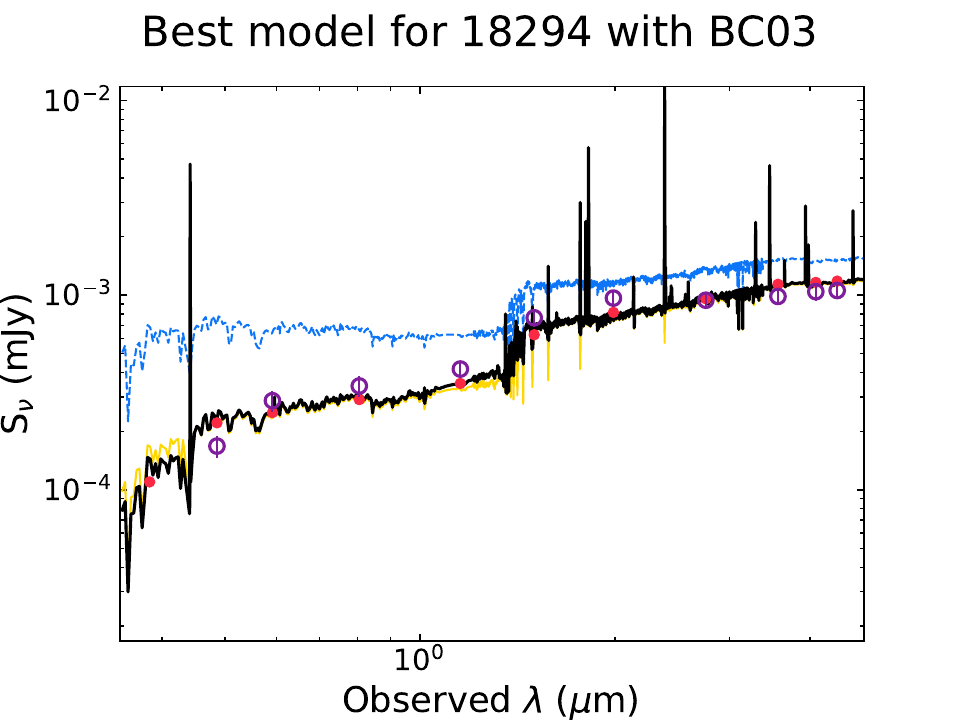}
\end{subfigure}
\begin{subfigure}{.33\linewidth}
  \centering
  \includegraphics[width=1.1\textwidth]{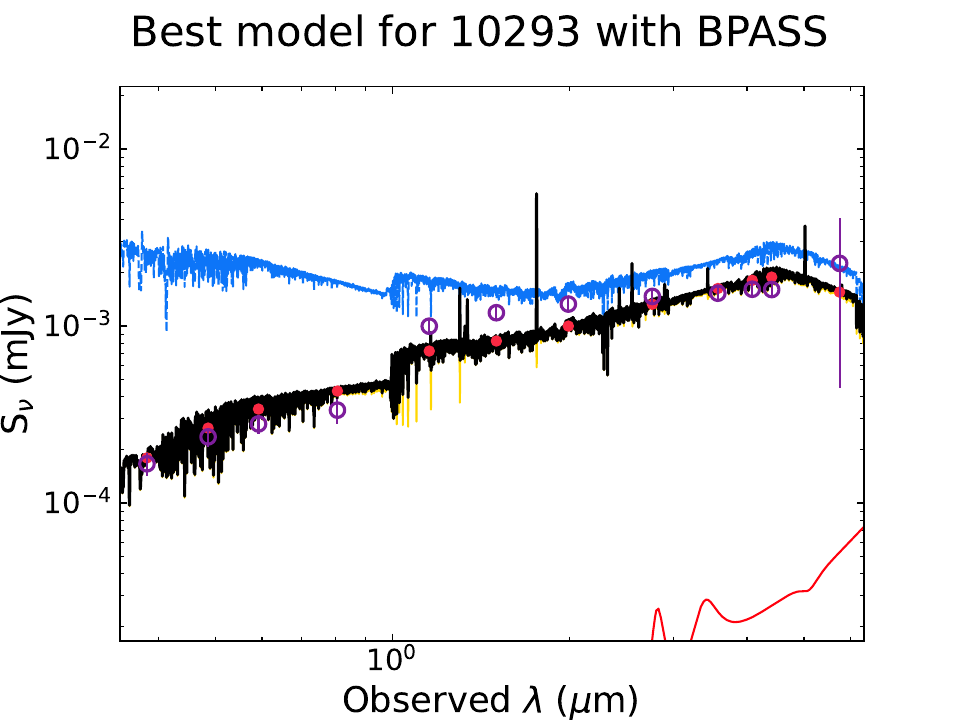}
\end{subfigure}%
\begin{subfigure}{.33\linewidth}
  \centering
  \includegraphics[width=1.1\textwidth]{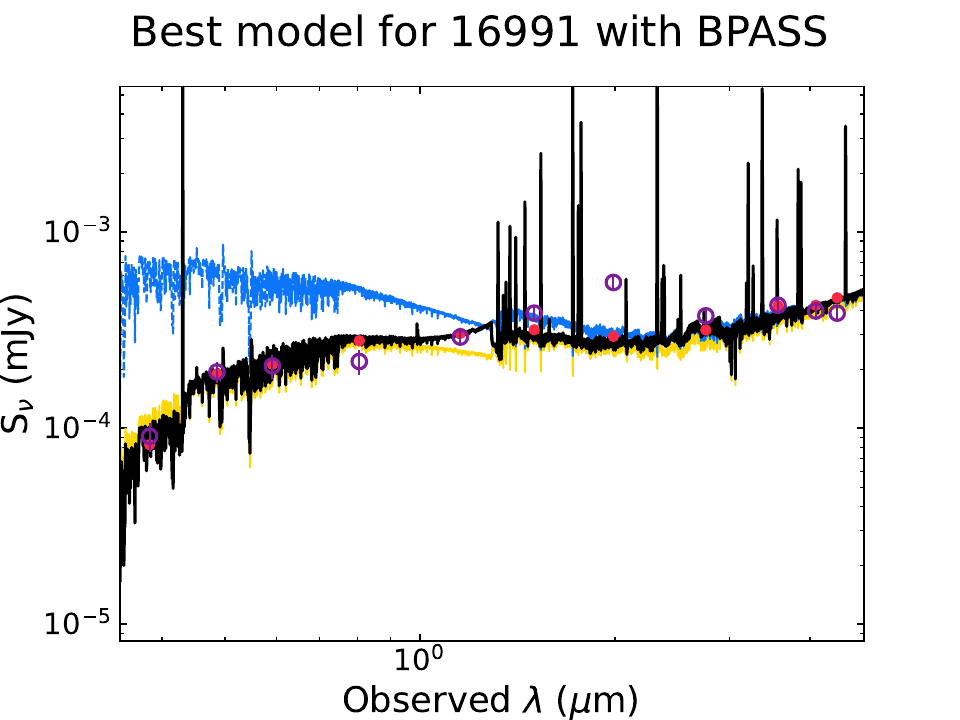}
\end{subfigure}
\begin{subfigure}{.33\linewidth}
  \centering
  \includegraphics[width=1.1\textwidth]{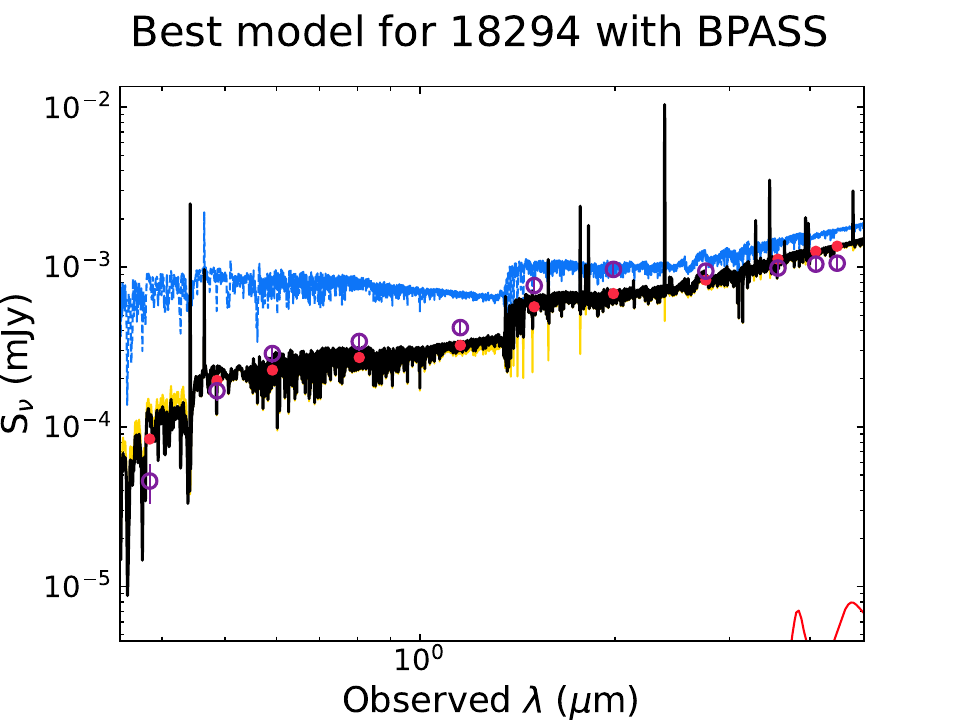}
\end{subfigure}
\caption{continued.}
\end{figure*}

\begin{figure*}\ContinuedFloat
\centering
\begin{subfigure}{.33\linewidth}
  \centering
  \includegraphics[width=1.1\textwidth]{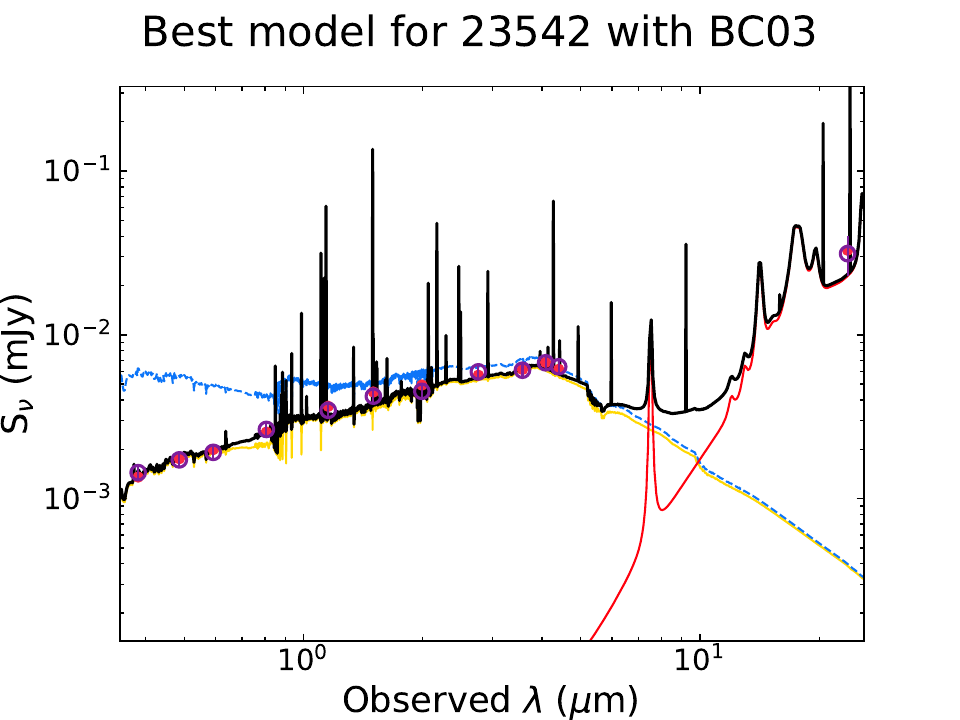}
\end{subfigure}%
\begin{subfigure}{.33\linewidth}
  \centering
  \includegraphics[width=1.1\textwidth]{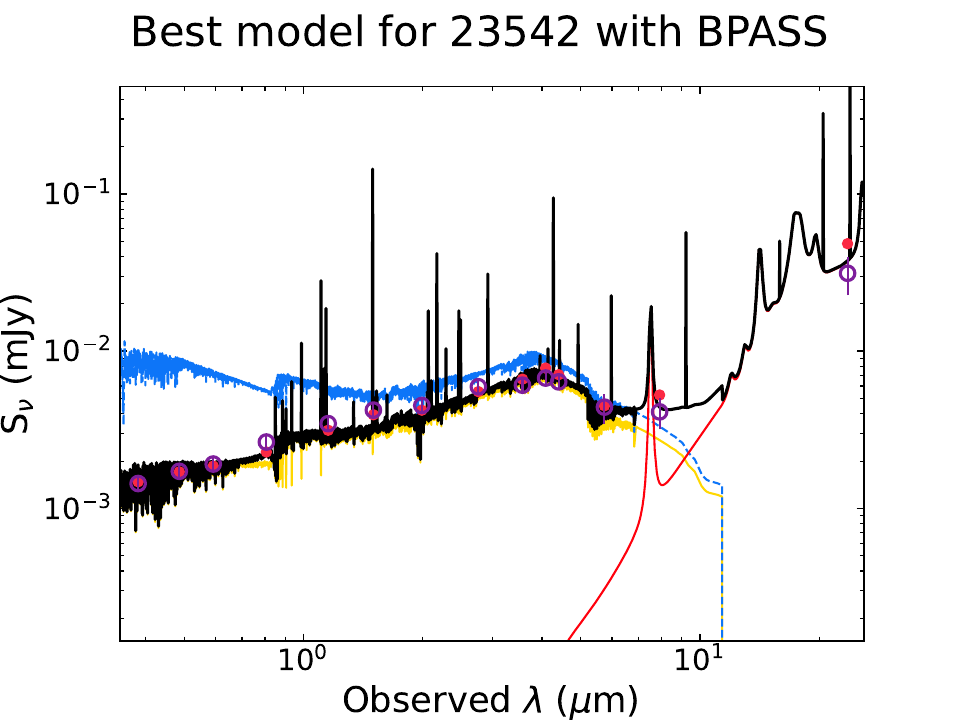}
\end{subfigure}
\caption{continued.}
\end{figure*}

\end{document}